\documentclass[useAMS,usenatbib]{mn2e}
\usepackage{graphicx,subfigure,color,afterpage}
\usepackage[total={17.8cm,24.0cm},centering]{geometry}
\usepackage{times}

\def\atlas{{{ATLAS}}$^{\rm 3D}$}
\def\kms{km s$^{-1}$}

\def\arcsec{$^{\prime \prime}$}
\definecolor{Mygrey}{gray}{0.75}

\newcommand{\gtsimeq}{\raisebox{-0.6ex}{$\,\stackrel{\raisebox{-.2ex}{$\textstyle >$}}{\sim}\,$}}
\newcommand{\farc}{\mbox{\ensuremath{.\!\!^{\prime\prime}}}}

\usepackage[compact]{titlesec}
\titlespacing{\section}{0pt}{*2}{*1}

\title[Molecular tracers of metallicity]{ISM chemistry in metal rich environments: molecular tracers of metallicity} 
\author[Timothy A. Davis et al.]{Timothy A. Davis$^{1}$\thanks{E-mail:\texttt{tdavis@eso.org}}, Estelle Bayet$^{2}$, Alison Crocker$^{3}$, Sel\c{c}uk Topal$^{2}$ and Martin Bureau$^{2}$
 \vspace{0.4cm}\\
\parbox{\textwidth}{$^{1}$European Southern Observatory, Karl-Schwarzschild-Str. 2, 85748 Garching, Germany\\
$^{2}$Sub-Dept. of Astrophysics, Dept. of Physics, University of Oxford, Denys Wilkinson Building, Keble Road, Oxford, OX1 3RH, UK\\
$^{3}$Dept. of Physics and Astronomy, University of Toledo, Mailstop 111, 2801 West Bancroft Street, Toledo, Ohio, 43606, USA }}

\begin{document}

\date{Accepted 2013 May 10. Received 2013 May 10; in original form 2013 March 21}

\pagerange{\pageref{firstpage}--\pageref{lastpage}} \pubyear{2013}

\maketitle

\label{firstpage}

\begin{abstract}

In this paper we use observations of molecular tracers in metal rich and $\alpha$-enhanced galaxies to study the effect of abundance changes on molecular chemistry. We selected a sample of metal rich spiral and star bursting objects from the literature, and present here new data for a sample of early-type galaxies (ETGs) previously studied by \cite{2012MNRAS.421.1298C}. We conducted the first survey of CS and methanol emission in ETGs, detecting 7 objects in at least one CS transition, and methanol emission in 5 ETGs. We find that ETGs whose gas is dominated by ionisation from star-formation have enhanced CS emission, compared to their HCN emission, supporting the hypothesis that CS is a better tracer of dense star-forming gas than HCN. We suggest that the methanol emission in these sources is driven by dust mantle destruction due to ionisation from high mass star formation in dense molecular clouds, but cannot rule out a component due to shocks dominating in some sources.
We construct rotation diagrams for each early-type source where at least two transitions of a given species were detected. The rotational temperatures we derive for linear molecules vary between 3 and 9~K, with the majority of sources having rotational temperatures around 5~K. Despite the large uncertainty inherent in this method, the derived source averaged CS and methanol column densities are similar to those found by other authors for normal spiral and starburst galaxies. This may suggest dense clouds are little affected by the differences between early and late type galaxies.
Finally we used the total column density ratios for both our ETG and literature galaxy sample to show for the first time that some molecular tracers do seem to show systematic variations that appear to correlate with metallicity, and that these variations roughly match those predicted by chemical models.
Using this fact, the chemical models of \cite{2012MNRAS.424.2646B}, and assumptions about the optical depth we are able to roughly predict the metallicity of our spiral and ETG sample, with a scatter of $\approx$0.3 dex. We provide the community with linear approximations to the relationship between the HCN and CS column density ratio and metallicity. Further study will clearly be required to determine if this, or any, molecular tracer can be used to robustly determine gas-phase metallically, but that a relationship exists at all suggests that in the future it may be possible to calibrate a metallicity indicator for the molecular interstellar medium.

\end{abstract}


\section{Introduction}

The interstellar medium (ISM) in most massive spiral and early-type galaxies (ETGs) is metal enriched relative to gas found in the solar neighbourhood \citep[e.g.][]{2005ApJ...621..673T}. 
Individual galaxies also have radial gradients in metallicity, in general having metal rich centres and metal poor outer regions \citep[e.g.][]{1990AJ....100.1091P,1994ApJ...420...87Z}. The Milky Way is no exception, the gas in the galactic centre is estimated to have up to 4 times the metallicity found in gas in the solar neighbourhood \citep{1994ApJ...430..236S,2002ApJ...570..671M}. In the next decade, as the power of millimetre wavelength facilities continues to grow we will be able to investigate in greater detail the chemistry of the ISM, and metallicity is likely to be revealed as a crucial parameter, which has so far not been explored in detail.

In the first paper of this series (\citealt{2012MNRAS.424.2646B}; hereafter B12) we presented an exploratory study of the chemistry of photon-dominated regions in high metallicity environments. We found that the molecular ISM is affected in various ways, for instance generally reaching chemical equilibrium at lower temperatures as metallicity increases. We also highlighted some molecules that were likely to be good tracers of high metallically environments, and some whose abundance was reduced by orders of magnitude as metallicity increases. This raises the possibility that molecular abundance ratios could be used to determine the metallicity of the cold molecular ISM.

In B12 we also identified molecules that respond to an increase in the fractional abundance of $\alpha$-elements. These $\alpha$-elements (e.g. Ne, Mg, Si, S, Ar, Ca, Ti) are mainly synthesized by $\alpha$-particle capture in the silicon fusing precursor state to core-collapse supernovae (SNe). Elements around the iron peak, however, are created by SNe type Ia. In environments where star-formation has been rapidly quenched (e.g. ETGs; \citealt{1999MNRAS.302..537T}) and at high redshifts, therefore, $\alpha$-elements are overabundant. How long this overabundance will persist in the gas phase has been little studied, however at high redshifts (before the first  SNe type I explode) the gas is likely to be significantly $\alpha$-enhanced. In the local universe, if stellar mass loss from $\alpha$-enriched stars is important in rebuilding the molecular gas reservoirs of ETGs \citep[see e.g.][]{2011MNRAS.417..882D} then the resultant molecular reservoirs in the local universe could also be  $\alpha$-element rich. Understanding the chemistry of the ISM in such conditions will thus be important for correctly interpreting results from future high redshift studies.

In this paper, we investigate observationally the ISM chemistry in high metallicity environments, with a focus on the common species identified as tracing metallicity in B12; carbon monoxide (CO), carbon monosulfide (CS) and hydrogen cyanide (HCN). These species were highlighted in B12 as having interesting behaviours in different metallicity and $\alpha$-enhancement regimes. Here we aim to determine empirically if any systematic variation in molecular abundances is seen in galaxies with different metallicites, and discuss the possibility of calibrating a metallicity indicator for the molecular ISM. We use a sample of data from the literature for high-metallicity galaxies where these species have been observed, and present our own data for a sample of molecular gas rich, $\alpha$-enhanced ETGs. 

In Section \ref{obs} of this work we introduce our sample of high metallicity galaxies drawn from the literature, and present our new observations. In Section \ref{results} we discuss this first ever survey of CS and methanol emission in ETGs. We then go on to derive estimates of column densities and rotational temperatures from all the molecular line detections presented in this paper, and compare the values with our literature sample. In Section \ref{metalpha} we determine if our derived molecular column density ratios show any trend with metallicity, and compare to the model predictions of B12. 
Finally in Section \ref{conclude} we present our conclusions and discuss prospects for the future.

\section{Observations}
\label{obs}
\subsection{Galaxy sample}
\label{samp}

In order to investigate chemistry in the molecular ISM of high-metallicity and $\alpha$-enhanced environments we require a sample of objects with measured metallicities/$\alpha$-enhancements and detections of CO, CS and HCN.
Such data exists for nearby spiral and star bursting galaxies which have been extensively studied in the literature. Selection of a suitable sample of galaxies is covered in Section \ref{lit_samp}. The molecular data we require, however, did not exist for early-type objects with significant $\alpha$-enhancements. This motivated us to obtain IRAM-30m telescope time to extend the work of  \cite{2012MNRAS.421.1298C} (who detected HCN emission in a sample of ETGs from the \atlas\ sample; \citealt{2011MNRAS.413..813C}) by attempting to detect transitions of CS in these objects. These observations are discussed in Section \ref{iram_obs}.

\subsection{Literature data}
\label{lit_samp}

We selected from the literature a sample of 6 well known nearby high-metallicity (Z$>$Z$_{\odot}$) spiral and star bursting galaxies which have been extensively studied in the millimetre regime. These have determinations of CO, CS and HCN fractional abundances available in the literature (see \citealt{2006ApJS..164..450M}), and measured gas-phase metallicities (mainly from strong-line methods, as tabulated in \citealt{2008ApJ...672..214G}).  Where more than one position in the galaxies has been measured, we always select the central pointing to match our observed data (see Section \ref{iram_obs}). The galaxies we selected, the values used for our analyses and their associated references are listed in Table \ref{lit_table}. The metallicities listed were converted to solar values assuming the  solar oxygen abundance is 12 + log(O/H) = 8.66 \citep{2004A&A...417..751A}. We selected 6 objects in order to match the number of ETGs with new observational data (presented in Section \ref{iram_obs}). This sample does not aim at being complete, but contains enough representative members for the purpose of this first exploratory study. We also include in our literature sample the nearby ETG Centaurus A, and a $\approx$3pc region of galactic centre region centred on Sagittarius A*(see Table \ref{lit_table}).

\begin{table*}
\caption{Data taken from the literature}
\begin{tabular*}{0.9\textwidth}{@{\extracolsep{\fill}}l l l l l l}
\hline
Galaxy/Region & log(Z/Z$_{\odot}$) & N(CO) & N(HCN) & N(CS) & Region size \\
 & (dex) & (10$^{18}$ cm $^{-2}$) & (10$^{14}$ cm $^{-2}$) & (10$^{14}$ cm $^{-2}$) &  \\
 (1) & (2) & (3) & (4) & (5) & (6)\\
 \hline
NGC253 & 0.34 & 6.70 & 3.36 & 2.67 & 0.3 kpc \\
IC342 & 0.24 & 2.20 & 0.70 & 0.88 & 0.36 kpc \\
M82 & 0.34 & 3.80 & 1.51 & 2.40 & 0.33 kpc \\
Maffei2  & 0.24 & 1.80 & 0.90 & 0.36 & 0.29 kpc \\
NGC6946  & 0.44 & 1.50 & 0.47 & 0.30 & 0.66 kpc  \\
NGC4945 & 0.14 & 6.40 & 3.21 & 2.55 & 0.35 kpc \\
Sgr~A* & 0.60 & 33.00 & 64.34 & 13.14 & 2.5 pc \\
Cen~A & 0.10 & 32.00 & 16.04 & 8.04 & 0.87 kpc \\
\hline
\end{tabular*}
\parbox[t]{0.9\textwidth}{\textit{Notes:} Literature data for molecular column densities in the central parts of the listed galaxies (Column 1). The metallicity (Column 2) is a gas phase measure from \cite{2008ApJ...672..214G}, converted to solar units assuming the solar oxygen abundance is 12 + log(O/H) = 8.66 \citep{2004A&A...417..751A}. A $\dagger$ means the metallicity has been estimated from the mass metallicity relation \citep{2004ApJ...613..898T}. The CO, HCN and CS total column densities in Columns 3-5 are calculated from Table 7 in \cite{2006ApJS..164..450M}. The metallicity of the Galactic centre (around Sgr~A*) is taken from \cite{1994ApJ...430..236S} and \cite{2002ApJ...570..671M}. HCN and CO column densities around Sgr~A* are estimated from the H$^{13}$CN and $^{13}$CO observations of \cite{2003JKAS...36..271L}, while CS column densities were estimated from the $^{13}$CS observations of \cite{2008ApJ...678..245M}, all assuming a $^{12}$C/$^{13}$C ratio of 27.8 from \cite{2008ApJ...678..245M}. Any uncertainty in this isotopic ratio assumed is removed in ratios of these quantities. The metallicity of the central parts of Centaurus A comes from \cite{2007ApJ...665..209M}, while the molecular column densities are estimated from absorption line data presented in \cite{1990ApJ...365..522E}. Column 6 contains the diameter of the region probed by the observations towards centre of these objects (where the distance to these objects has been taken from the NASA/IPAC Extragalactic Database).}
\label{lit_table}
\end{table*}

\subsection{IRAM-30m telescope observations}
\label{iram_obs}

\begin{table*}
\caption{Properties of ETG sample observed with the IRAM-30m}
\begin{tabular*}{0.9\textwidth}{@{\extracolsep{\fill}}l r r r r r r r}
\hline
Galaxy & T Type & V$_{\rm sys}$ & Dist. & R$_{\rm e}$ &  log$_{10}$(M$_{\rm H_2}$) & Z$_{\rm Re/8}$ & [OIII]/H$\beta$  \\
 & &  (\kms) & (Mpc) & (") & (M$_{\odot}$) & (dex) &  \\
 (1) & (2) & (3) & (4) & (5) & (6) & (7) & (8)\\
 \hline
NGC2764 & -2.0 & 2706  & 39.6 & 12.3 & 9.19 & -0.68 & 0.44 \\
NGC3032 & -1.9 & 1562  &  21.4 & 13.2 &  8.41 & 0.03 & 0.31 \\
NGC3665 & -2.1 & 2069  &  33.1 & 30.9 & 8.91 & 0.05 & 0.61 \\
NGC4526 & -1.9 & 617   & 16.4 &  44.7 & 8.59 & 0.20 & 0.81 \\
NGC4710 & -0.9 & 1102 &   16.5 & 30.2 & 8.69 & -0.09 & 1.13 \\
NGC5866 & -1.3 & 755 & 14.9 & 36.3 & 8.47 & 0.05 &  0.50\\
NGC6014 & -1.9 & 2381  &  35.8 & 22.4 & 8.8 & -0.27 & 0.45 \\
UGC09519 & -1.9 & 1631  & 27.6  & 7.4 & 8.8 &  0.25 & 1.14 \\
\hline
\end{tabular*}
\parbox[t]{0.9\textwidth}{\textit{Notes:} Table listing the galaxies observed in this work, and some of their properties. Column 1 lists the galaxy name, and Column 2 the Hubble T-type of the system. Columns 3-5 list the systemic velocity of the galaxy, its distance, and the effective radius of the galaxy (the radius enclosing half the light). All of these quantities are taken from \citealt{2011MNRAS.413..813C}. Column 6 lists the H$_2$ mass present in the central 22\arcsec\ of the galaxies, from \cite{2011MNRAS.414..940Y}. Column 7 shows the metallicity derived from stellar population fits to the inner regions of these galaxies, within Re/8 (from McDermid et al., in prep). Column 8 shows the [OIII]/H$\beta$ line ratio within the CO rich region, taken from \cite{2012MNRAS.421.1298C}. }
\label{obssamp_table}
\end{table*}

The IRAM 30-m telescope at Pico Veleta, Spain, was used during February 2012 to observe CS emission in the sample of gas rich ETGs detected in HCN by \cite{2010MNRAS.407.2261K} and  \cite{2012MNRAS.421.1298C} (and originally detected in CO by \citealt{2007MNRAS.377.1795C} and \citealt{2011MNRAS.414..940Y}).  See Table \ref{obssamp_table} for a list of the properties of these objects. 
 We aimed to detect CS(2-1) and CS(3-2) simultaneously in the 3mm and 2mm atmospheric windows.
The beam full width at half-maximum (FWHM) of the IRAM-30m at these frequencies is 25\farc7 and 17\farc1, respectively. The EMIR receiver was used for observations in the wobbler switching mode, with reference positions offset by $\pm$100\arcsec\ in azimuth. The FTS back-end gave an effective total bandwidth of $\approx$4 GHz per window, and a raw spectral resolution of 200 kHz ($\approx$0.6 \kms\ at $\lambda=$3mm, $\approx$0.4 \kms\  at $\lambda=$2mm). We were able to observe methanol lines in the same spectral window as CS in both frequencies. When the weather was too bad for 2mm observations we were able to simultaneously observe $^{13}$CO and C$^{18}$O in the upper-sideband of the 3mm receiver.

The system temperatures ranged between 90 and 200 K at 3~mm and between 100 and 220 K at 1.3~mm.
 The pointing was checked every 2 hours on a nearby planet or bright quasar, and the focus was checked at the beginning of each night as well as after sunrise or more often if a suitable planet was available. The time on source ranged from 30 to 400 min, being weather-dependent, and was interactively adjusted to try and ensure detection of molecular emission.
 
The individual $\approx$6 minute scans were inspected, and baselined, either with a simple constant (zero-order) baseline, a linear (first-order) baseline, or a second order polynomial, depending on the scan. Scans with poor quality baselines or other problems were discarded.
The good scans were averaged together, weighted by the inverse square of the system temperature. When the integrated intensity of a line transition is greater than three times its own uncertainty (calculated as in \citealt{2012MNRAS.421.1298C}, including the uncertainty from estimating the baseline level), we include the line as a detection.  Integrated intensities for each detected line in each galaxy were computed by fitting a Gaussian profile in the \texttt{\small CLASS} package of \texttt{\small GILDAS}\footnote{http://www.iram.fr/IRAMFR/GILDAS/ - accessed 14/01/13}, or summing the spectrum over the velocity (where the profile shape is non-gaussian). Both methods produce consistent results, and here we present the values derived from the sum. 
Table \ref{obstable} lists the RMS noise levels in each spectrum, the velocity width summed over, and the line integrated intensities (in K \kms). 

We convert the spectra from the observed antenna temperature (T$_a^*$) to main beam temperature (T$_{\rm mb}$) by dividing by the ratio of the beam and forward efficiencies ($\eta=$  B$_{\rm eff}$/F$_{\rm eff}$) as tabulated on the IRAM website\footnote{http://www.iram.es/IRAMES/mainWiki/EmirforAstronomers - accessed 23/04/2013}. Where the exact frequency has no measured efficiency available we linearly interpolated the values required.

The detected lines for the first three sources (selected alphabetically) are displayed in Figure \ref{spec1}, and the rest are presented in Appendix A.

\begin{figure*}
\begin{center}
\subfigure[NGC2764]{\includegraphics[height=7cm,angle=0,clip,trim=0.0cm 0cm 0cm 0.0cm]{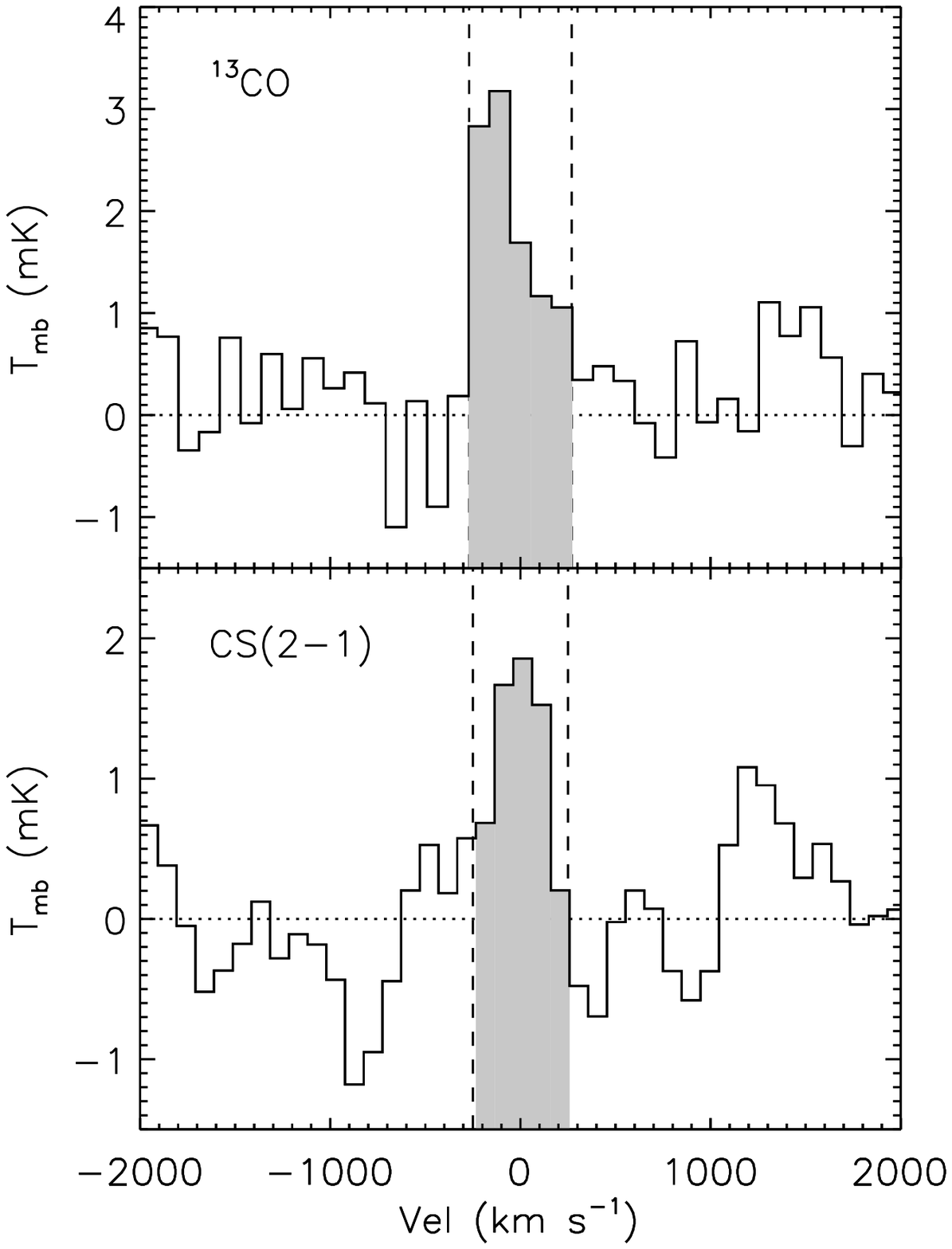}}
\subfigure[NGC3032]{\includegraphics[height=7cm,angle=0,clip,trim=0.0cm 0cm 0cm 0.0cm]{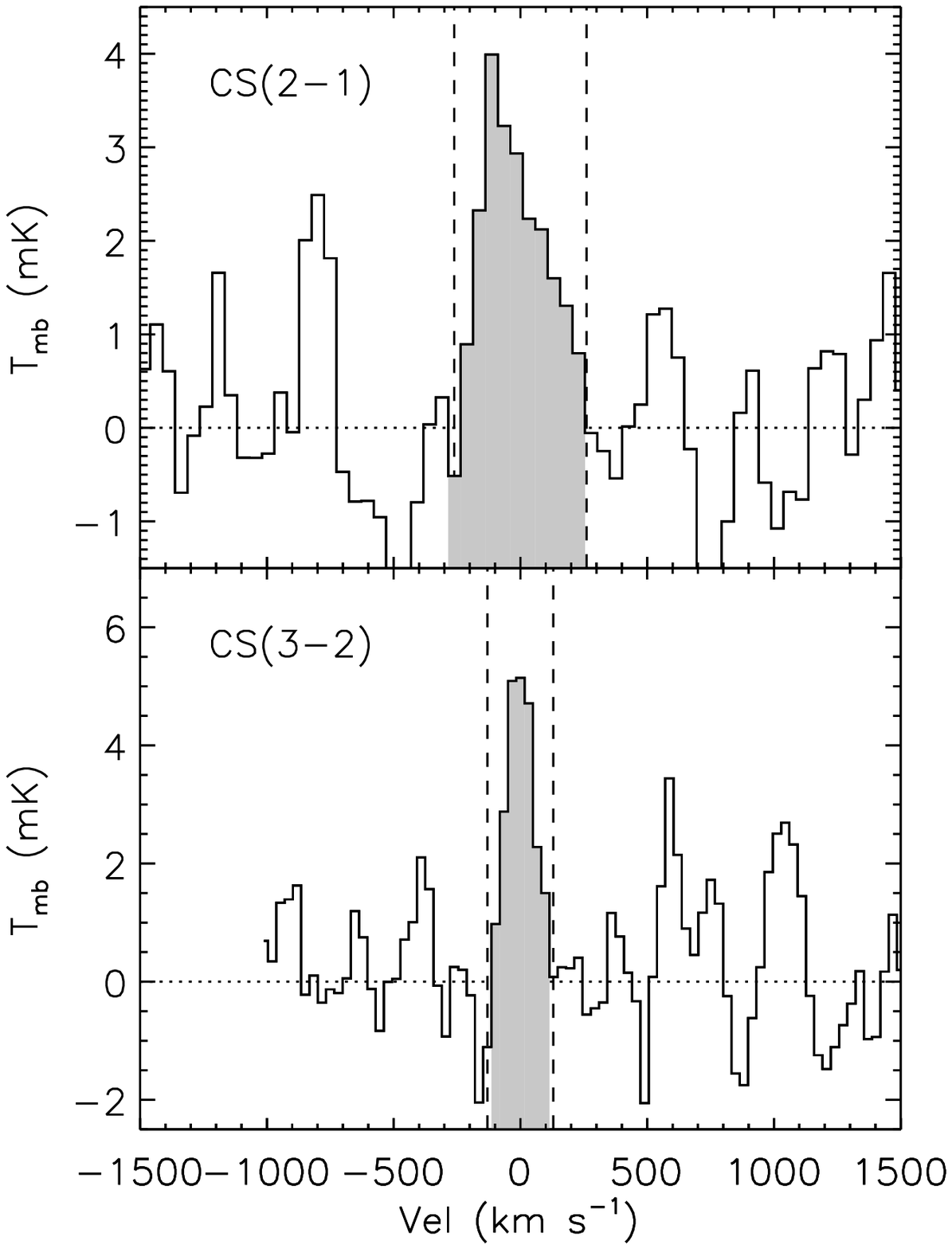}}
\subfigure[NGC3665]{\includegraphics[height=7cm,angle=0,clip,trim=0.0cm 0cm 0cm 0.0cm]{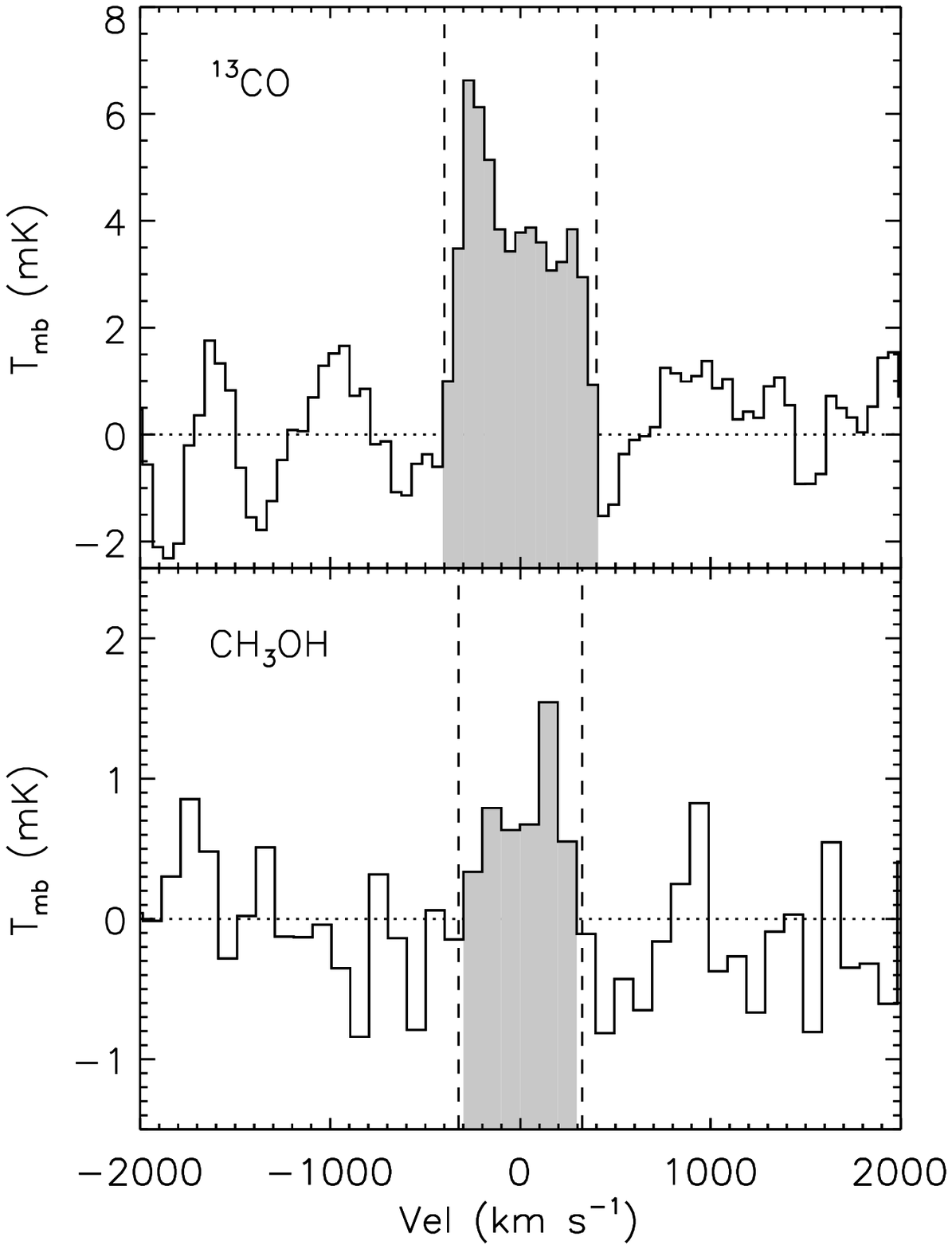}}
 \end{center}
 \caption{Example molecular line detections for three of the ETGs observed for this paper, selected alphabetically (panel a,b and c show the lines detected in NGC2764, NGC3032 and NGC3665, respectively). 
 The rest of the detections for the other sources are presented in Appendix A.
  The observed transition is indicated in the figure legend. The x-axis shows velocity offset from the line centre, corrected for the systemic velocity of the galaxy (listed in Table \ref{obssamp_table}). Average velocity binning in these plots is $\approx$100 \kms, 40 \kms\ and 80 \kms, for sources a, b and c respectively, and in general varies between 25 \kms\ and 115 \kms. The intensity is in units of main beam temperature (T$_{\rm mb}$). A short-dashed line shows the zero level, and long-dashes shown the measured velocity width.  }
 \label{spec1}
 \end{figure*}

\begin{table*}
\caption{Molecular transitions detected in the observed ETGs.}
\begin{tabular*}{1\textwidth}{@{\extracolsep{\fill}}l r r r c r r r r r r r r}
\hline
Galaxy & R$_{\rm gas}$ &Transition & Freq  & Peak & RMS & $\delta$V & $\Delta V$ & $\int$T$_{\rm mb}$dV  & $\int$T$_{\rm B}$dV & N$\mu/$g$\mu$ & Ref \\
 & (\arcsec) &  &(GHz) & (mK)  & (mK)  & (\kms) & (\kms) & (K \kms) & (K \kms) & (10$^{12}$ cm $^{-2}$) & \\
 (1) & (2) & (3) & (4) & (5) & (6) & (7) & (8) & (9) & (10) & (11) & (12) \\
 \hline
NGC2764 &    8.7 & CS(2-1) &   97.98 & 2.25 & 1.03 &    100 &    500 $\pm$     56 &    0.59 $\pm$    0.16 & 5.7 $\pm$ 1.6 &    1.27 &    1 \\
 & & $^{13}$CO(1-0) &   110.20 & 3.28 & 0.50 &    100 &    540 $\pm$     59 &    1.08 $\pm$    0.15 & 8.5 $\pm$ 1.2 &   528.00 &    1,2 \\
 & & HCN &   88.63 & - & - & - & - & - & 3.3 $\pm$ 0.9 &    0.23 &    2 \\
 & & HCO+ &   89.19 & - & - & - & - & - & 5.4 $\pm$ 0.9 &    0.67 &    2 \\
 & & CO(1-0) &   115.27 & - & - & - & - & - & 120 $\pm$ 3.5 &  14100.00 &    3 \\
 & & CO(2-1) &   230.55 & - & - & - & - & - & 63 $\pm$ 1.8 &  1890.00 &    3 \\
 & & $^{13}$CO(2-1) &   220.40 & - & - & - & - & - & 5.5 $\pm$ 0.4 &   85.00 &    2 \\
\hline
NGC3032 &    6.6 & CS(2-1) &   97.98 & 3.58 & 1.60 &     40 &    262 $\pm$     62 &    1.03 $\pm$    0.19 & 17 $\pm$ 3.1 &    3.70 &    1 \\
 & & CS(3-2) &   146.96 & 6.72 & 1.97 &     40 &    115 $\pm$     27 &    0.74 $\pm$    0.14 & 5.7 $\pm$ 1.1 &    0.58 &    1 \\
 & & HCN &   88.63 & - & - & - & - & - & 5.3 $\pm$ 0.8 &    0.37 &    2 \\
 & & CO(1-0) &   115.27 & - & - & - & - & - & 99 $\pm$ 2.7 &  11900.00 &    3 \\
 & & CO(2-1) &   230.55 & - & - & - & - & - & 25 $\pm$ 1.0 &   738.00 &    3 \\
 & & $^{13}$CO(1-0) &   110.20 & - & - & - & - & - & 11 $\pm$ 0.9 &   676.00 &    2 \\
 & & $^{13}$CO(2-1) &   220.40 & - & - & - & - & - & 4.9 $\pm$ 0.4 &   76.30 &    2 \\
\hline
NGC3665 &    10.2 & CH$_3$OH &   96.73 & 1.02 & 0.34 &     80 &    650 $\pm$     97 &    0.44 $\pm$    0.13 & 3.3 $\pm$ 1.0 &    4.69$^\dagger$ &    1 \\
 & & $^{13}$CO(1-0) &   110.20 & 5.49 & 1.16 &     80 &    720 $\pm$     35 &    3.37 $\pm$    0.16 & 20 $\pm$ 1.0 &  1250.00 &    1,2 \\
 & & HCN &   88.63 & - & - & - & - & - & 4.3 $\pm$ 0.7 &    0.30 &    2 \\
 & & CO(1-0) &   115.27 & - & - & - & - & - & 67 $\pm$ 3.6 &  7980.00 &    3 \\
 & & CO(2-1) &   230.55 & - & - & - & - & - & 31 $\pm$ 1.8 &   920.00 &    3 \\
 & & $^{13}$CO(2-1) &   220.40 & - & - & - & - & - & 13 $\pm$ 0.4 &   197.00 &    2 \\
 & & CO(3-2) &   345.82 & - & - & - & - & - & 29 $\pm$ 1.0 &   388.00 &    4 \\
\hline
NGC4526 &    2.6 & CS(2-1) &   97.98 & 0.83 & 0.34 &    100 &    980 $\pm$    145 &    0.60 $\pm$    0.13 & 59 $\pm$ 13 &   13.10 &    1 \\
 & & CH$_3$OH &   96.73 & 0.90 & 0.34 &    100 &    800 $\pm$    107 &    0.43 $\pm$    0.10 & 44 $\pm$ 10. &   62.20$^\dagger$ &    1 \\
 & & $^{13}$CO(1-0) &   110.20 & 7.00 & 0.60 &     20 &    733 $\pm$     5 &    5.63 $\pm$    0.13 & 440 $\pm$ 10. &  27200.00 &    1,2 \\
 & & C$^{18}$O(1-0) &   109.78 & 1.61 & 0.60 &    100 &    800 $\pm$     80 &    1.37 $\pm$    0.16 & 110 $\pm$ 13 &  13400.00 &    1 \\
 & & HCN &   88.63 & - & - & - & - & - & 210 $\pm$ 14 &   14.80 &    2 \\
 & & CO(1-0) &   115.27 & - & - & - & - & - & 1500 $\pm$ 71 & 185000.00 &    3 \\
 & & CO(2-1) &   230.55 & - & - & - & - & - & 600 $\pm$ 17 &  18000.00 &    3 \\
 & & $^{13}$CO(2-1) &   220.40 & - & - & - & - & - & 140 $\pm$ 6.9 &  2130.00 &    2 \\
 & & CO(3-2) &   345.82 & - & - & - & - & - & 560 $\pm$ 17 &  7410.00 &    4 \\
\hline
NGC4710 &    17.2 & CS(2-1) &   97.98 & 1.83 & 0.55 &     50 &    702 $\pm$     64 &    0.71 $\pm$    0.11 & 2.3 $\pm$ 0.4 &    0.51 &    1 \\
 & & CH$_3$OH &   96.73 & 2.55 & 0.55 &     50 &    352 $\pm$     40 &    0.46 $\pm$    0.08 & 1.5 $\pm$ 0.3 &    2.15$^\dagger$ &    1 \\
 & & CS(3-2) &   146.96 & 1.84 & 0.53 &    100 &    500 $\pm$     92 &    0.79 $\pm$    0.16 & 1.6 $\pm$ 0.3 &    0.16 &    1 \\
 & & CH$_3$OH &   145.10 & 5.43 & 0.53 &    100 &    328 $\pm$     22 &    0.82 $\pm$    0.14 & 1.6 $\pm$ 0.3 &    0.78$^\dagger$ &    1 \\
 & & HCN &   88.63 & - & - & - & - & - & 5.3 $\pm$ 0.4 &    0.37 &    2 \\
 & & HCO+ &   89.19 & - & - & - & - & - & 3.3 $\pm$ 0.4 &    0.40 &    2 \\
 & & CO(1-0) &   115.27 & - & - & - & - & - & 83 $\pm$ 1.9 &  9870.00 &    3 \\
 & & CO(2-1) &   230.55 & - & - & - & - & - & 57 $\pm$ 0.9 &  1690.00 &    3 \\
 & & $^{13}$CO(1-0) &   110.20 & - & - & - & - & - & 13 $\pm$ 0.4 &   818.00 &    2 \\
 & & $^{13}$CO(2-1) &   220.40 & - & - & - & - & - & 8.8 $\pm$ 0.3 &   137.00 &    2 \\
\hline
NGC5866 &    17.3 & CS(2-1) &   97.98 & 1.47 & 0.39 &     50 &    640 $\pm$     49 &    0.72 $\pm$    0.08 & 2.3 $\pm$ 0.3 &    0.51 &    1 \\
 & & CH$_3$OH &   96.73 & 3.96 & 0.39 &     50 &    450 $\pm$     49 &    1.47 $\pm$    0.09 & 4.8 $\pm$ 0.3 &    6.81$^\dagger$ &    1 \\
 & & CS(3-2) &   146.96 & 1.19 & 0.28 &    100 &    500 $\pm$     94 &    0.38 $\pm$    0.07 & 0.8 $\pm$ 0.1 &    0.08 &    1 \\
 & & CH$_3$OH &   145.10 & 0.82 & 0.20 &    100 &    640 $\pm$     94 &    0.39 $\pm$    0.08 & 0.8 $\pm$ 0.2 &    0.37$^\dagger$ &    1 \\
 & & HCN &   88.63 & - & - & - & - & - & 3.1 $\pm$ 0.3 &    0.22 &    2 \\
 & & HCO+ &   89.19 & - & - & - & - & - & 2.1 $\pm$ 0.3 &    0.26 &    2 \\
 & & CO(1-0) &   115.27 & - & - & - & - & - & 56 $\pm$ 1.0 &  6680.00 &    3 \\
 & & CO(2-1) &   230.55 & - & - & - & - & - & 24 $\pm$ 1.0 &   725.00 &    3 \\
 & & $^{13}$CO(1-0) &   110.20 & - & - & - & - & - & 9.0 $\pm$ 0.4 &   557.00 &    2 \\
 & & $^{13}$CO(2-1) &   220.40 & - & - & - & - & - & 5.5 $\pm$ 0.3 &   84.70 &    2 \\
 & & CO(3-2) &   345.82 & - & - & - & - & - & 39 $\pm$ 1.0 &   524.00 &    4 \\
 \hline
\end{tabular*}
\parbox[t]{1 \textwidth}{ \textit{Notes:} Column one contains the name of the object, and column two contains the geometric mean CO(1-0) source size estimated from the interferometric observations of \cite{2012arXiv1210.5524A,2013MNRAS.429..534D}. Column 3 lists the detected transition, and the rest frequency is listed in Column 4.
The telescope beam size at the frequency of CS(2-1) is 25\farc7, 22\farc8 at the frequency of $^{13}$CO/C$^{18}$O, and 17\farc1 for CS(3-2). For the methanol blend at 96.73 GHz we use a telescope beam size of 26\farc0, and 17\farc3 for the blend at 145.10GHz. The telescope beam sizes for the literature detections are taken from the source listed in Column 11.   
Columns 5 and 6 show the peak brightness temperature of the line, and the RMS around the baseline calculated in line free regions in the spectrum with channel width $\delta$V, tabulated in Column 7. The beam temperatures in Columns 5 and 6  are main beam temperatures (T$_{\mathrm{mb}}$). Column 8 shows the line width and the statistical error in this quantity returned by the fit. All lines are consistent with arising from gas centred around the systemic velocity of the system (which is listed in Table \ref{obssamp_table}). The observed integrated line flux is shown in Column 9. The flux corrected using Equation \ref{tbeameq} is shown in Column 10. Column 11 shows the derived column density in the upper level of the observed transition, assuming LTE (and that the emission is optically thin for all moleculars but $^{12}$CO. Transitions marked with a $\dagger$ in this column are blends, and are treated separately (see Section \ref{dataredux}). Column 12 shows the data reference: (1) this work, (2)  \cite{2012MNRAS.421.1298C}, (3) \cite{2011MNRAS.414..940Y}, (4) \cite{2010MNRAS.407.2261K}, (5) \cite{2012arXiv1212.2630B}.}
\label{obstable}
\end{table*}

\begin{table*}
\begin{tabular*}{1\textwidth}{@{\extracolsep{\fill}}l r r r c r r r r r r r r}
\hline
Galaxy & R$_{\rm gas}$ &Transition & Freq  & Peak & RMS & $\delta$V & $\Delta V$ & $\int$T$_{\rm mb}$dV  & $\int$T$_{\rm B}$dV & N$\mu/$g$\mu$ & Ref \\
 & (\arcsec) &  &(GHz) & (mK)  & (mK)  & (\kms) & (\kms) & (K \kms) & (K \kms) & (10$^{12}$ cm $^{-2}$) & \\
 (1) & (2) & (3) & (4) & (5) & (6) & (7) & (8) & (9) & (10) & (11) & (12) \\
\hline
NGC6014 &    3.6 & CS(2-1) &   97.98 & 5.71 & 1.27 &     25 &    220 $\pm$     21 &    0.60 $\pm$    0.09 & 31 $\pm$ 4.7 &    6.93 &    1 \\
 & & CH$_3$OH &   96.73 & 5.11 & 1.27 &     50 &    340 $\pm$     31 &    0.55 $\pm$    0.14 & 29 $\pm$ 7.4 &   41.60$^\dagger$ &    1 \\
 & & $^{13}$CO(1-0) &   110.20 & - & - & - & - & - & 28 $\pm$ 3.3 &  1760.00 &    2 \\
 & & HCN &   88.63 & - & - & - & - & - & 13 $\pm$ 3.2 &    0.89 &    2 \\
 & & CO(1-0) &   115.27 & - & - & - & - & - & 280 $\pm$ 14 &  33700.00 &    3 \\
 & & CO(2-1) &   230.55 & - & - & - & - & - & 110 $\pm$ 3.8 &  3290.00 &    3 \\
 & & $^{13}$CO(2-1) &   220.40 & - & - & - & - & - & 17 $\pm$ 1.1 &   265.00 &    2 \\
 & & CO(3-2) &   345.82 & - & - & - & - & - & 140 $\pm$ 7.4 &  1840.00 &    4 \\
\hline
UGC09519 &    5.3 & CS(2-1) &   97.98 & 2.84 & 0.59 &     25 &    450 $\pm$     56 &    0.68 $\pm$    0.14 & 17 $\pm$ 3.4 &    3.70 &    1 \\
 & & CH$_3$OH &   96.73 & 2.64 & 0.59 &     25 &    400 $\pm$     46 &    0.53 $\pm$    0.11 & 13 $\pm$ 2.8 &   18.90$^\dagger$ &    1 \\
 & & $^{13}$CO(1-0) &   110.20 & - & - & - & - & - & 11 $\pm$ 2.3 &   685.00 &    2 \\
 & & $^*$C$^{18}$O(1-0) &   109.78 & 0.83 & 0.36 &     50 &    231 $\pm$     96 &    0.20 $\pm$    0.08 & 3.9 $\pm$ 1.6 &   490.00 &    1 \\
 & & HCO+ &   89.19 & - & - & - & - & - & 8.5 $\pm$ 2.1 &    1.05 &    2 \\
 & & CO(1-0) &   115.27 & - & - & - & - & - & 230 $\pm$ 6.8 &  27100.00 &    3 \\
 & & CO(2-1) &   230.55 & - & - & - & - & - & 73 $\pm$ 1.5 &  2190.00 &    3 \\
 & & $^{13}$CO(2-1) &   220.40 & - & - & - & - & - & 5.0 $\pm$ 1.0 &   77.10 &    2 \\
 & & CO(3-2) &   345.82 & - & - & - & - & - & 95 $\pm$ 4.1 &  1260.00 &    4 \\
  \hline
 \end{tabular*}
\parbox[t]{1 \textwidth}{Table \protect \ref{obstable} continued.}
\end{table*}

\subsection{Comparison with literature fluxes}

For three of these objects (NGC2764, NGC3665 and NGC4526) where our IRAM-30m were taken in bad weather we re-detected $^{13}$CO(1-0) lines which had been previously observed. 

NGC2764 was detected in $^{13}$CO by  \cite{2012MNRAS.421.1298C}, and they report a main beam brightness temperature of 1.41$\pm$0.1 K \kms. Here we find a main beam brightness temperature of 1.08$\pm$0.15 K \kms.
NGC3665 was also detected by   \cite{2012MNRAS.421.1298C} who report a main beam brightness temperature of 3.7$\pm$0.19 K \kms. Our measurement is 3.37$\pm$0.16 K \kms.
$^{13}$CO was detected in NGC4526 by \cite{2010MNRAS.407.2261K}, who found a main beam brightness temperature of 5.95$\pm$0.28 K \kms. In this paper we report a flux of 5.63$\pm$0.13 K \kms. All these measurements are consistent within the quoted errors, especially considering the nominal flux scale accuracy of the IRAM-30m ($\approx$10\%\footnote{http://www.iram.es/IRAMES/mainWiki/EmirforAstronomers - accessed 23/04/2013}).

\subsection{Data products}
\label{dataredux}

In order to directly compare our observed transitions to literature data, and to the models of B12, we need to convert the observed line intensities into total column densities of each molecule. We outline the process we used to do this below, following the procedure also used by \cite{2006ApJS..164..450M,2011A&A...535A..84A,2013A&A...549A..39A}.

 The source-averaged brightness temperature (T$_{\rm B}$) can be estimated from the measured main beam brightness temperature (T$_{\rm mb}$).
 In the approximation of a Gaussian source distribution of size $\theta_{s}$ observed with a Gaussian beam of size $\theta_{b}$ the below equations correct for the dilution effect due to the coupling between the source and the telescope beam \citep{2006ApJS..164..450M}.

\begin{equation}
T_{\rm B} = \frac{T_{\rm mb}}{\mu_{\rm bf}},
\label{tbeameq}
\end{equation}
where
\begin{equation}
\mu_{\rm bf} = \frac{\theta_{s}^2}{\theta_{s}^2+\theta_{b}^2}
\end{equation}

Here we use the geometric mean of the minimum and maximum extent of the molecular gas distribution in these ETGs, mapped in CO(1-0) by \cite{2012arXiv1210.5524A}, as tabulated in \cite{2013MNRAS.429..534D}. The resulting source size we assume in the rest of this paper is listed in Table \ref{obstable}, along with the beam corrected integrated intensity ($\int${T$_{\rm B}$dV). We also show in Table \ref{obstable} the line detections from  \cite{2012MNRAS.421.1298C}, \cite{2011MNRAS.414..940Y} and \cite{2012arXiv1212.2630B}, converted to T$_{\rm B}$ as described here.
The real extent (and distribution) of each emitting component is likely to be different  \citep[e.g.,][]{2012ApJ...755..104M} and this is one of the main uncertainties in our resulting quantities.

\subsubsection{Rotation Diagrams}
\label{rotdiag_method}

From the observed beam corrected brightness temperature of the
measured molecular lines one can directly estimate the column density of these molecules in the upper level of the observed transition (in an LTE approximation). All the necessary spectroscopic information required to derive these parameters (e.g., Einstein coefficients, upper and lower energy levels) were extracted from the Cologne Database for Molecular Spectroscopy ({\small CDMS})\footnote{http://www.astro.uni-koeln.de/cdms/ - accessed 14/01/2013} \citep{2005JMoSt.742..215M}. Table \ref{obstable} includes an estimate of the column density in the upper level of the observed transition (N$\mu/$g$\mu$; where g$\mu$ is the degeneracy of the state), calculated as in Equation \ref{eqnnmugmu}  (see \citealt{1999ApJ...517..209G} and appendix B1 of \citealt{2006ApJS..164..450M}  for detailed discussion of this technique).

\begin{equation}
\frac{N_\mu}{g_\mu} =  \left(\frac{1.667\times10^{17}}{{S\mu^2}}\right) \left( \frac{\nu}{\rm MHz}\right)^{-1}\left( \frac{\tau}{1-e^{-\tau}}\right)\left(\frac{\int T_b\, \delta V}{\rm K\,km\,s^{-1}}\right) 
\label{eqnnmugmu}
\end{equation}

where S$\mu^2$ is the (temperature independent) product of the line strength and the molecule dipole moment (in units of nm$^{2}$ MHz; extracted from the CDMS, $\nu$ is the frequency of the transition in MHz, $\tau$ is the optical depth (if this is not known then this relationship is only valid in the optically thin limit), and $\int T_b\, \delta V$ is the integrated intensity of the beam corrected brightness temperature, in units of K \kms, as tabulated in Table \ref{obstable}. Equation \ref{eqnnmugmu} assumes that the rotation temperature is large compared to the temperature of the background radiation field (usually dominated by the 2.7 K cosmic microwave background). If this were not the case, then failure to take into account the contribution of the absorption of the 2.7K background would make the derived ${N_\mu}/{g_\mu}$ a lower limit.

In order to convert the column densities
in the observed states to a total column density for a given molecule, more than one transition has to be detected, in order to give an estimate for the rotation temperature (T$_{\rm rot}$). 
Figure \ref{NGC4710rotdiag} shows an example (for the source where we detect the most transitions) of the rotation diagrams derived for each molecule where we detect at least two lines of that species. Rotation diagrams for each galaxy are presented in Appendix B (Fig \ref{NGC2764rotdiag}). 
The rotational temperature and column density are determined from the slope [-(log e)/T$_{\rm rot}$] and intercept [log(N/Q$_{\rm rot}$) at Eu = 0], where Q$_{\rm rot}$ is the partition function for that molecule evaluated at the determined rotational temperature (the values for the partition functions used here were interpolated from the data available at the Cologne Database for Molecular Spectroscopy). The fitted parameters are tabulated in Table \ref{trot_table}. The derived T$_{\rm rot}$ will be a lower limit for the kinetic temperature (T$_{\rm kin}$) if the lines are sub-thermally excited, e.g. if the H$_2$ densities are not high enough to counterbalance spontaneous decay of the excited levels.

If the emission from these molecular species is optically thick then in order to derive the correct column density we need an estimate of the optical depth (obtained through detection of an optically thin isotopologue) to feed into Equation \ref{eqnnmugmu}. $^{13}$CO(1-0) and $^{13}$CO(2-1) have been observed for all these objects, and so we are able to calculate the $^{12}$CO optical depth of each transition, assuming a  $^{12}$C/$^{13}$C ratio.  The true $^{12}$C/$^{13}$C ratio may vary between objects, but as we lack the information to constrain this ratio directly, we here use the Milky Way of $\approx$70 \citep{1994ARA&A..32..191W}, applicable for approximately solar metallicity gas. Lower values of this ratio have been reported towards the galactic centre, and if we instead used these values our optical depths (and thus our total column density estimates) would be lower.
In addition, if $^{13}$CO is also optically thick then we will underestimate the optical depth of $^{12}$CO. In NGC4526 we also detect C$^{18}$O(1-0), and so have a second independent measure of the optical depth \citep[which we calculate assuming a Milky Way $^{12}$C/$^{13}$C ratio of $\approx$250;][]{1994ARA&A..32..191W}. Both estimates of the $^{12}$CO optical depth are consistent ($\tau_{13}$=19.8$\pm$0.9, $\tau_{18}$=17.4$\pm$2.1), suggesting that $^{13}$CO is not significantly optically thick in this object. We do not have this information for the other ETGs considered in this work, so work under the assumption that $^{13}$CO is also optically thin in these other objects.
  For molecular species other than CO we do not detect isotopologues, and thus cannot correct for optical depth effects. The total column density estimates we derive are thus formally lower limits. We are however able to set limits on the optical depths of some of these species, as described in Section \ref{metcomp_empirical}.

 In addition, with the observations we possess here we are only able to derive a single rotational temperature for each gas tracer. In reality several gas components with different rotation temperatures are often required to fit observational data \citep[e.g.][]{2009ApJ...707..126B}.
 This is a second large source of uncertainty, but only in future works with a larger number of observations we will be able to tackle this issue in detail. 
 The colder components of the ISM that emit in low $J$ lines (which we observe here for most species) usually dominate the total column density estimate, suggesting the uncertainty in total column densities when one only has a few low-$J$ lines should be smaller than the uncertainty in the mean rotation temperature. 
A few sources have CO(3-2) detections from \cite{2012arXiv1212.2630B}, but no detection of the $J$=3-2 transition of any CO isotopes.  When we use the CO(3-2) in a na\"ive way (i.e. in rotation diagrams without optical depth correction) we do require two gas components with different rotational temperatures to fit the data, but this second, higher rotational temperature component contributes $<$10\% of the total column density in all cases.
We include the existence of these components in our error budget for all molecules, by including an additional 30\% error in the column density, to attempt to account for the presence of any additional higher rotational temperature components of the ISM we are unable to detect.

We also detect methanol emission in various sources in our galaxy sample. The methanol lines at 96.73 GHz and 145.1 GHz are blended (4 times and 7 times, respectively). In order to create the rotation diagram and calculate the total column density for methanol we use a statistical decomposition (from Appendix B2 in \citealt{2006ApJS..164..450M}), which is based on the relative ratios of the Einstein coefficients. One must assume an internal rotation temperature in order to split the levels, and here (as in \citealt{2006ApJS..164..450M}) we use a value that minimises the difference between the slope derived from a single set of blended lines, and the slope derived from both sets of lines. This statistical method for splitting blended lines should be treated with caution, as the error on the derived quantities may be much higher than reflected in the formal error bars, especially given that the few transitions observed represent only a small fraction of the emitting column density. Given the large uncertainties we present the values derived for completeness, but do not use them further.

In order to estimate the column density of the cases where only one transition was detected an educated
guess of the rotation temperature is made by using the measured T$_{\rm rot}$ from the molecule closest in critical density (of the ground state) to the molecule in question. This technique has been used by other authors \citep[e.g.][]{2013A&A...549A..39A}, however it is clear that the column densities derived in this way should be treated with caution. Table \ref{trot_table} shows the derived total column densities and rotation temperatures for the detected species in our newly observed galaxies.

\begin{figure}
\begin{center}
\includegraphics[width=0.23\textwidth,angle=0,clip,trim=1.0cm 0cm 0.8cm 1.0cm]{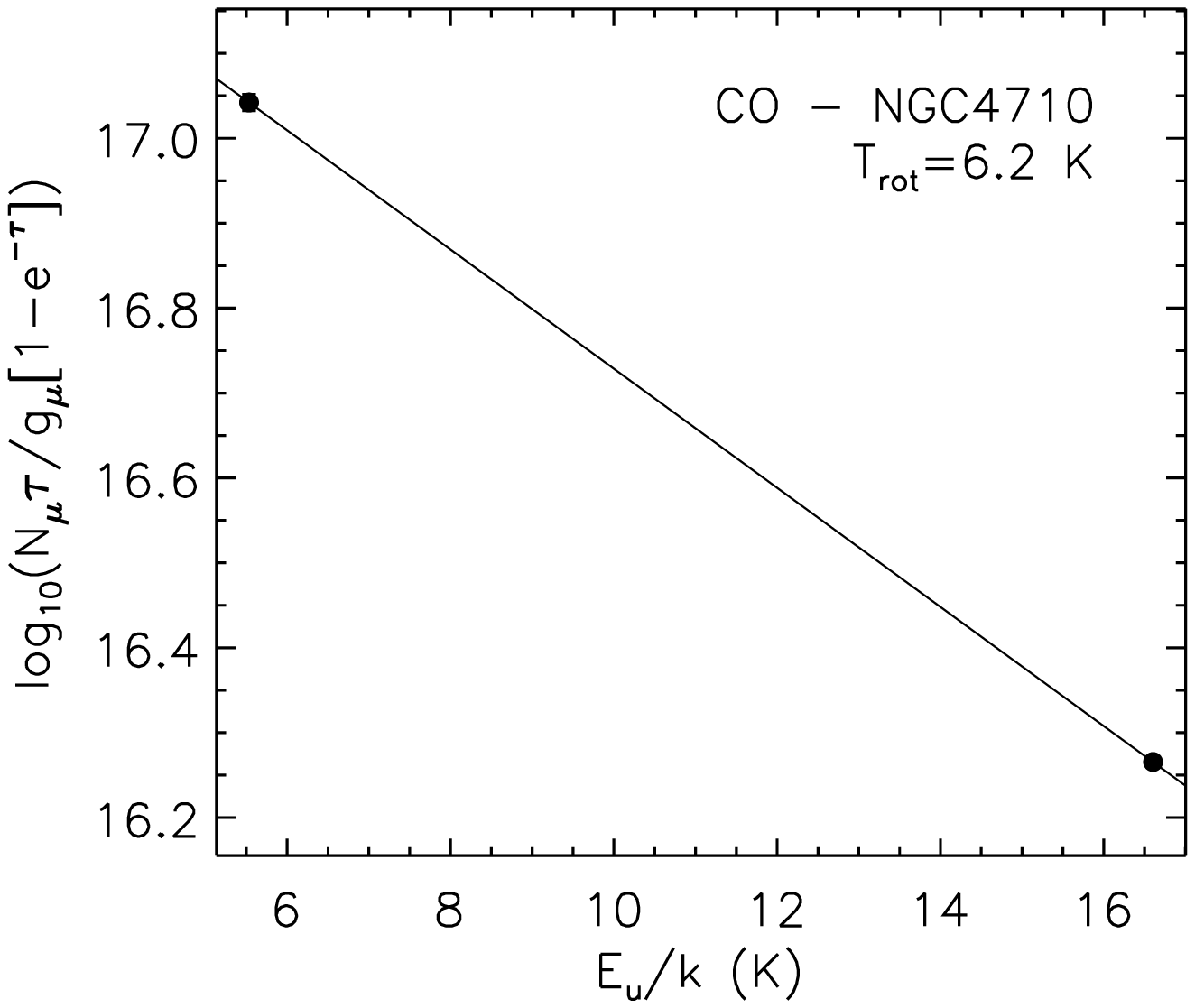}
\includegraphics[width=0.23\textwidth,angle=0,clip,trim=1.0cm 0cm 0.8cm 1.0cm]{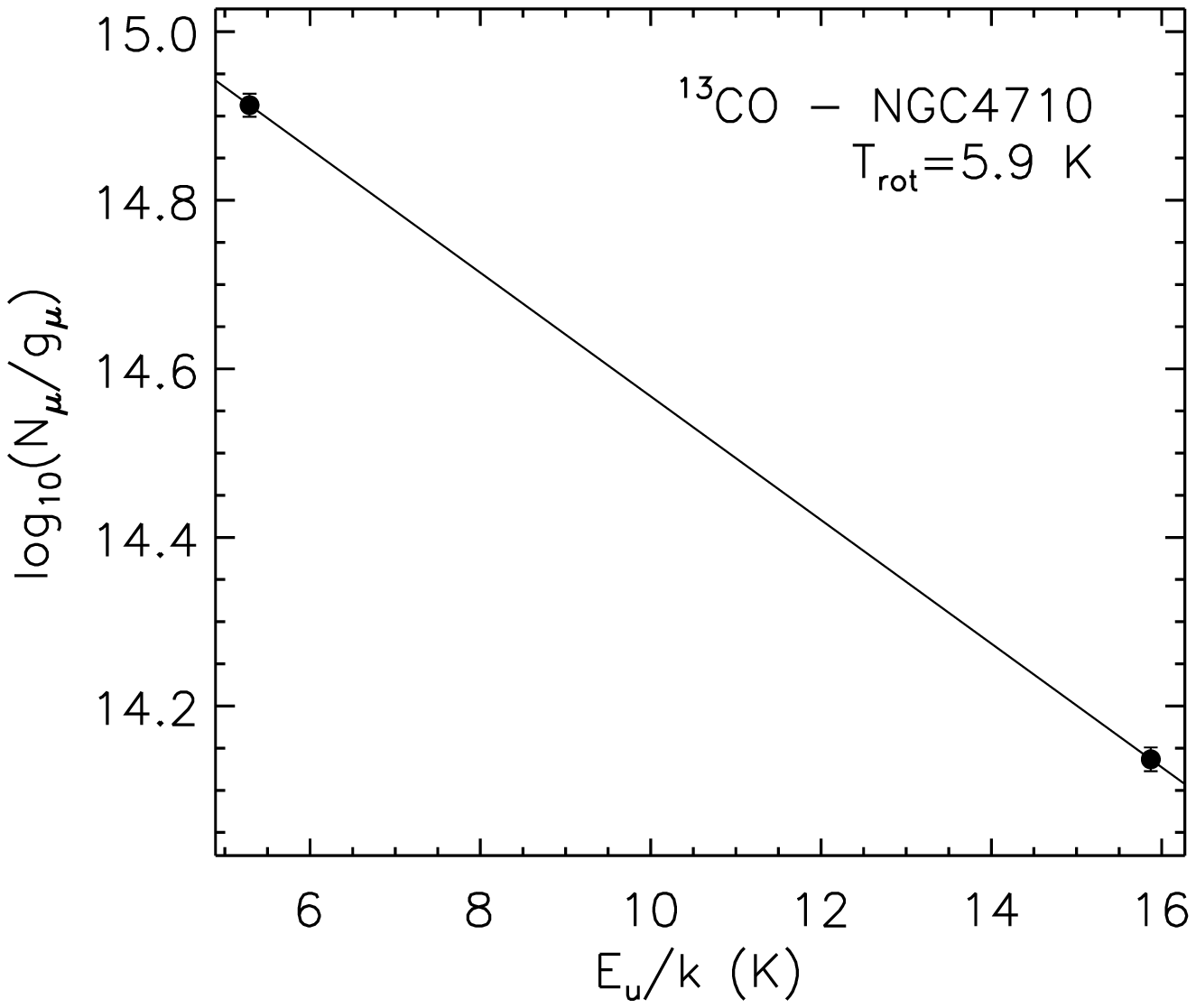}\\
\includegraphics[width=0.23\textwidth,angle=0,clip,trim=1.0cm 0cm 0.8cm 1.0cm]{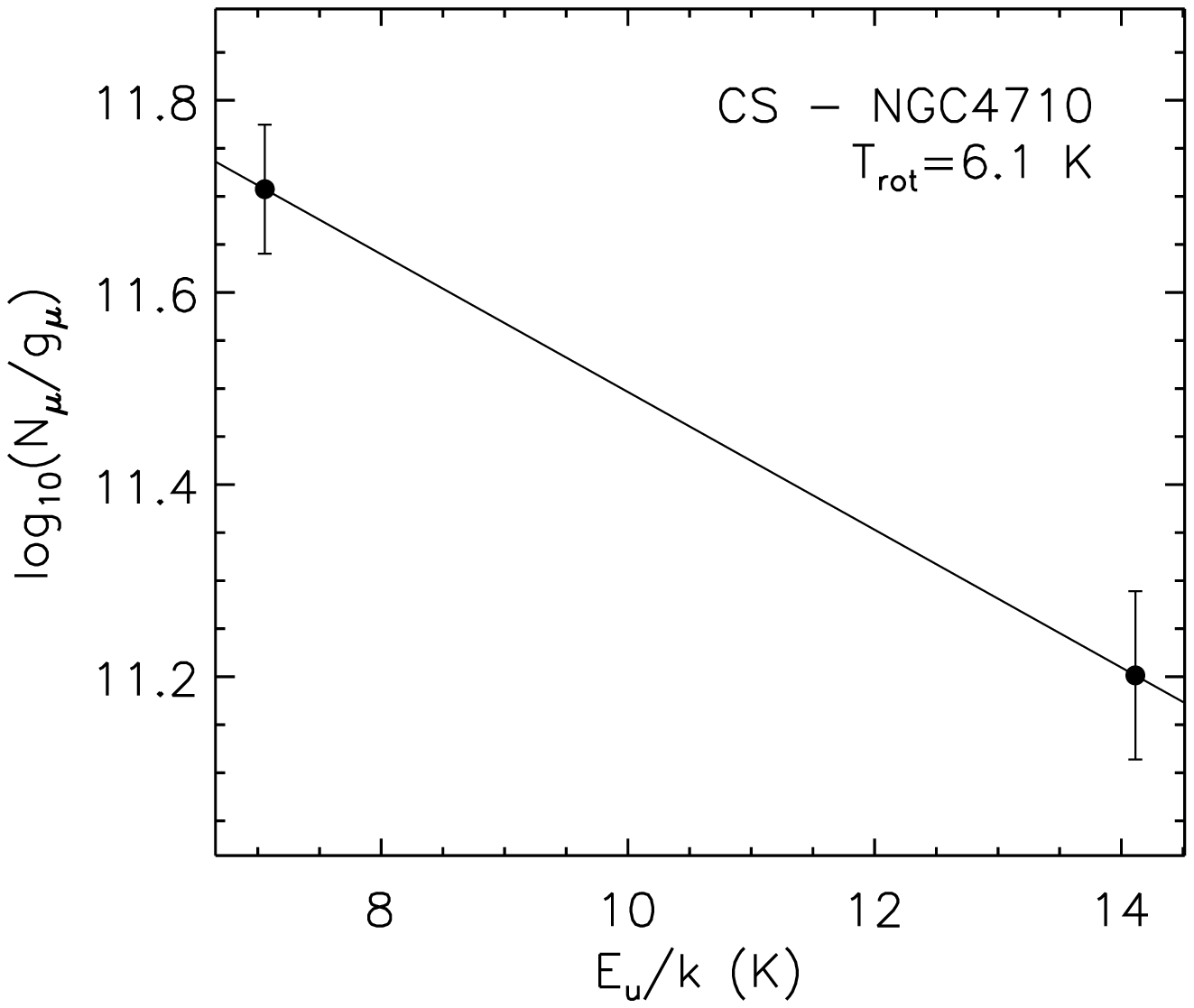}
\includegraphics[width=0.23\textwidth,angle=0,clip,trim=1.0cm 0cm 0.8cm 1.0cm]{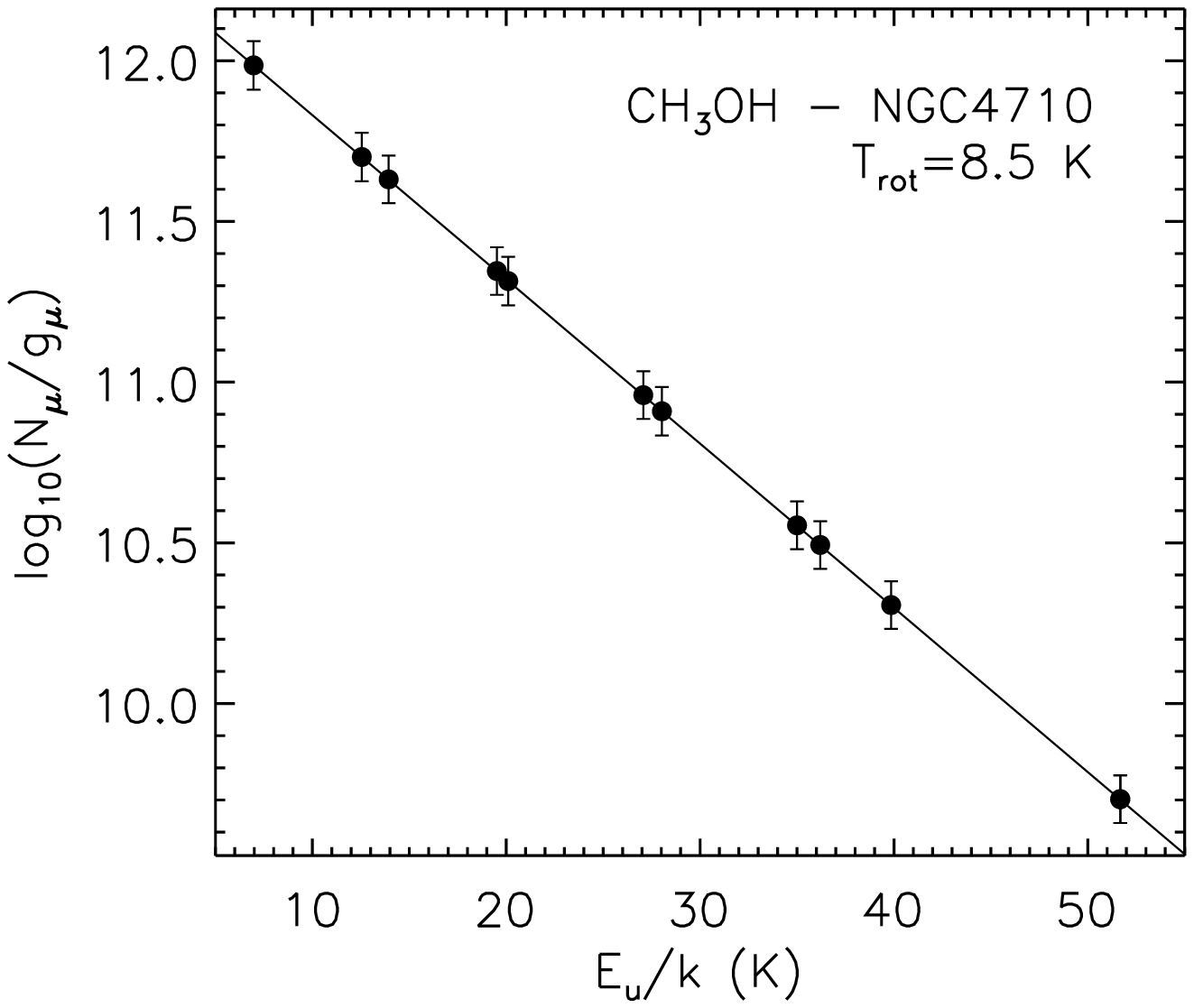}
 \end{center}
 \caption{Rotation diagrams for NGC\,4710 (the source where we detect the most transitions). The energy of the upper level is plotted against the derived column density in the upper level for each observed transition (black points), as tabulated in Table \ref{obstable}. These data points are fitted with a line and the best fit, corresponding to the rotation temperature indicated in the legend, is displayed (solid line). We split methanol blends statistically, as described in Section \ref{dataredux}. Rotation diagrams for the other galaxies are presented in Appendix B (Figs \ref{NGC2764rotdiag}-\ref{UGCrotdiag}).}
 \label{NGC4710rotdiag}
 \end{figure}

\begin{table*}
\caption{Total column densities and rotation temperatures derived from rotational diagrams. }
\begin{tabular*}{0.9\textwidth}{@{\extracolsep{\fill}}l l r r r r}
\hline
Galaxy & Transition & Gradient & Intercept & Rot. Temp & N(x) \\
 & & (K$^{-1}$) & log$_{10}$(cm $^{-2}$)  & (K) & (10$^{14}$ cm $^{-2}$)  \\
 (1) & (2) & (3) & (4) & (5) & (6) \\
\hline
NGC2764 & CO & -0.072 & 17.3 & 6.1 $\pm$ 0.1 & 2790 \\ 
 & $^{13}$CO & -0.075 & 15.1 & 5.8 $\pm$ 0.5 & 41.3 \\ 
 & CS(2-1) & - & - & 5.8$^{*}$ &0.2 \\
 & HCN(1-0) & - & - & 5.8$^{*}$ &0.04 \\
 & HCO$^+$(1-0) & - & - & 5.8$^{*}$ &0.04 \\
\hline
NGC3032 & CO & -0.086 & 17.4 & 5.1 $\pm$ 0.1 & 3850 \\ 
 & $^{13}$CO & -0.090 & 15.3 & 4.9 $\pm$ 0.2 & 57.3 \\ 
 & CS & -0.114 & 13.4 & 3.8 $\pm$ 0.5 & 0.92 \\ 
 & HCN(1-0) & - & - & 3.8$^{*}$ &0.08 \\
\hline
NGC3665 & CO & -0.073 & 17.6 & 6.0 $\pm$ 0.3 & 6610 \\ 
 & $^{13}$CO & -0.076 & 15.5 & 5.7 $\pm$ 0.2 & 98.1 \\ 
 & HCN(1-0) & - & - & 5.8$^{*}$ &0.06 \\
\hline
NGC4526 & CO & -0.100 & 19.1 & 4.3 $\pm$ 0.09 & 173000 \\ 
 & $^{13}$CO & -0.105 & 17.0 & 4.2 $\pm$ 0.09 & 2580 \\ 
 & CS(2-1) & - & - & 4.3$^{*}$ &2.9 \\
 & C$^{18}$O(1-0) & - & - & 4.3$^{*}$ &620 \\
 & HCN(1-0) & - & - & 4.3$^{*}$ &3.0 \\
\hline
NGC4710 & CO & -0.070 & 17.4 & 6.2 $\pm$ 0.10 & 4270 \\ 
 & $^{13}$CO & -0.073 & 15.3 & 5.9 $\pm$ 0.2 & 63.5 \\ 
 & CS & -0.072 & 12.2 & 6.1 $\pm$ 1.3 & 0.087 \\ 
 & CH$_3$OH & -0.051 & 12.3 & 8.5 $\pm$ 3.2 & 0.38 \\ 
 & HCN(1-0) & - & - & 6.1$^{*}$ &0.07 \\
 & HCO$^+$(1-0) & - & - & 6.1$^{*}$ &0.03 \\
\hline
NGC5866 & CO & -0.074 & 17.3 & 5.9 $\pm$ 0.1 & 2950 \\ 
 & $^{13}$CO & -0.077 & 15.2 & 5.6 $\pm$ 0.2 & 44.0 \\ 
 & CS & -0.118 & 12.5 & 3.7 $\pm$ 0.4 & 0.13 \\ 
 & CH$_3$OH & -0.167 & 13.8 & 2.6 $\pm$ 0.1 & 5.36 \\ 
 & HCN(1-0) & - & - & 3.7$^{*}$ &0.05 \\
 & HCO$^+$(1-0) & - & - & 3.7$^{*}$ &0.02 \\
\hline
NGC6014 & CO & -0.074 & 17.8 & 5.8 $\pm$ 0.2 & 9350 \\ 
 & $^{13}$CO & -0.078 & 15.7 & 5.6 $\pm$ 0.4 & 139 \\ 
 & CS(2-1) & - & - & 5.7$^{*}$ &1.2 \\
 & HCN(1-0) & - & - & 5.7$^{*}$ &0.2 \\
\hline
UGC09519 & CO & -0.087 & 17.5 & 5.0 $\pm$ 0.08 & 4050 \\ 
 & $^{13}$CO & -0.090 & 15.3 & 4.8 $\pm$ 0.7 & 58.1 \\ 
 & CS(2-1) & - & - & 4.9$^{*}$ &0.7 \\
 & HCO$^+$(1-0) & - & - & 4.9$^{*}$ &0.07 \\
\hline
\end{tabular*}
\parbox[t]{0.9 \textwidth}{ \textit{Notes:} Column 1 lists the galaxy name, and Column 2 the transition name. The rotational quantum numbers are indicated in the cases where we only detect one transition. When CO(3-2) data exists we include the parameters derived from the higher rotation temperature component in a second row.  Column 3 and 4 contain the gradient and intercept (at E$_{\rm u}$/k=0) of the best fit line on the rotation diagrams displayed in Figure \ref{NGC2764rotdiag}. Column 5 shows the derived rotation temperature. Where the value is marked with a star the temperature is assumed from the molecule with the best matched critical density, or the mean if more than one such molecule exists. Column 6 is the derived total source-averaged column density of this species, calculated as described in Section \ref{dataredux}.}
\label{trot_table}
\end{table*}

\section{Results and Analysis}
\label{results}
\subsection{Gas properties}
\label{gasprop}

This paper presents the first systematic survey of CS and methanol in early-type galaxies, and as such we discuss the relative intensity of these transitions, and can search for correlations between the properties of these early-type galaxies and the ratios of the different molecular tracers. 
We find the CS(2-1) emission line in ETGs is generally between 10 and 40 times weaker than CO(1-0).   
In some cases it can be brighter than $^{13}$CO(1-0), with the detected lines having fluxes between 0.1 and 1.5 times that of $^{13}$CO. Methanol emission at 3mm is generally of a similar brightness to CS.
Table \ref{obssamp_table} (and Table 1 of  \citealt{2012MNRAS.421.1298C}) lists various physical properties of these systems (derived as part of the \atlas\ project; \citealt{2011MNRAS.413..813C}). We have searched for correlations between the ratios of the detected molecular tracers and these physical properties, and display the clearest ones here.

The top panel of Figure \ref{propertyplot}(a) shows the CS(2-1) to HCN(1-0) integrated intensity ratio for each galaxy of our ETG sample (where detected), plotted against the log of the [OIII]/H$\beta$ ratio. This ionised gas line ratio is often used to determine if the mechanism powering the ionised gas emission is star formation (in which case line ratios $\log_{10}(\mathrm{[OIII]/H\beta})<$-0.2 are expected), or other energy sources (AGN/shocks).  
Three galaxies have CS(2-1)/HCN(1-0) ratios greater than unity, and all are located at low [OIII]/H$\beta$ ratios, suggesting that ongoing massive star-formation dominates the gas ionisation. The critical density required to excite CS(2-1) emission is similar to that of HCN(1-0) \citep{1999ARA&A..37..311E}. However, sulfur-bearing species are shown to be particularly enhanced during strong massive star formation (as the accelerated collapse of cores enhances its formation; \citealt{2005ApJ...620..795L}). These three objects have normal HCN/$^{13}$CO ratios (bottom panel of Figure  \ref{propertyplot}(a)), while their CS/$^{13}$CO ratios (not shown) are enhanced. This supports the hypothesis that CS is a better tracer of dense, star-forming gas than HCN \citep{2005MNRAS.360.1527L,2009ApJ...707..126B}.  

\begin{figure}
\begin{center}
\subfigure[CS enhancement]{\includegraphics[width=0.45\textwidth,angle=0,clip,trim=1.0cm 0.3cm 0.5cm 0.0cm]{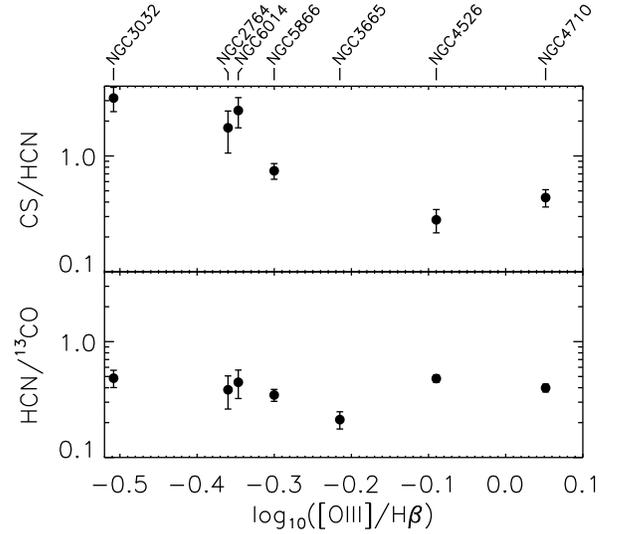}}
\subfigure[Sources of methanol]{\includegraphics[width=0.45\textwidth,angle=0,clip,trim=1.0cm 0.4cm 0.5cm 0.0cm]{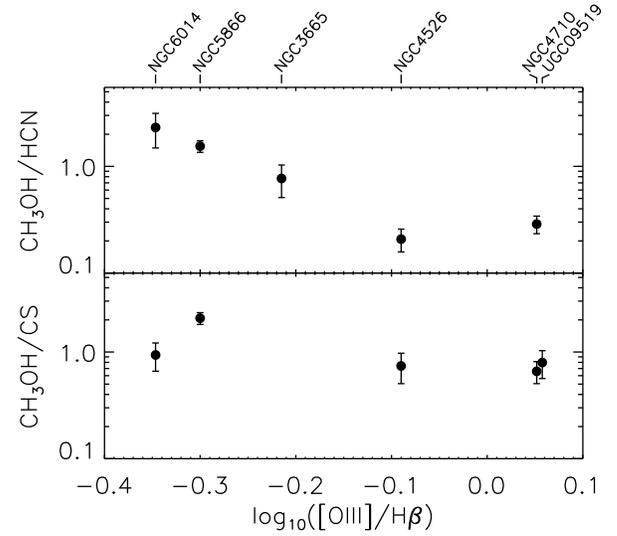}}
 \end{center}
 \caption{Figure 3(a) shows the ratio of the integrated intensity of CS(2-1) and HCN(1-0) (top panel) and HCN(1-0) and $^{13}$CO(1-0) (bottom panel) for the ETGs observed in this work, plotted against the [OIII]/H$\beta$ ratio. Figure 3(b) shows the ratio of the integrated intensity of methanol blend at 96.73 GHz to HCN(1-0) (top) and CS(2-1) (bottom) for the ETGs observed in this work, again plotted against the [OIII]/H$\beta$ line ratio. All the line intensities presented have been corrected for beam dilution effects. Only galaxies where both transitions have been detected are included. The galaxy names are indicated above the top panel of each sub-figure, to aid discussion in Section \ref{gasprop}.}
 \label{propertyplot}
 \end{figure}
 
In Figure \ref{propertyplot}(b) we plot the integrated intensity of the methanol blend at 96.73 GHz against the [OIII]/H$\beta$ ratio introduced above.
Methanol is a grain species, formed in icy grain mantles (no known gas phase pathways can explain the observed galactic abundances; e.g. \citealt{1991ApJ...369..147M}). High gas phase abundances can thus only result for a significant release of mantle ices into the gas phase by some heating mechanism, generally attributed to either shocks or ionisation from high mass star formation. 
Figure \ref{propertyplot}(b) shows that these objects all have low [OIII]/H$\beta$ ratios, which (as explained above), suggests that the methanol enhancement in these sources may be driven by star-formation. Furthermore, the bottom right panel of Figure \ref{propertyplot}(b) shows that (with one exception) the ratio of methanol to CS is not enhanced in these objects, strongly suggesting that high mass star-formation in dense molecular clouds could be the cause of the enhancement in methanol emission.  

We are, of course, unable to completely rule out the presence of shocks accompanying the star formation in these galaxies \citep{2005ApJ...618..259M}. \cite{2012ApJ...755..104M}, for instance, found methanol traces hydrodynamical bar shocks in Maffei 2. Bars, spiral structure, and the presence of AGN can all cause shocks, and enhance star-formation at the same time. NGC5866 has a higher ratio of methanol emission with respect to all the other tracers studied here, and thus we suggest that shocks could be important in this object. This object has a classical bulge, but molecular gas kinematics reveal it may well be barred \citep{2013MNRAS.429..534D}.
Detection of pure shock tracers (such as SiO), or spatially resolving the emission of these tracers would be required to enable us to put stronger constraints on the cause of the methanol emission in early-type galaxies.

The spectral profile of each molecular line is governed by the velocity field of the galaxy coupled with the spatial distribution of the gas emitting in that line. Changes in spectral shape from line to line can indicate different molecular gas properties at different locations within a galaxy. 
The line widths of the  species detected here are not all consistent (some varying by a factor $\approx$2), suggesting that in some galaxies (as discussed in \citealt{2012MNRAS.421.1298C}), different gas lines trace gas components with different spatial extents. For instance NGC3032 has a $\approx$50\% smaller line width at CS(3-2) than at CS(2-1). In this case, however, \cite{2012MNRAS.421.1298C} find a line width more similar to CS(3-2) for other lines in this object. This suggests that profile differences may be caused by noise in this case. NGC4526 has a similar line width in all transitions, but the detected methanol emission line peaks much more strongly around the systemic velocity than the other lines. NGC4710 also has centrally peaked methanol emission, in this case with a smaller line width than the other transitions. This could indicate that some additional mechanism (e.g. shocks) is liberating additional methanol in the central parts of these objects (which both have bars). Spatially resolved observations of these galaxies will clearly be required to determine if this is indeed the case.

\subsubsection{Rotational Diagrams}

Figure \ref{NGC4710rotdiag} and Figure \ref{NGC2764rotdiag} in Appendix B present the rotational diagrams derived for each molecule where we detect at least two lines of that species, for each galaxy. The derived gas rotation temperatures, and total column densities are presented in Table \ref{trot_table}. The temperatures we derive for the simple unblended molecules vary between 3.7 and 8.5 K, with the majority of sources having rotation temperatures around 6 K. Such low temperatures may arise because that the assumption of a single source size for all transitions may be flawed, and/or that the low-$J$ emitting gas may be sub-thermally excited in this system.
This analysis also assumes all transitions (other than CO) are optically thin, which makes it somewhat surprising that between different tracers the derived temperatures (for the low-$J$ lines) in each galaxy vary little.

Given the above caveats (and the large uncertainty induced for blended lines), the derived rotation temperatures from the methanol blends are somewhat different from those derived from the simpler unblended molecules. In NGC4710 the methanol transitions suggest a rotation temperature of  $\approx$8.5 K, larger than the CO/CS rotation temperatures. In NGC5866 the trend is reversed, with the methanol transitions suggesting a rotation temperature of  $\approx$2.6 K, half the CO rotation temperatures of $\approx$5.5 K (formally rotation temperatures lower than the cosmic microwave background temperature are forbidden, suggesting that the real distribution/extent of the methanol emitting component in NGC5866 is very different from that assumed).
Methanol which has been liberated from grains by star-formation is usually present in warm star-forming cores, while methanol released by shocks is generally found in the fast cooling post-shock ISM \citep[e.g.][]{2001A&A...370.1017V}. NGC5866 was identified as an object that may have an enhanced methanol abundance due to shocks in Section \ref{gasprop}.
The methanol emission has much cooler methanol rotation temperatures, as expected from the post-shock cooled ISM. Thus, although some of these differences may arise from the increased uncertainty inherent in a statistical decomposition of blended lines and/or changes in the beam dilution factors, we may also be seeing real differences linked to the mechanisms liberating methanol from grain surfaces (see Section \ref{gasprop}).

The total source averaged column densities of CS derived here range from $\approx$1$\times$10$^{13}$ - 3$\times$10$^{14}$ cm$^{-2}$. These values are in good agreement with the values found for spiral galaxies by \cite{2009ApJ...707..126B}, suggesting this dense gas may be little affected by any differences between spiral and ETGs.
Where available our derived methanol column densities are also reasonable, being similar to those found in NGC253 \citep{1997A&A...326...59H}, and to individual molecular clouds in the Milky Way \citep{2007A&A...466..215L}. In all of our ETG sources (apart from NGC4526) the total column density of HCN is lower than that of CS, ranging from $\approx$1$\times$10$^{13}$ - 6$\times$10$^{14}$ cm$^{-2}$, similar or slightly lower than found in Seyfert galaxies \citep{2007A&A...476..177P}.

\section{Metallicity and $\alpha$-enhancment tracers}
\label{metalpha}

In our previous paper (B12) we considered the effect that a super-solar metallicity or $\alpha$-enhanced ISM has on molecular abundances. We highlighted some molecules that were likely to be good tracers of these types of environments, and in this section we test if these tracers do show any significant trends with respect to metallicity (and $\alpha$-enhancement). 
We first do this empirically in Section \ref{metcomp_empirical}, and then compare with the models of B12 in Section \ref{metcomp_model}.
It is worth bearing in mind that we are forced to treat the effects of metallicity and $\alpha$-enhancement as separable, while in reality there is degeneracy between the two. Further studies of the joint effect of metallicity and $\alpha$-enhancement are certainly required if we wish to set strong constraints on the gas chemistry in these sorts of environments. 

%\clearpage

\subsection{Metallicity}
\label{metcomp_empirical}

In Figure \ref{met_comp_nomod} we show the column density ratio of CS and HCN (which was predicted in B12 to show systematic trends as a function of metallicity), as a function of the measured galaxy metallicity. 

Here we unfortunately have had to use stellar metallicities for the ETGs (from the \atlas\ survey; McDermid et al., in prep), and gas-phase metallicities for the literature galaxies (as detailed in Table \ref{lit_table}). In actively star-forming spiral galaxies the stellar and gas phase metallicities are likely to be closely linked. In ETGs, however, the light is dominated by a metal rich old population, meaning the true gas phase metallicity could be very different (especially if the gas has an external origin; \citealt{2011MNRAS.417..882D}). 
This situation is not ideal, but in this work we continue with the data available. We here use stellar metallicities derived from the very inner parts of these objects (within R$_e$/8), which provides the maximum possible sensitivity to any such changes in the metallicity of the young population, but caution the reader that the error on the metallicity of the ETGs may be larger than indicated by the formal error bar.
 
As described in Section \ref{rotdiag_method}, our estimates of the column density of CS and HCN are lower limits (by a factor $\tau/(1-e^{-\tau})$; \citealt{1999ApJ...517..209G}), as we have no information on the optical depth in these transitions (this is not the case in the Galactic centre and Centaurus A, where the molecular column density estimates we use come from optically thin tracers). An upper limit can be set to the CS optical depth using our IRAM-30m observations, which included C$^{34}$S within the bandpass. Assuming a $^{32}$S/$^{34}$S ratio of 8 \citep[derived in NGC253, which has a high gas-phase metallicity;][]{2006ApJS..164..450M}, the upper limits to the brightness temperature of C$^{34}$S imply CS optical depths of $<5$ in the early-type objects. These estimates are consistent with those of \cite{2013A&A...549A..39A}, who found $\tau_{\rm CS(2-1)}$=1.53 and  $\tau_{\rm HCN(1-0)}$=1.55 in NGC1068, and  \cite{2006ApJS..164..450M} who found $\tau_{\rm CS(3-2)}$=3.2 in NGC253. We proceed from this point assuming that $\tau_{\rm CS(2-1,3-2)}$ and $\tau_{\rm HCN(1-0)}$ are $< 5$ in these sources, and (as their critical densities are similar- Nc$_{\rm HCN(1-0)}$ = 2.6$\times$10$^6$, Nc$_{\rm CS(3-2)}$ = 1.3$\times$10$^6$; \citealt{1999ARA&A..37..311E}) additionally that the optical depth in these two species should be similar (within a factor of $\approx$2). These uncertainties are included in the error bars presented in Figure \ref{met_comp_nomod}.  
 
Despite these large uncertainties an empirical correlation is observed between the measured molecular column density ratio and the measured metallicity in Figure \ref{met_comp_nomod} (with a Spearmans rank correlation coefficient of 0.64). The best fit linear relationship is shown below:

\begin{equation}
[Z/H]=(0.79\pm0.17) \times \log_{10}\left(\frac{N(\mathrm{HCN})}{N(\mathrm{CS})}\right)+(0.08\pm0.09)
 \label{met_eqn_empirical}
\end{equation}

We caution however that because of the mix of abundance measurements used, and the systematics inherent in converting line intensities to column densities,  the uncertainties on the best fit line could be larger still. Despite the 0.3 dex scatter around this best fit line, the fact that a rough relation exists suggests that molecular column density ratios do change systematically as a function of metallicity. This raises the intriguing possibility  that one could eventually estimate the metallicity of the ISM from observations in the millimetre regime.

\begin{figure}
\begin{center}
\includegraphics[width=0.5\textwidth,angle=0,clip,trim=0.8cm 0.0cm 0.0cm 0.0cm]{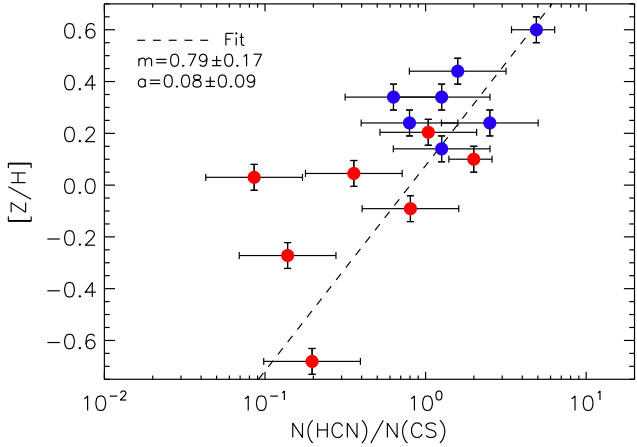}
 \end{center}
 \caption{Total column density ratio of HCN and CS, plotted against measured metallicity. Points based on literature data for spiral and starburst galaxies are coloured blue, while ETGs are coloured red. The error bars reflect the range induced if the HCN and CS optical depths are different from each other by a factor of two. A fit to the points is indicated with the dashed line, and the best fit gradient and slope are indicated in the top left of the figure.}
 \label{met_comp_nomod}
 \end{figure}

\subsubsection{Comparison with models}
\label{metcomp_model}
Figure \ref{met_comp_diag} shows the fractional abundance ratios for CO, HCN and CS as a function of metallicity, predicted from the models of B12.  We include lines for varying optical depths between 1 and 20 visual magnitudes (A$_{\rm V}$), and shade the region between the A$_{\rm V}$=3 and A$_{\rm V}$=20 lines.
In the PDR models of B12 the optical depth at $V$-band is computed assuming a gas-to-dust ratio of 100, and a mix of silicon and carbon dust grains. It is the main parameter in these models, as it controls how much of the incident UV light is able to penetrate and drive chemical reactions in the cloud. For full details see \cite{2006ApJS..164..506L}.

We provide here linear approximations to the N(HCN)/H(CS) vs [Z/H] relations (as a function of A$_{\rm V}$) in the hope they will prove useful to the community. These approximations are valid in the metallicity range 0.5-3 $Z_{\odot}$ only (as the N(HCN)/N(CS) vs metallicity relation is bimodal at low metallicities). Table \ref{met_rel_table} contains the linear fit coefficients and the formal errors in the fitted parameters, and should be used in the following functional form.
\begin{equation}
[Z/H]=m \times \log_{10}\left(\frac{N(\mathrm{HCN})}{N(\mathrm{CS})}\right)+c
 \label{met_eqn}
\end{equation}

 In the models of B12 fractional abundance ratios are directly proportional to column density ratios. Assuming the same is true in our observations we are able to overplot the column density ratios derived from observations of our spiral (blue points) and ETGs (red points). The models of B12 go up to a metallicity of 3 times solar, and here we cross-hatch the region between metallicites of 3 and 5 times solar, based on an extrapolation of the model trends. We caution against drawing any conclusions in this region. 
Within their error bars (calculated as described above, and including additional error terms to account for our lack of knowledge of the optical depth) all of the observed data points lie within the model regions predicted by B12. Although the range of allowed column densities is large, this at least gives us confidence that the models produce reasonable solutions. 
The diagrams all give reasonably consistent results, with every point (within its errors) lying between an A$_{\rm v}$ of 1 and 5. These are typical visual extinctions we would expect for PDRs \citep{2009ApJ...706.1323M}, which sit at the edges of molecular clouds. 

Given this, Figure \ref{met_relation} shows the metallicity we would predict for each galaxy based on the N(HCN)/N(CS) ratio (which was predicted by B12 to be the most sensitive to metallicity) and assuming we are probing gas at an A$_{\rm v}$ of 3, as expected for PDR regions \citep{2009ApJ...706.1323M}. We compare the predicted metallicity against the measured values, and find that there is a relation. The points do not follow the one-to-one line however, as at high metallicites simply using the model relations (with an assumed A$_{\rm v}$ of 3) causes us to underestimate the measured metallicity of the gas. This is to be expected, given that the gradient of the relation between N(HCN)/N(CS) and metallicity predicted by the model (0.93$\pm$0.11 dex$^{-1}$) is somewhat steeper than the empirical relation derived in Section \ref{metcomp_empirical} (0.79$\pm$0.17 dex$^{-1}$).  
The cause of this difference in slope is unclear, but this could arise due to issues in the assumption of an A$_{\rm V}$, or calibration offsets between the different metallicity measures used to create this plot.
We also caution that models of B12 are for photon-dominated regions only. Galaxies where the gas energy balance is dominated by X-ray irradiation or cosmic-rays will likely not follow these relations. Further studies of the response of gas chemistry in high metallicity environments will clearly be required before any such method can be calibrated, and applied robustly.

\begin{figure}
\begin{center}
\includegraphics[width=0.45\textwidth,angle=0,clip,trim=1.0cm 2.0cm 4.5cm 1.0cm]{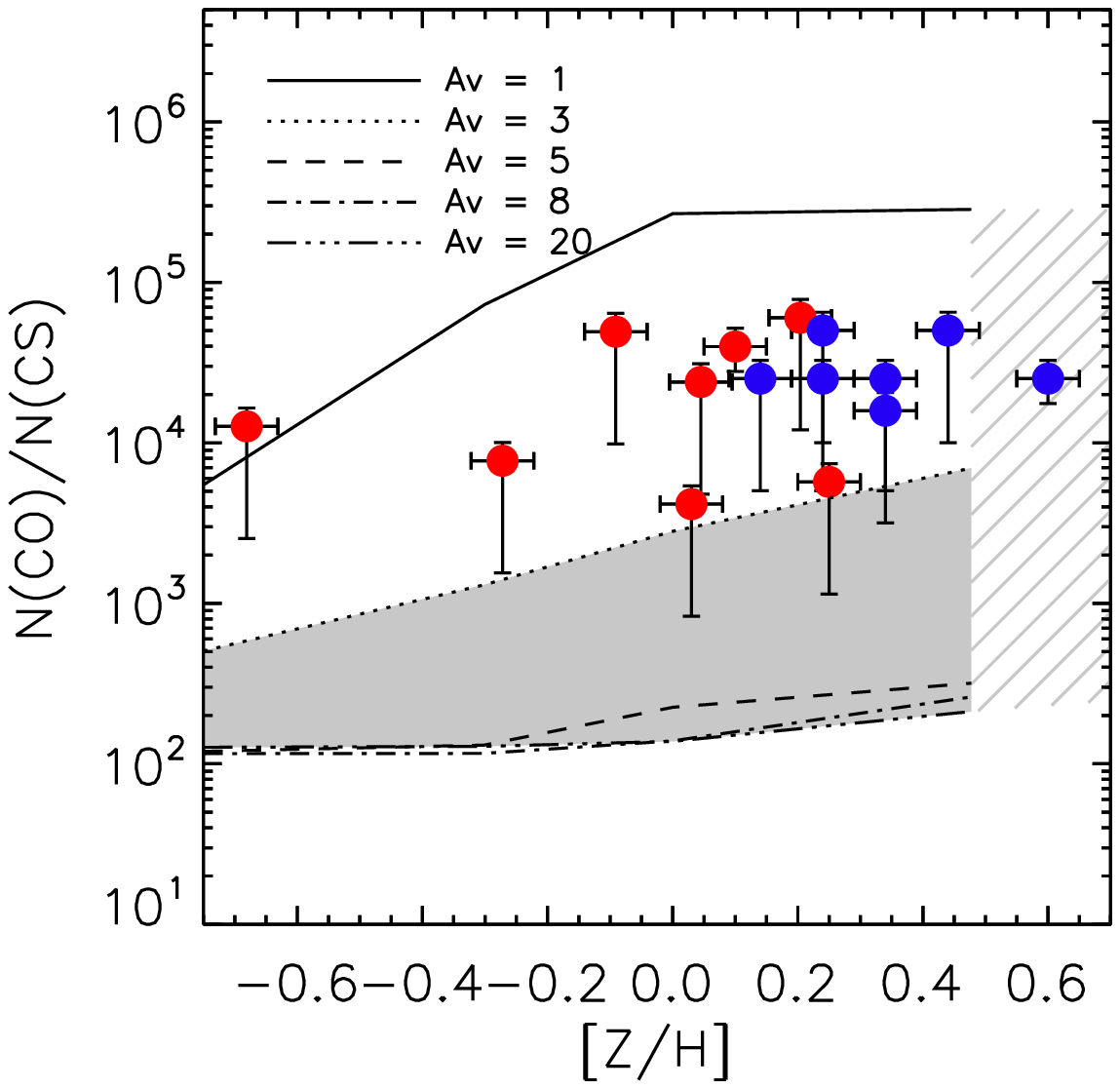}
\includegraphics[width=0.45\textwidth,angle=0,clip,trim=1.0cm 2.0cm 4.5cm 0.9cm]{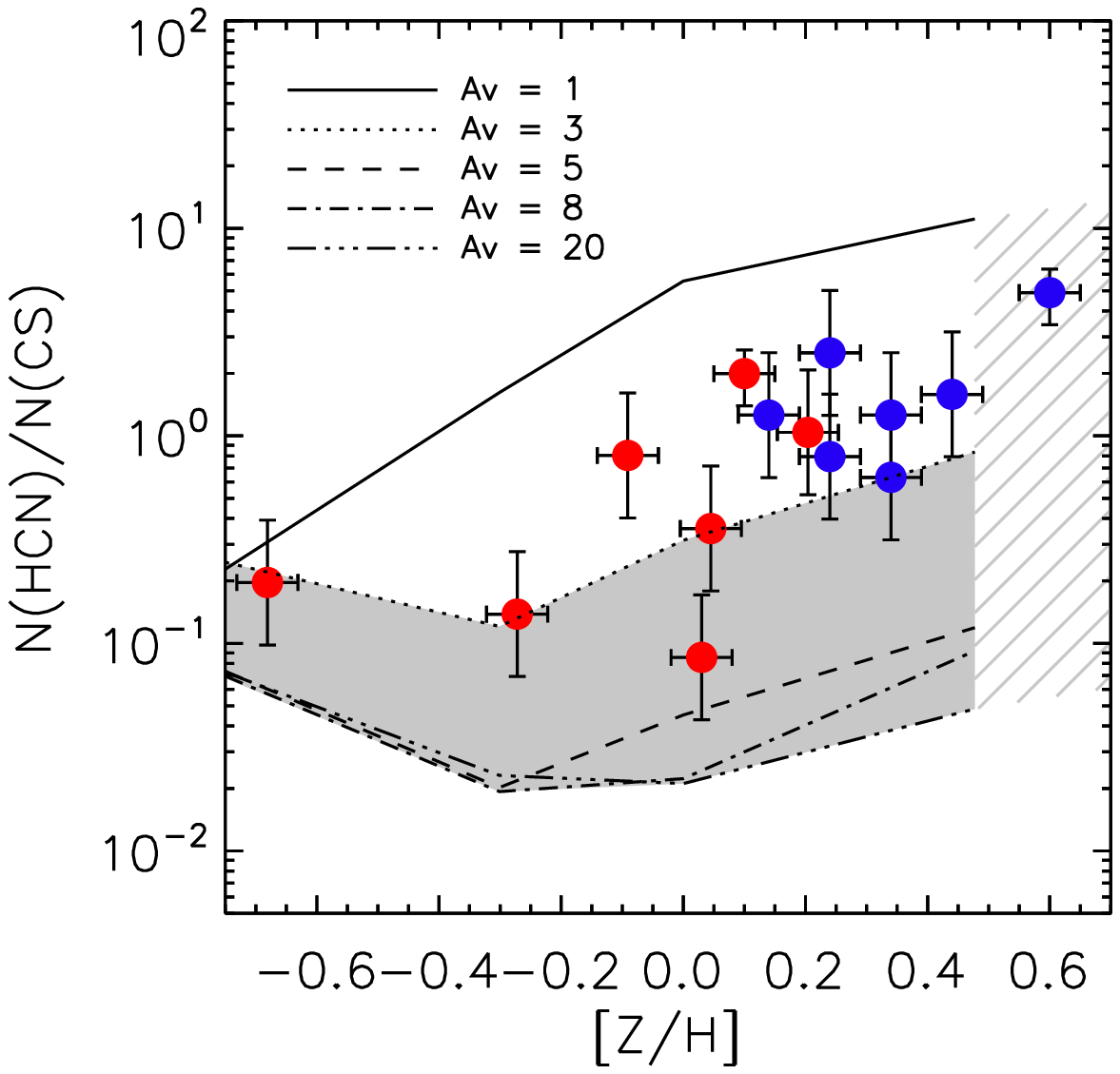}
\includegraphics[width=0.45\textwidth,angle=0,clip,trim=1.0cm 0.35cm 4.5cm 0.9cm]{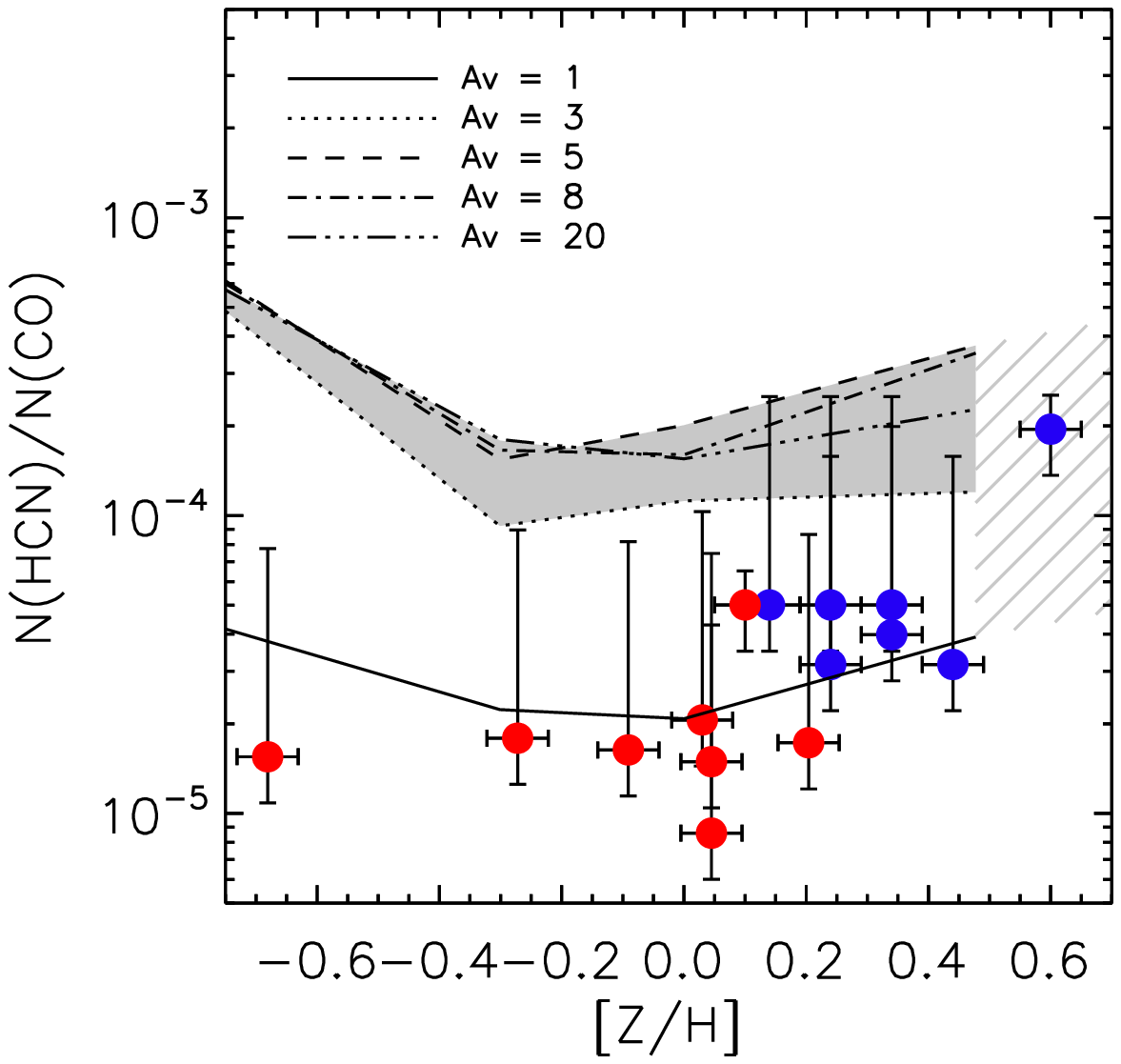}
 \end{center}
 \caption{Total column density ratios of our molecular tracers plotted against metallicity. Points based on literature data for spiral and starburst galaxies are coloured blue, while ETGs are coloured red. The model grids from B12 are included, we here show relations for varying optical depths between 1 and 20 A$_{\rm v}$, and shade the region between the A$_{\rm v}$=3 and A$_{\rm v}$=20 relations. We cross-hatch the region between 3 and 5 times solar metallicity, based on an extrapolation of the model trends.  The error bars in the top and bottom panels include the effect of varying the unknown optical depth of CS and HCN between 0 and 5. In the middle panel the error bars reflect the range induced if the HCN and CS optical depths are different from each other by a factor of two.}
 \label{met_comp_diag}
 \end{figure}
 
 \begin{figure}
\begin{center}
\includegraphics[width=0.45\textwidth,angle=0,clip,trim=0.5cm 0cm 0.5cm 0.8cm]{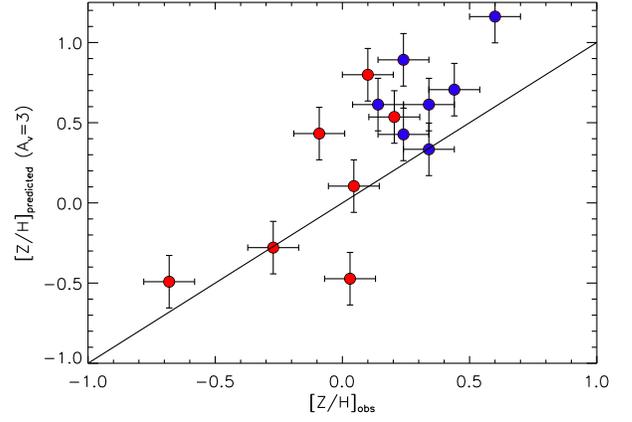}
 \end{center}
 \caption{Metallicity predicted from the N(HCN)/N(CS) ratio assuming an A$_{\rm v}$ of 3, plotted against the observed metallicity. The line of equality is shown as a guide to the eye. The scatter around the one-to-one relation is 0.36 dex. Spirals are shown as blue-points, and ETGs are shown in red. Maffei 2 is not included here as the observed line ratio lies well outside the model grid for A$_{\rm v}$=3.}
 \label{met_relation}
 \end{figure}

\begin{table}
\caption{Linear fits to the model N(HCN)/N(CS) vs [Z/H] relations}
\begin{tabular*}{0.45\textwidth}{@{\extracolsep{\fill}}r r r}
\hline
A$_{\rm v}$ & m & c \\
(mag) & (dex$^{-1}$) & (dex)  \\
 (1) & (2) & (3) \\
\hline
 1 &      1.06 $\pm$      0.24 &      -0.69 $\pm$      0.19\\
 3 &      0.93 $\pm$      0.11 &       0.52 $\pm$      0.07\\
 5 &      1.01 $\pm$      0.08 &       1.40 $\pm$      0.11\\
 8 &      1.00 $\pm$      0.32 &       1.52 $\pm$      0.48\\
 20 &      1.75 $\pm$      0.94 &       2.76 $\pm$      1.46\\
\hline
\end{tabular*}
\parbox[t]{0.45 \textwidth}{ \textit{Notes:} This table contains linear approximations to the N(HCN)/H(CS) vs [Z/H] relations as a function of optical depth (column 1). Columns 2 and 3 contain the gradient and intercept (respectively) of a linear fit (and their formal fitting errors), parameterized as in Equation \ref{met_eqn}. These approximations are valid in the metallicity range 0.5-3 $Z_{\odot}$ only. }

\label{met_rel_table}
\end{table}

\subsection{$\alpha$-enhancement}

Figure \ref{alpha_comp_diag} shows the observed HCN, CO and CS column density ratios plotted against the \textit{stellar} $\alpha$-enhancements for the \atlas\ ETGs. We do not include our spiral sample here, as we have no information on the abundance of $\alpha$-elements in these systems. We again show the model grid from B12 (for the \citealt{1998A&A...335..943S} $\alpha$-enhancement scenario), for varying optical depths between 1 and 20 A$_{\rm v}$. The $\alpha$-enhanced models of B12 go up to an enhancement of 0.4 dex, and here we cross-hatch the region between $\alpha$-enhancements of 0.4 and 0.5 dex, which are based on an extrapolation of the model trends. We caution against drawing any conclusions in this region. 

The observed data points lie within the model grids. However, if one believes that the models of B12 are correct, and again assumes a mean A$_{\rm V}$ of 3 magnitudes, in the N(CO)/N(CS) and N(HCN)/N(CS) diagrams almost every galaxy would be consistent with having no \textit{gas-phase} $\alpha$-enhancement. For the N(HCN)/N(CO) panel (identified as the most sensitive by B12) the majority of points could be consistent with the measured stellar $\alpha$-enhancements, but the uncertainties are large. 

If the gas in these objects was formed through stellar mass loss from the $\alpha$-enhanced stellar populations (and we assume that the models are correct), seeing no gas-phase $\alpha$-enhancement would imply that the gas has been around long enough for SNe Ia to enrich the ISM with iron peak elements (diluting the $\alpha$-enhancement). If alternatively the gas has been obtained from external sources this would imply that these sources were not $\alpha$-enhanced (as one would expect from minor merger scenarios; e.g. \citealt{2011MNRAS.411.2148K}). \cite{2011MNRAS.417..882D} argue over half the gas in local ETGs has been accreted, and that Virgo objects (which are unable to accrete new gas) have had this gas several gigayears. In this context, it is thus not too surprising that little evidence is found for $\alpha$-enhancements in the gas-phase in these objects. Of course, one should be cautious when drawing any firm conclusions from these results, as the lack of correlation could also be because of the large uncertainties in our measurements, because the models do not include sufficient physics to capture the effect of $\alpha$-enhancements on the ISM, or that the degeneracies with respect to changing metallicity are strong. 

A study of isolated ETGs, where the gas is expected to have recently cooled from stellar mass loss, may be required to determine if these molecular diagnostics truly are able to trace $\alpha$-enhancements. Alternatively gas in the vicinity of core-collapse SNe in our own galaxy may show such a signature. 
In the ALMA (Atacama Large Millimeter/submillimeter Array) era it will also be possible to probe molecular gas at high redshifts (z\gtsimeq6), where SNe Ia have had insufficient time to provide any iron enrichment, and these diagnostics may prove useful there.

\begin{figure}
\begin{center}
\includegraphics[width=0.45\textwidth,angle=0,clip,trim=1.0cm 2.0cm 4.5cm 1.0cm]{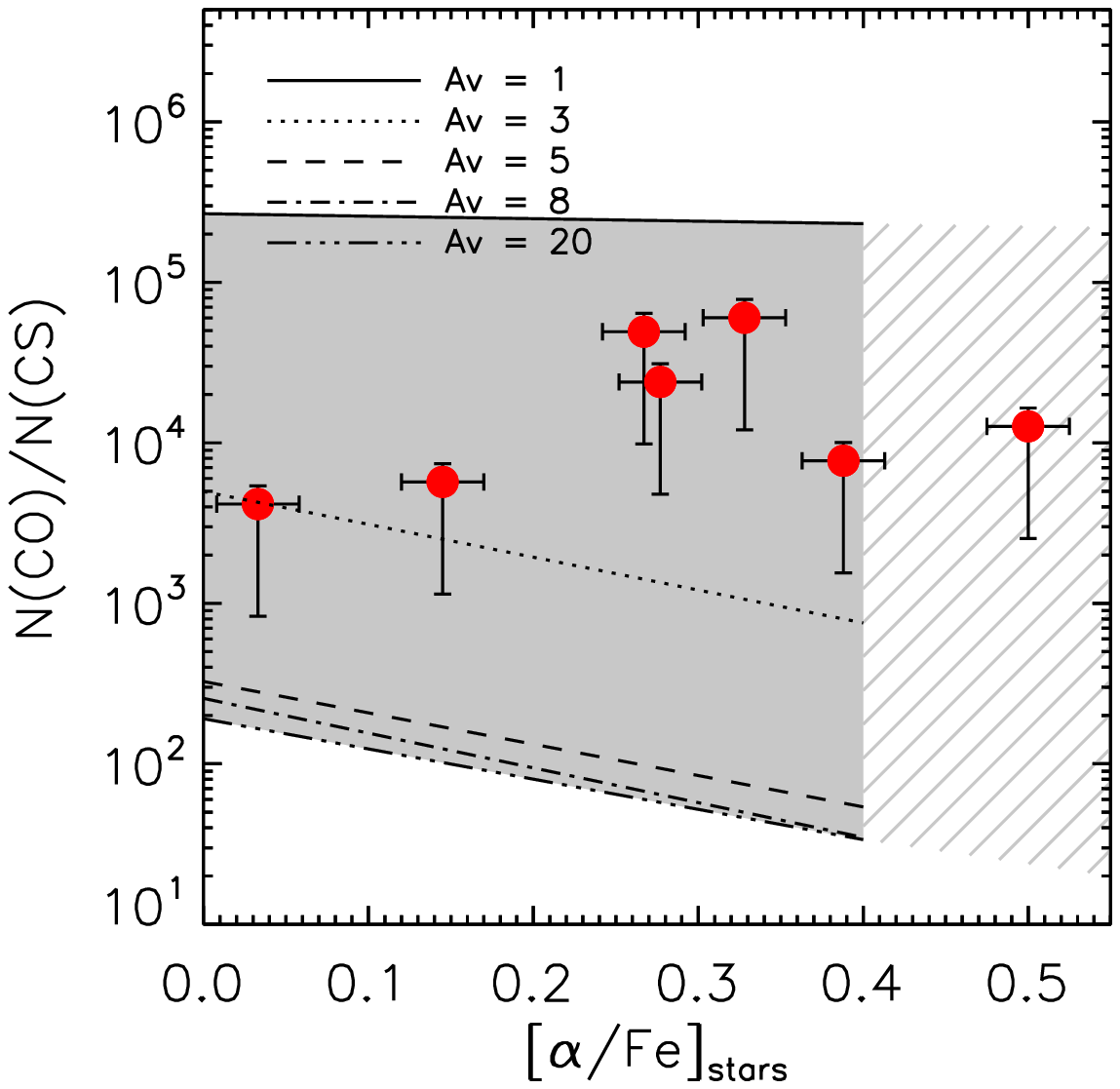}
\includegraphics[width=0.45\textwidth,angle=0,clip,trim=1.0cm 2.0cm 4.5cm 0.9cm]{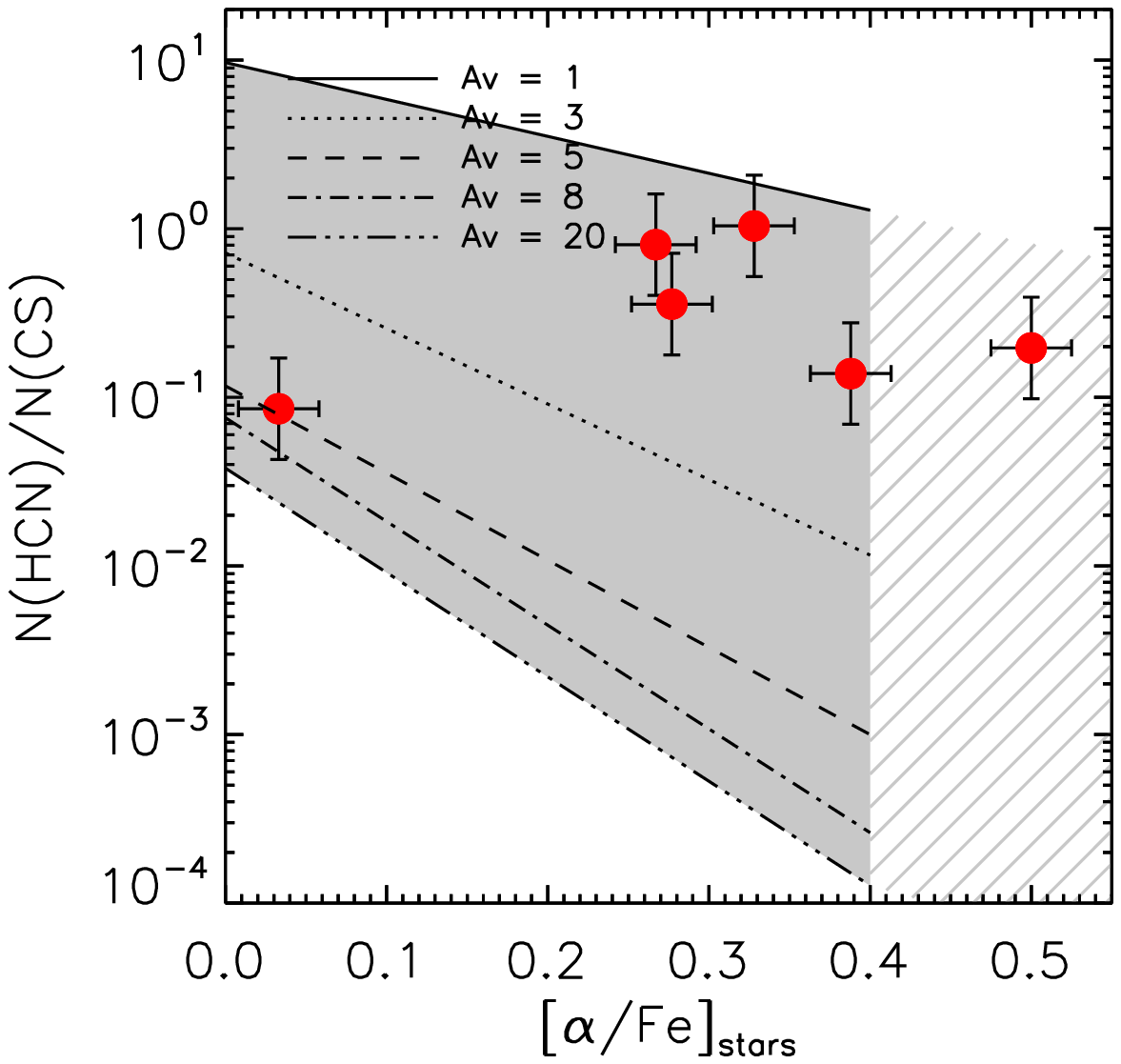}
\includegraphics[width=0.45\textwidth,angle=0,clip,trim=1.0cm 0.5cm 4.5cm 0.9cm]{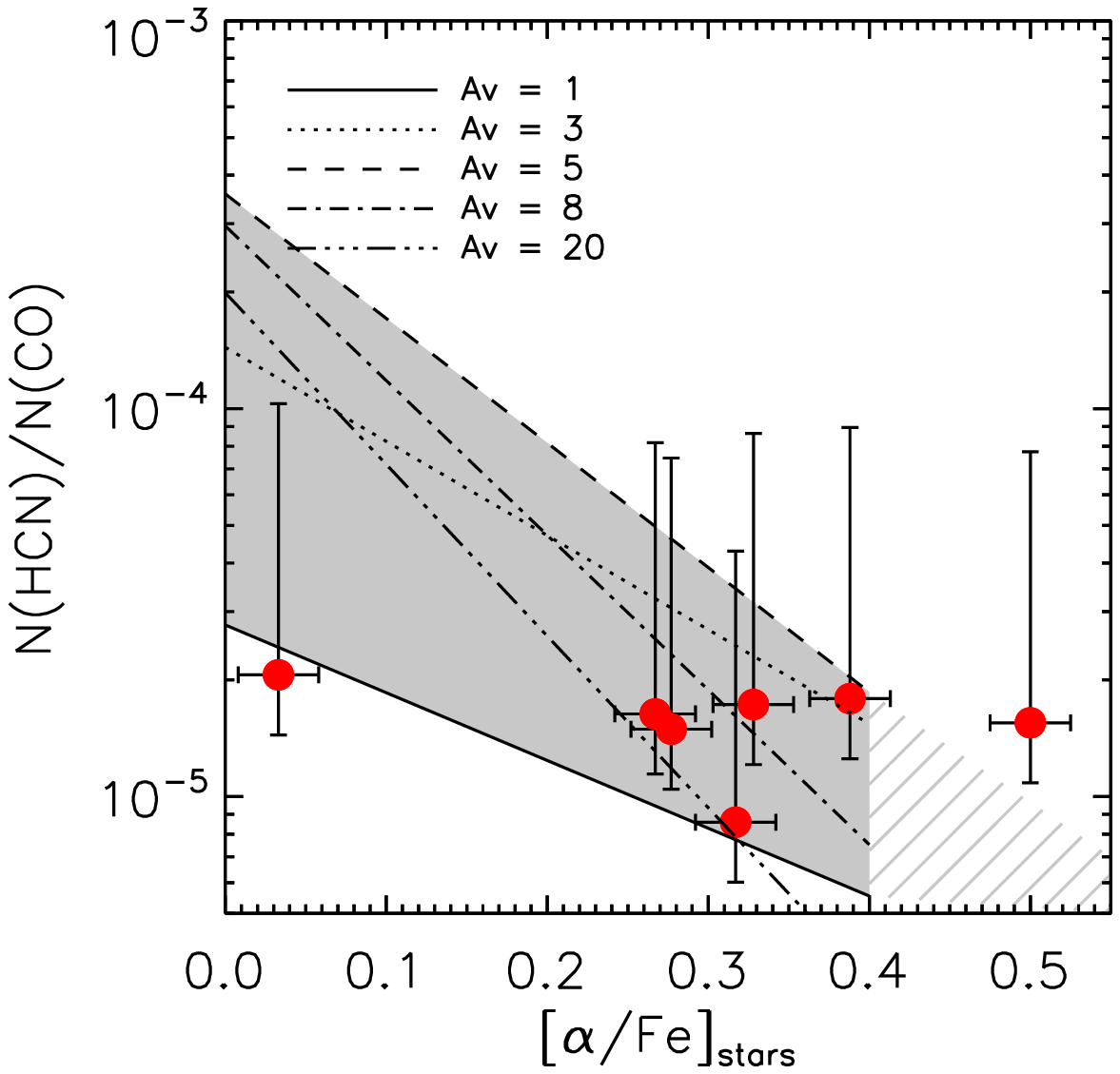}
 \end{center}
 \caption{Total column density ratios of our molecular tracers plotted against stellar $\alpha$-element enhancement. Our new observations of ETGs are coloured red. We do not include our spiral sample, as we have no information on their $\alpha$-element abundances. The model grids from B12 are included, we here show relations for varying optical depths between 1 and 20 A$_{\rm v}$. We cross-hatch the region between $\alpha$-enhancements of 0.4 and 0.5 dex, based on an extrapolation of the model trends. The error bars in the top and bottom panels include the effect of varying the unknown optical depth of CS and HCN between 0 and 5. In the middle panel the error bars reflect the range induced if the HCN and CS optical depths are different from each other by a factor of two.}
 \label{alpha_comp_diag}
 \end{figure}

\section{Conclusions}
\label{conclude}

In this paper we have used observations of molecular tracers in metal rich and $\alpha$-enhanced galaxies to present an exploratory study of the effect of abundance changes on molecular chemistry. We selected a sample of metal rich spiral and star-bursting objects from the literature (and additionally included observations of Sgr~A* and Cen. A), and present here new data for a sample of ETGs previously studied by  \cite{2012MNRAS.421.1298C}. We conducted the first survey of CS emission in ETGs, detecting at least one transition in 7 objects. We also obtained detections of methanol emission in 5 ETGs.

We compared flux ratios of the CS and methanol lines to those already observed (HCN, CO) to determine if and how the presence of these emission lines is linked to galaxy properties. We find that the systems where star-formation is the dominant gas ionisation mechanism have enhanced CS emission, compared to their HCN emission, supporting the hypothesis that CS is a better tracer of dense star-forming gas than HCN. 

We have shown evidence that the methanol enhancement in these sources may be driven by high mass star-formation in dense molecular clouds, consistent with the lack of methanol enhancement when compared to CS emission.  NGC5866 has a higher ratio of methanol emission with respect to all the other tracers, and cooler methanol rotation temperatures, and thus shocks may be more important in this source. Detection of pure shock tracers (such as SiO), and/or spatially resolving the emission of these tracers would be required to enable us to put stronger constraints on the cause of the methanol emission in early-type galaxies.

We constructed rotation diagrams for each early-type source where at least two molecules of a given species were detected. This is the first time rotation temperatures have been studied for a sample of ETGs. The temperatures we derive vary between 3 and 10~K, with the majority of sources having rotation temperatures around 6~K. Despite the large uncertainties (arising from our assumptions of LTE, identical extents for the regions emitting in different transitions, a single rotational temperature, etc) the derived source averaged total column densities are similar to those found by other authors for normal spiral and starburst galaxies, suggesting dense clouds may be little affected by the differences between early and late type galaxies.
 
Finally we used our ETG and literature sample to determine if these objects show any systematic change in total column density ratios that correlate with the metallicity of the source. Despite the large uncertainties we do find a correlation, with a scatter of $\approx$0.3 dex. This suggests that with further study it may be possible to calibrate a metallicity tracer for the molecular ISM in this way. 

We additional attempt to verify if the correlation observed matches that predicted by the chemical models of B12. We find that the observed datapoints almost always lie within the model grid from B12, providing some confidence that the model predictions are valid for these types of sources. 

As shown in B12, understanding the mean optical depths at which the molecular emission lines arise is crucial to get an accurate determination of the metallicity of the gas. However we show here that, for this selection of sources, crudely assuming a mean optical depth of 3 magnitudes results in a reasonable determination of the metallicity, with a scatter of $\approx$0.35 dex. The relation predicted by the model is steeper than that found empirically, which may be due to the large uncertainties, or a break down in our simplifying assumptions.  We provide the community with linear approximations to both the observed and predicted relationship between the N(HCN)/N(CS) ratio and metallicity. Further study will clearly be required to determine if this, or any, molecular tracer can be used to robustly determine metallicity. That a relationship exists at all, given the large uncertainties are crude assumptions made here, is promising. 

When we compared our data to the models of $\alpha$-enhancement from B12 we find that the molecular measures show little correlation with stellar $\alpha$-enhancement. This could suggest that the gas has an external origin, or has been present in the galaxy long enough for SNe Ia to enrich it. Alternatively the lack of a correlation could be due to our large uncertainties, or because current models do not fully treat the degeneracy between metallicity and $\alpha$-enhancement. Other evidence suggests that the gas could really have different $\alpha$-abundances to the stars, but we are unable to determine which explanation is correct with our current data.

This study, along with B12, represents a first tentative step towards understanding the chemistry of molecular gas in high metallicity environments. We show that some molecular tracers do indeed show systematic variations that appear to correlate with metallicity, and that these variations can be roughly reproduced by chemical models. With the power of the next generation of mm-facilities (e.g. ALMA, Large Millimeter Telescope, Cornell Caltech Atacama Telescope) we will be able to study such variations in detail (both in our own Milky Way, and other galaxies), shedding light on the physics and chemistry of the ISM, and perhaps providing a new method of estimating metallicity from the molecular gas phase.

\noindent \textbf{Acknowledgments}

We thank the referee for useful suggestions which improved this paper, and R. McDermid (and the \atlas\ team) for letting us use their metallicity estimates. 
The research leading to these results has received funding from the European
Community's Seventh Framework Programme (/FP7/2007-2013/) under grant agreement
No 229517. This paper is based on observations carried out with the IRAM Thirty Meter Telescope. IRAM is supported by INSU/CNRS (France), MPG (Germany) and IGN (Spain).

\bsp
\bibliographystyle{mn2e}
\bibliography{bibCSMETH}

\begin{thebibliography}{44}
\expandafter\ifx\csname natexlab\endcsname\relax\def\natexlab#1{#1}\fi

\bibitem[{Aladro {et~al}\mbox{.}(2011)Aladro, Mart{\'\i}n, Mart{\'\i}n-Pintado,
  Mauersberger, Henkel, Oca{\~n}a~Flaquer, \&
  Amo-Baladr{\'o}n}]{2011A&A...535A..84A}
Aladro R., Mart{\'\i}n S., Mart{\'\i}n-Pintado J., Mauersberger R., Henkel C.,
  Oca{\~n}a~Flaquer B., Amo-Baladr{\'o}n M.~A., 2011, Astronomy and
  Astrophysics, 535, 84

\bibitem[{Aladro {et~al}\mbox{.}(2013)Aladro, Viti, Bayet, Riquelme,
  Mart{\'\i}n, Mauersberger, Mart{\'\i}n-Pintado, Requena-Torres, Kramer, \&
  Wei{\ss}}]{2013A&A...549A..39A}
Aladro R. {et~al.}, 2013, Astronomy and Astrophysics, 549, 39

\bibitem[{Alatalo {et~al}\mbox{.}(2012)Alatalo, Davis, Bureau, Young, Blitz,
  Crocker, Bayet, Bois, Bournaud, Cappellari, Davies, de~Zeeuw, Duc, Emsellem,
  Khochfar, Krajnovic, Kuntschner, Lablanche, Morganti, McDermid, Naab,
  Oosterloo, Sarzi, Scott, Serra, \& Weijmans}]{2012arXiv1210.5524A}
Alatalo K. {et~al.}, 2012, arXiv, 5524

\bibitem[{Asplund {et~al}\mbox{.}(2004)Asplund, Grevesse, Sauval,
  Allende~Prieto, \& Kiselman}]{2004A&A...417..751A}
Asplund M., Grevesse N., Sauval A.~J., Allende~Prieto C., Kiselman D., 2004,
  Astronomy and Astrophysics, 417, 751

\bibitem[{Bayet {et~al}\mbox{.}(2009)Bayet, Aladro, Mart{\'\i}n, Viti, \&
  Mart{\'\i}n-Pintado}]{2009ApJ...707..126B}
Bayet E., Aladro R., Mart{\'\i}n S., Viti S., Mart{\'\i}n-Pintado J., 2009, The
  Astrophysical Journal, 707, 126

\bibitem[{Bayet {et~al}\mbox{.}(2012{\natexlab{a}})Bayet, Bureau, Davis, Young,
  Crocker, Alatalo, Blitz, Bois, Bournaud, Cappellari, Davies, de~Zeeuw, Duc,
  Emsellem, Khochfar, Krajnovic, Kuntschner, McDermid, Morganti, Naab,
  Oosterloo, Sarzi, Scott, Serra, \& Weijmans}]{2012arXiv1212.2630B}
Bayet E. {et~al.}, 2012{\natexlab{a}}, arXiv, 2630

\bibitem[{Bayet {et~al}\mbox{.}(2012{\natexlab{b}})Bayet, Davis, Bell, \&
  Viti}]{2012MNRAS.424.2646B}
Bayet E., Davis T.~A., Bell T.~A., Viti S., 2012{\natexlab{b}}, Monthly Notices
  of the Royal Astronomical Society, 424, 2646

\bibitem[{Cappellari {et~al}\mbox{.}(2011)Cappellari, Emsellem, Krajnovic,
  McDermid, Scott, Verdoes~Kleijn, Young, Alatalo, Bacon, Blitz, Bois,
  Bournaud, Bureau, Davies, Davis, de~Zeeuw, Duc, Khochfar, Kuntschner,
  Lablanche, Morganti, Naab, Oosterloo, Sarzi, Serra, \&
  Weijmans}]{2011MNRAS.413..813C}
Cappellari M. {et~al.}, 2011, Monthly Notices of the Royal Astronomical
  Society, 413, 813

\bibitem[{Combes, Young \& Bureau(2007)Combes, Young, \&
  Bureau}]{2007MNRAS.377.1795C}
Combes F., Young L.~M., Bureau M., 2007, Monthly Notices of the Royal
  Astronomical Society, 377, 1795

\bibitem[{Crocker {et~al}\mbox{.}(2012)Crocker, Krips, Bureau, Young, Davis,
  Bayet, Alatalo, Blitz, Bois, Bournaud, Cappellari, Davies, de~Zeeuw, Duc,
  Emsellem, Khochfar, Krajnovic, Kuntschner, Lablanche, McDermid, Morganti,
  Naab, Oosterloo, Sarzi, Scott, Serra, \& Weijmans}]{2012MNRAS.421.1298C}
Crocker A. {et~al.}, 2012, Monthly Notices of the Royal Astronomical Society,
  421, 1298

\bibitem[{Davis {et~al}\mbox{.}(2013)Davis, Alatalo, Bureau, Cappellari, Scott,
  Young, Blitz, Crocker, Bayet, Bois, Bournaud, Davies, de~Zeeuw, Duc,
  Emsellem, Khochfar, Krajnovic, Kuntschner, Lablanche, McDermid, Morganti,
  Naab, Oosterloo, Sarzi, Serra, \& Weijmans}]{2013MNRAS.429..534D}
Davis T.~A. {et~al.}, 2013, Monthly Notices of the Royal Astronomical Society,
  429, 534

\bibitem[{Davis {et~al}\mbox{.}(2011)Davis, Alatalo, Sarzi, Bureau, Young,
  Blitz, Serra, Crocker, Krajnovic, McDermid, Bois, Bournaud, Cappellari,
  Davies, Duc, de~Zeeuw, Emsellem, Khochfar, Kuntschner, Lablanche, Morganti,
  Naab, Oosterloo, Scott, \& Weijmans}]{2011MNRAS.417..882D}
Davis T.~A. {et~al.}, 2011, Monthly Notices of the Royal Astronomical Society,
  417, 882

\bibitem[{Eckart {et~al}\mbox{.}(1990)Eckart, Cameron, Genzel, Jackson,
  Rothermel, Stutzki, Rydbeck, \& Wiklind}]{1990ApJ...365..522E}
Eckart A., Cameron M., Genzel R., Jackson J.~M., Rothermel H., Stutzki J.,
  Rydbeck G., Wiklind T., 1990, Astrophysical Journal, 365, 522

\bibitem[{Evans(1999)}]{1999ARA&A..37..311E}
Evans N. J.~I., 1999, Annual Review of Astronomy and Astrophysics, 37, 311

\bibitem[{Galliano, Dwek \& Chanial(2008)Galliano, Dwek, \&
  Chanial}]{2008ApJ...672..214G}
Galliano F., Dwek E., Chanial P., 2008, The Astrophysical Journal, 672, 214

\bibitem[{Goldsmith \& Langer(1999)}]{1999ApJ...517..209G}
Goldsmith P.~F., Langer W.~D., 1999, The Astrophysical Journal, 517, 209

\bibitem[{H{\"u}ttemeister, Mauersberger \& Henkel(1997)H{\"u}ttemeister,
  Mauersberger, \& Henkel}]{1997A&A...326...59H}
H{\"u}ttemeister S., Mauersberger R., Henkel C., 1997, Astronomy and
  Astrophysics, 326, 59

\bibitem[{Kaviraj {et~al}\mbox{.}(2011)Kaviraj, Tan, Ellis, \&
  Silk}]{2011MNRAS.411.2148K}
Kaviraj S., Tan K.-M., Ellis R.~S., Silk J., 2011, Monthly Notices of the Royal
  Astronomical Society, 411, 2148

\bibitem[{Krips {et~al}\mbox{.}(2010)Krips, Crocker, Bureau, Combes, \&
  Young}]{2010MNRAS.407.2261K}
Krips M., Crocker A.~F., Bureau M., Combes F., Young L.~M., 2010, Monthly
  Notices of the Royal Astronomical Society, 407, 2261

\bibitem[{Le~Petit {et~al}\mbox{.}(2006)Le~Petit, Nehm{\'e}, Le~Bourlot, \&
  Roueff}]{2006ApJS..164..506L}
Le~Petit F., Nehm{\'e} C., Le~Bourlot J., Roueff E., 2006, The Astrophysical
  Journal Supplement Series, 164, 506

\bibitem[{Lee \& Lee(2003)}]{2003JKAS...36..271L}
Lee C.~W., Lee H.~M., 2003, Journal of the Korean Astronomical Society, 36, 271

\bibitem[{Leurini {et~al}\mbox{.}(2007)Leurini, Schilke, Wyrowski, \&
  Menten}]{2007A&A...466..215L}
Leurini S., Schilke P., Wyrowski F., Menten K.~M., 2007, Astronomy and
  Astrophysics, 466, 215

\bibitem[{Lintott {et~al}\mbox{.}(2005{\natexlab{a}})Lintott, Viti, Rawlings,
  Williams, Hartquist, Caselli, Zinchenko, \& Myers}]{2005ApJ...620..795L}
Lintott C.~J., Viti S., Rawlings J. M.~C., Williams D.~A., Hartquist T.~W.,
  Caselli P., Zinchenko I., Myers P., 2005{\natexlab{a}}, The Astrophysical
  Journal, 620, 795

\bibitem[{Lintott {et~al}\mbox{.}(2005{\natexlab{b}})Lintott, Viti, Williams,
  Rawlings, \& Ferreras}]{2005MNRAS.360.1527L}
Lintott C.~J., Viti S., Williams D.~A., Rawlings J. M.~C., Ferreras I.,
  2005{\natexlab{b}}, Monthly Notices of the Royal Astronomical Society, 360,
  1527

\bibitem[{Maeda {et~al}\mbox{.}(2002)Maeda, Baganoff, Feigelson, Morris, Bautz,
  Brandt, Burrows, Doty, Garmire, Pravdo, Ricker, \&
  Townsley}]{2002ApJ...570..671M}
Maeda Y. {et~al.}, 2002, The Astrophysical Journal, 570, 671

\bibitem[{Markowitz {et~al}\mbox{.}(2007)Markowitz, Takahashi, Watanabe,
  Nakazawa, Fukazawa, Kokubun, Makishima, Awaki, Bamba, Isobe, Kataoka,
  Madejski, Mushotzky, Okajima, Ptak, Reeves, Ueda, Yamasaki, \&
  Yaqoob}]{2007ApJ...665..209M}
Markowitz A. {et~al.}, 2007, The Astrophysical Journal, 665, 209

\bibitem[{Mart{\'\i}n, Mart{\'\i}n-Pintado \& Viti(2009)Mart{\'\i}n,
  Mart{\'\i}n-Pintado, \& Viti}]{2009ApJ...706.1323M}
Mart{\'\i}n S., Mart{\'\i}n-Pintado J., Viti S., 2009, The Astrophysical
  Journal, 706, 1323

\bibitem[{Mart{\'\i}n {et~al}\mbox{.}(2006)Mart{\'\i}n, Mauersberger,
  Mart{\'\i}n-Pintado, Henkel, \& Garc{\'\i}a-Burillo}]{2006ApJS..164..450M}
Mart{\'\i}n S., Mauersberger R., Mart{\'\i}n-Pintado J., Henkel C.,
  Garc{\'\i}a-Burillo S., 2006, The Astrophysical Journal Supplement Series,
  164, 450

\bibitem[{Mart{\'\i}n {et~al}\mbox{.}(2008)Mart{\'\i}n, Requena-Torres,
  Mart{\'\i}n-Pintado, \& Mauersberger}]{2008ApJ...678..245M}
Mart{\'\i}n S., Requena-Torres M.~A., Mart{\'\i}n-Pintado J., Mauersberger R.,
  2008, The Astrophysical Journal, 678, 245

\bibitem[{Meier \& Turner(2005)}]{2005ApJ...618..259M}
Meier D.~S., Turner J.~L., 2005, The Astrophysical Journal, 618, 259

\bibitem[{Meier \& Turner(2012)}]{2012ApJ...755..104M}
Meier D.~S., Turner J.~L., 2012, The Astrophysical Journal, 755, 104

\bibitem[{Millar, Herbst \& Charnley(1991)Millar, Herbst, \&
  Charnley}]{1991ApJ...369..147M}
Millar T.~J., Herbst E., Charnley S.~B., 1991, Astrophysical Journal, 369, 147

\bibitem[{M{\"u}ller {et~al}\mbox{.}(2005)M{\"u}ller, Schl{\"o}der, Stutzki, \&
  Winnewisser}]{2005JMoSt.742..215M}
M{\"u}ller H. S.~P., Schl{\"o}der F., Stutzki J., Winnewisser G., 2005, Journal
  of Molecular Structure, 742, 215

\bibitem[{Peletier {et~al}\mbox{.}(1990)Peletier, Davies, Illingworth, Davis,
  \& Cawson}]{1990AJ....100.1091P}
Peletier R.~F., Davies R.~L., Illingworth G.~D., Davis L.~E., Cawson M., 1990,
  Astronomical Journal (ISSN 0004-6256), 100, 1091

\bibitem[{P{\'e}rez-Beaupuits, Aalto \& Gerebro(2007)P{\'e}rez-Beaupuits,
  Aalto, \& Gerebro}]{2007A&A...476..177P}
P{\'e}rez-Beaupuits J.~P., Aalto S., Gerebro H., 2007, Astronomy and
  Astrophysics, 476, 177

\bibitem[{Salaris \& Wei{\ss}(1998)}]{1998A&A...335..943S}
Salaris M., Wei{\ss} A., 1998, Astronomy and Astrophysics, 335, 943

\bibitem[{Shields \& Ferland(1994)}]{1994ApJ...430..236S}
Shields J.~C., Ferland G.~J., 1994, The Astrophysical Journal, 430, 236

\bibitem[{Thomas, Greggio \& Bender(1999)Thomas, Greggio, \&
  Bender}]{1999MNRAS.302..537T}
Thomas D., Greggio L., Bender R., 1999, Monthly Notices of the Royal
  Astronomical Society, 302, 537

\bibitem[{Thomas {et~al}\mbox{.}(2005)Thomas, Maraston, Bender, \& Mendes~de
  Oliveira}]{2005ApJ...621..673T}
Thomas D., Maraston C., Bender R., Mendes~de Oliveira C., 2005, The
  Astrophysical Journal, 621, 673

\bibitem[{Tremonti {et~al}\mbox{.}(2004)Tremonti, Heckman, Kauffmann,
  Brinchmann, Charlot, White, Seibert, Peng, Schlegel, Uomoto, Fukugita, \&
  Brinkmann}]{2004ApJ...613..898T}
Tremonti C.~A. {et~al.}, 2004, The Astrophysical Journal, 613, 898

\bibitem[{Viti {et~al}\mbox{.}(2001)Viti, Caselli, Hartquist, \&
  Williams}]{2001A&A...370.1017V}
Viti S., Caselli P., Hartquist T.~W., Williams D.~A., 2001, Astronomy and
  Astrophysics, 370, 1017

\bibitem[{Wilson \& Rood(1994)}]{1994ARA&A..32..191W}
Wilson T.~L., Rood R., 1994, Annual Review of Astronomy and Astrophysics, 32,
  191

\bibitem[{Young {et~al}\mbox{.}(2011)Young, Bureau, Davis, Combes, McDermid,
  Alatalo, Blitz, Bois, Bournaud, Cappellari, Davies, de~Zeeuw, Emsellem,
  Khochfar, Krajnovi{\'c}, Kuntschner, Lablanche, Morganti, Naab, Oosterloo,
  Sarzi, Scott, Serra, \& Weijmans}]{2011MNRAS.414..940Y}
Young L.~M. {et~al.}, 2011, Monthly Notices of the Royal Astronomical Society,
  688

\bibitem[{Zaritsky, Kennicutt \& Huchra(1994)Zaritsky, Kennicutt, \&
  Huchra}]{1994ApJ...420...87Z}
Zaritsky D., Kennicutt R. C.~J., Huchra J.~P., 1994, Astrophysical Journal,
  420, 87

\end{thebibliography}
\bibdata{bibCSMETH}
\bibstyle{mn2e}

\label{lastpage}

\clearpage
\appendix

 \begin{figure}
 \begin{minipage}[b!]{\textwidth}
 \section{Detected Lines}
\begin{center}
\subfigure[NGC4526]{\includegraphics[height=14cm,angle=0,clip,trim=0.0cm 0cm 0cm 0.0cm]{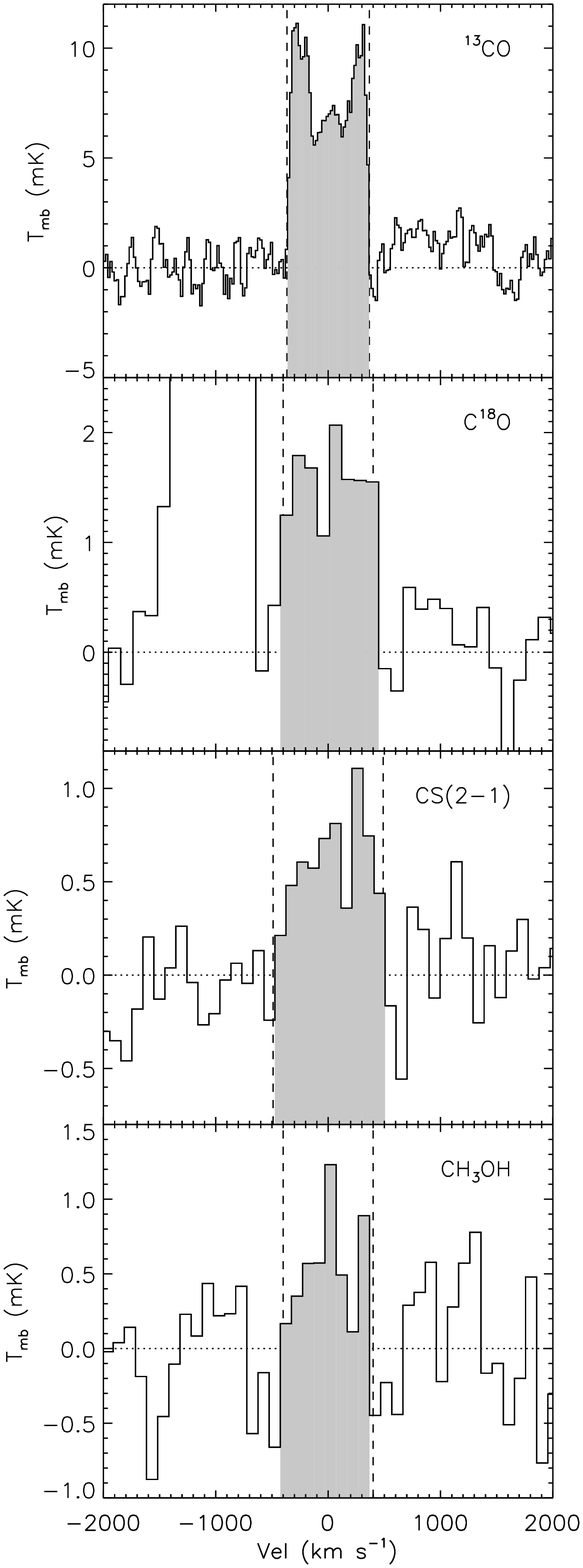}}
\subfigure[NGC4710]{\includegraphics[height=14cm,angle=0,clip,trim=0.0cm 0cm 0cm 0.0cm]{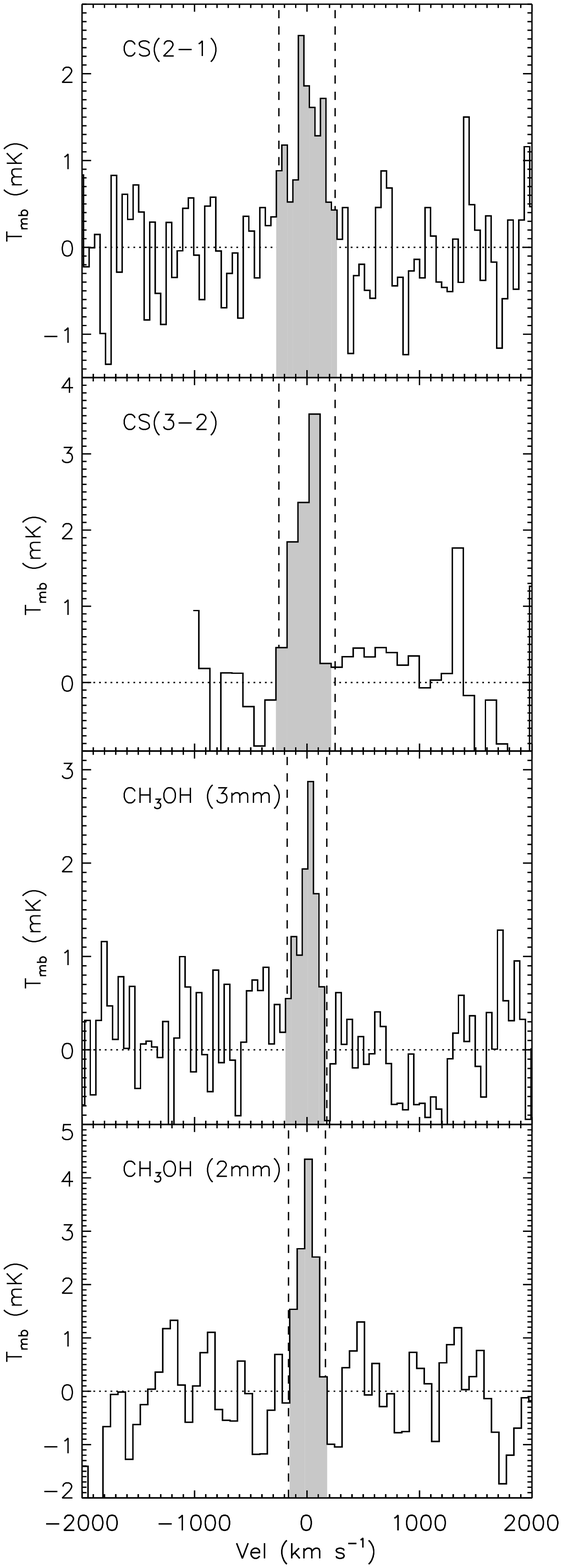}}
\subfigure[NGC5866]{\includegraphics[height=14cm,angle=0,clip,trim=0.0cm 0cm 0cm 0.0cm]{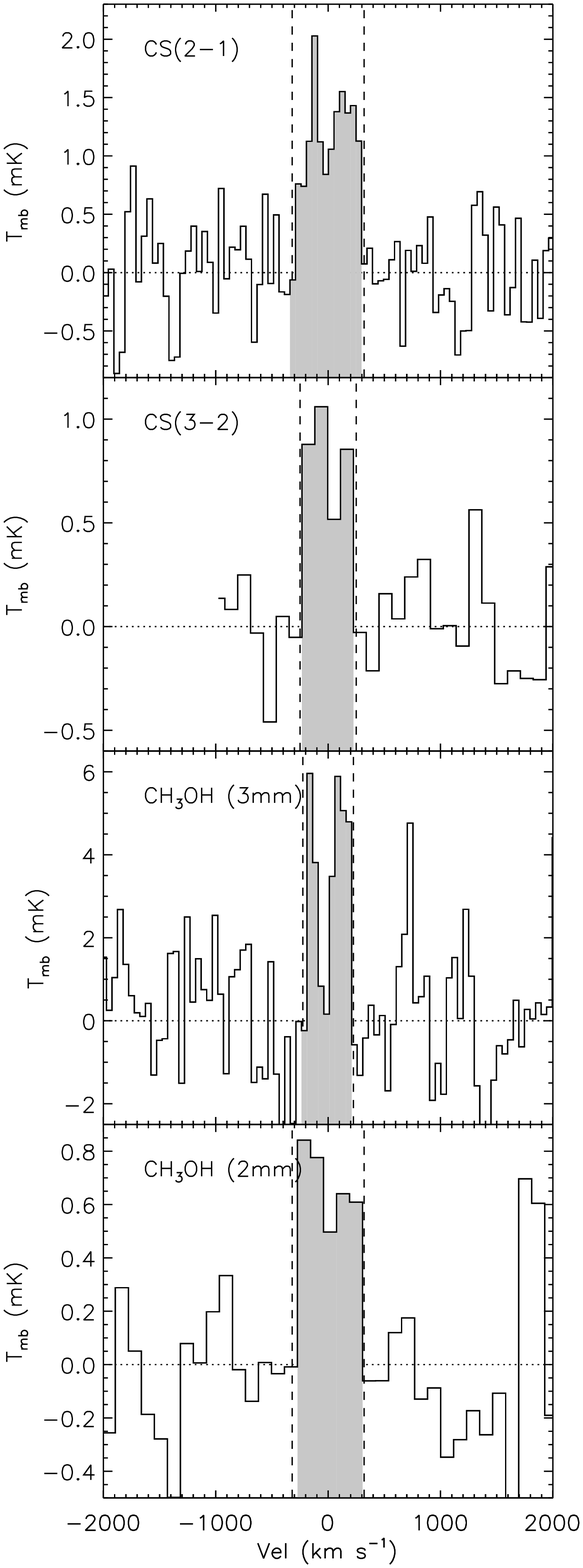}}
 \end{center}
\caption{Molecular line detections reported in this paper, from our IRAM-30m observations. The observed transition is indicated in the figure legend. The x-axis shows velocity offset from the line centre, corrected for the systemic velocity of the galaxy (listed in Table \ref{obssamp_table}). For NGC4526 the plotted $^{13}$CO detection has a channel width of 20 \kms, and the other lines $\approx$100 \kms. The plots for NGC4710 and NGC5866 use channel widths of $\approx$50 \kms\ at 3mm and $\approx$100 \kms\ at 2mm.  The intensity is in units of main beam temperature (T$_{\rm mb}$). A short-dashed line shows the zero level, and long-dashes shown the derived velocity width.}
\label{spec2}
 \end{minipage}
 \end{figure}

 \clearpage
  \begin{figure}
\begin{center}
\subfigure[NGC6014]{\includegraphics[height=10.5cm,angle=0,clip,trim=0.0cm 0cm 0cm 0.0cm]{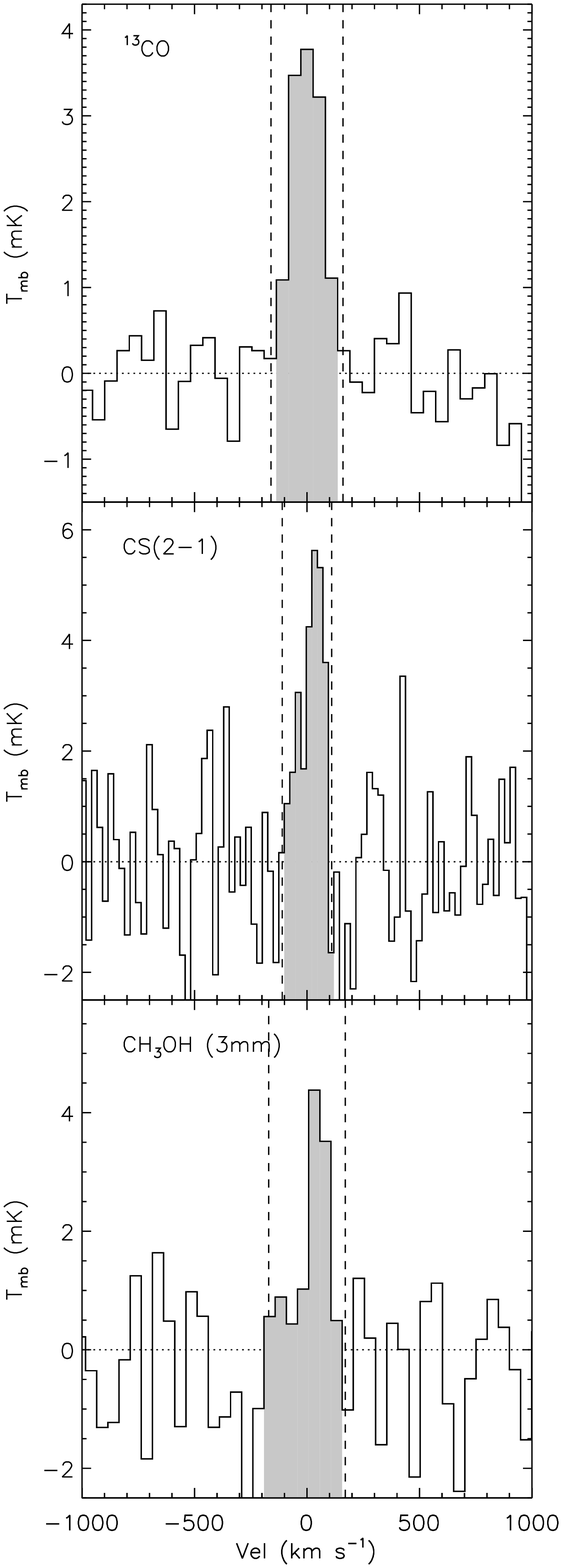}}
\subfigure[UGC09519]{\includegraphics[height=10.5cm,angle=0,clip,trim=0.0cm 0cm 0cm 0.0cm]{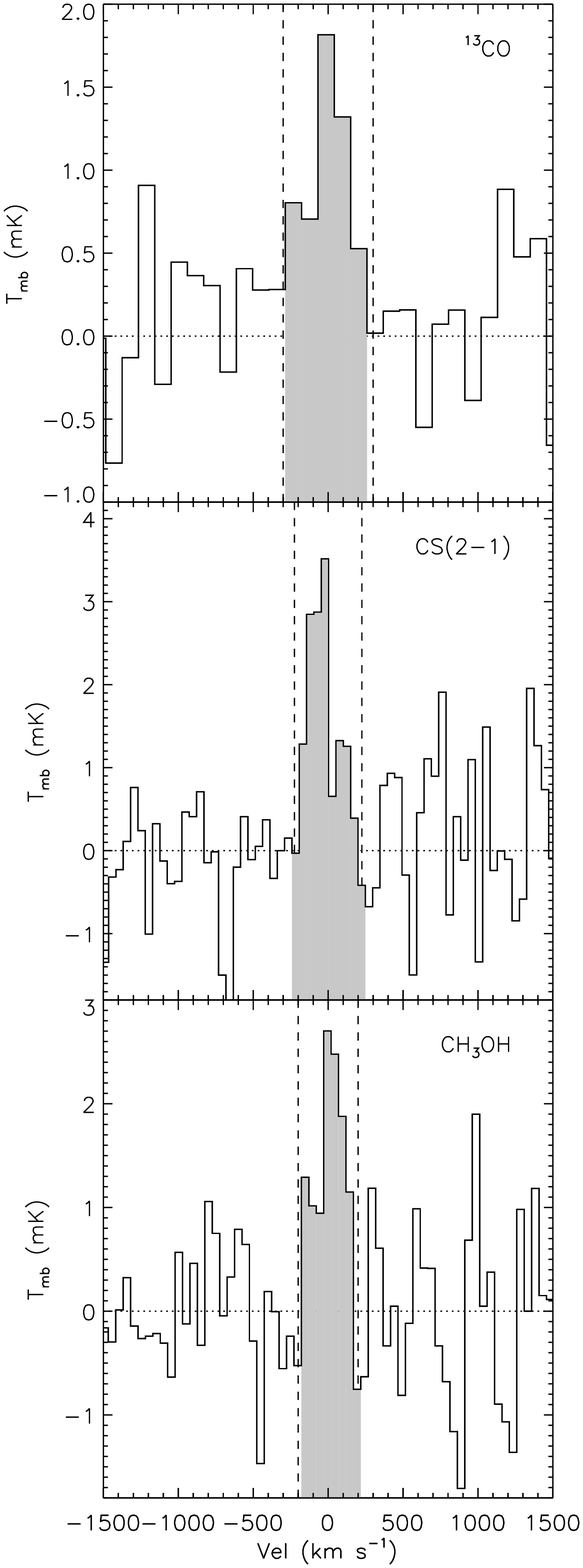}}
\contcaption{:- The plots for NGC6014 used a channel width of $\approx$25 \kms\ for CS(2-1), and $\approx$50 \kms\ for the other lines. The plots for UGC09519 used a channel width of $\approx$50 \kms\ for $^{13}$CO(1-0), and $\approx$25 \kms\ for the other lines. }{}
 \end{center}

 \end{figure}

\begin{figure}
\section{Rotation Diagrams}
\begin{center}
\includegraphics[width=0.23\textwidth,angle=0,clip,trim=1.0cm 0cm 0.8cm 1.0cm]{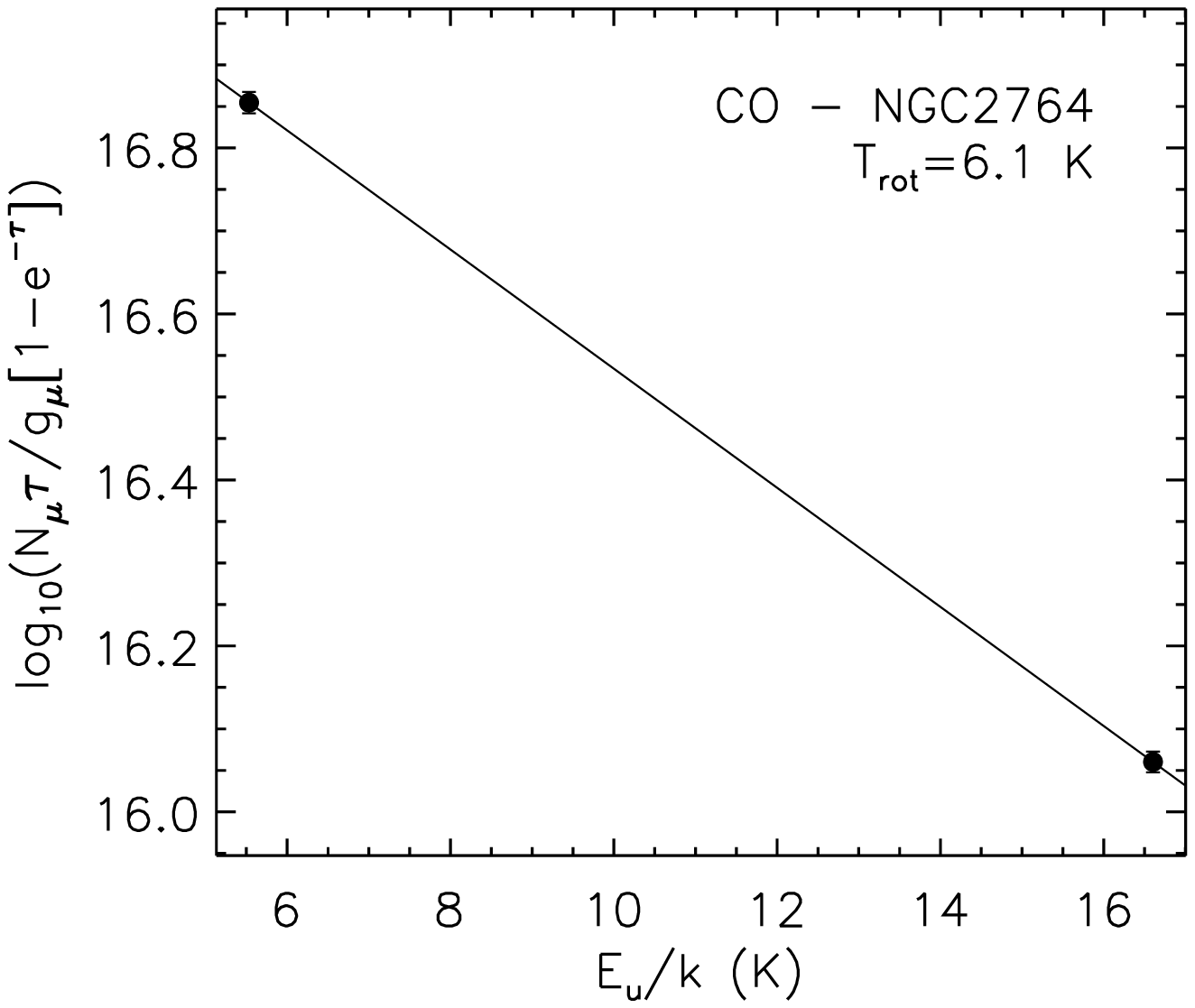}
\includegraphics[width=0.23\textwidth,angle=0,clip,trim=1.0cm 0cm 0.8cm 1.0cm]{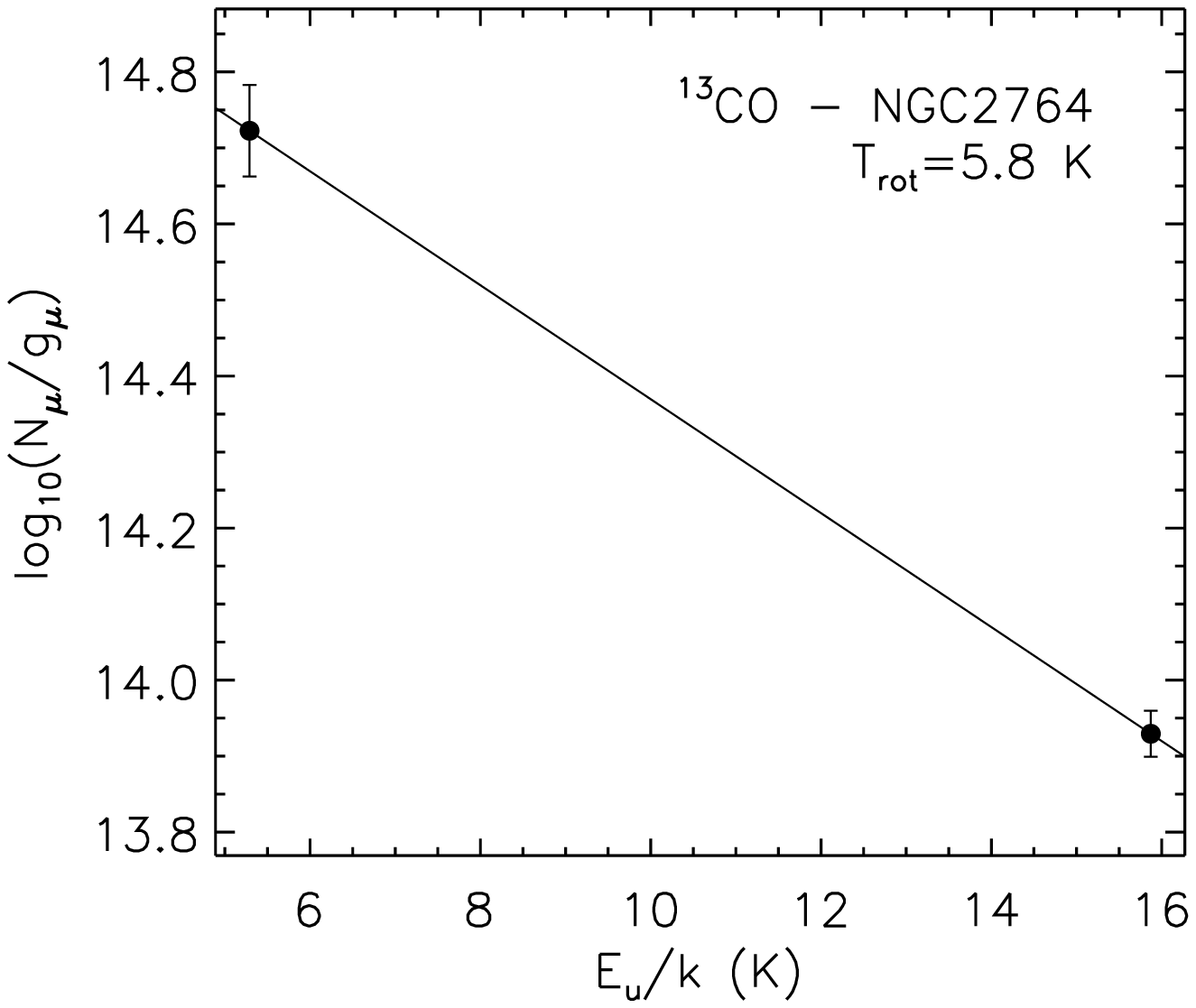}
 \end{center}
 \caption{Rotation diagrams for NGC\,2764. The energy of the upper level is plotted against the derived column density in the upper level for each observed transition (black points), as tabulated in Table \ref{obstable}. These data points are fitted with a line and the best fit, corresponding to the rotation temperature indicated in the legend, is displayed (solid line). }
 \label{NGC2764rotdiag}
 \end{figure}

\begin{figure}
\begin{center}
\includegraphics[width=0.23\textwidth,angle=0,clip,trim=1.0cm 0cm 0.8cm 1.0cm]{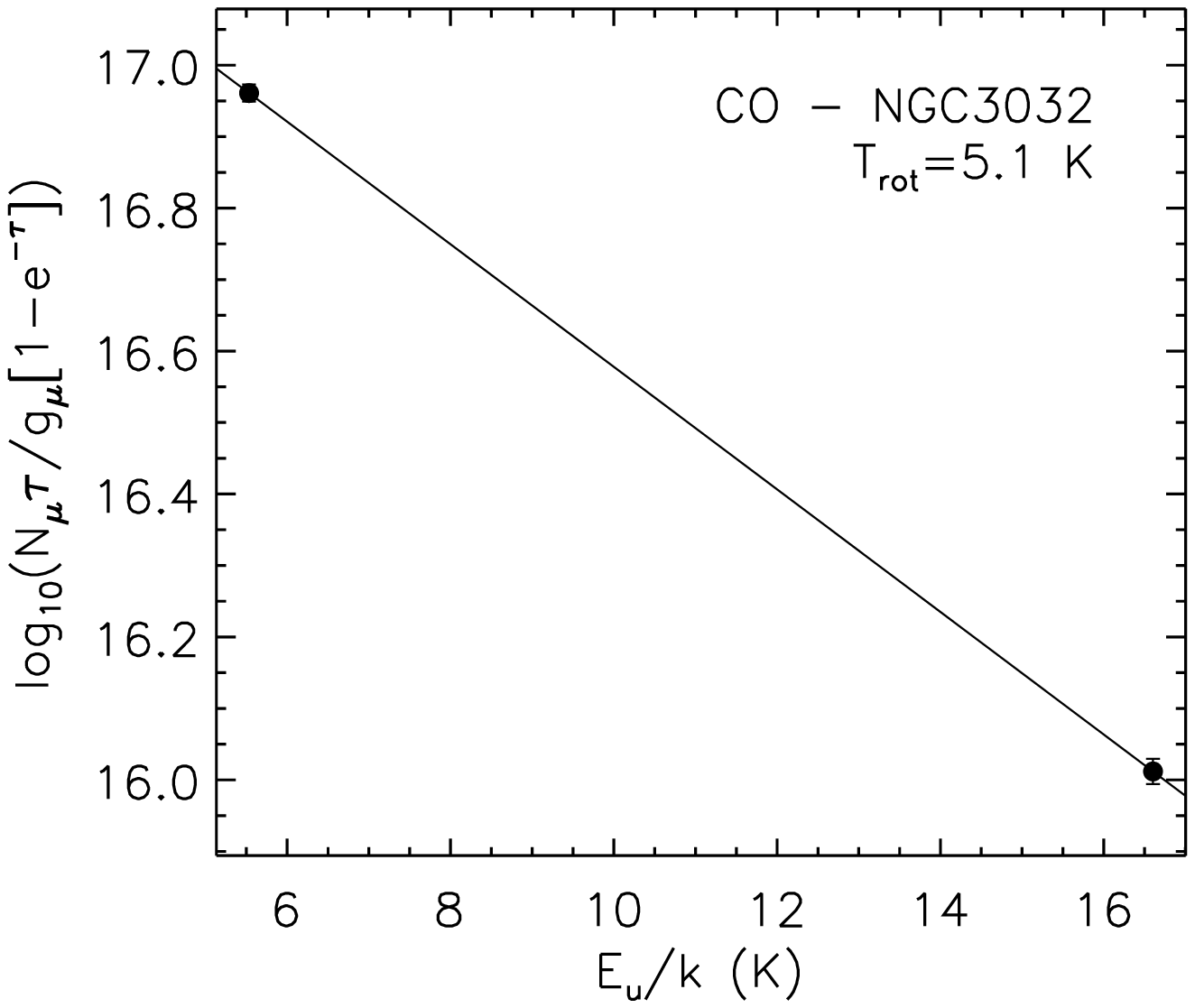}
\includegraphics[width=0.23\textwidth,angle=0,clip,trim=1.0cm 0cm 0.8cm 1.0cm]{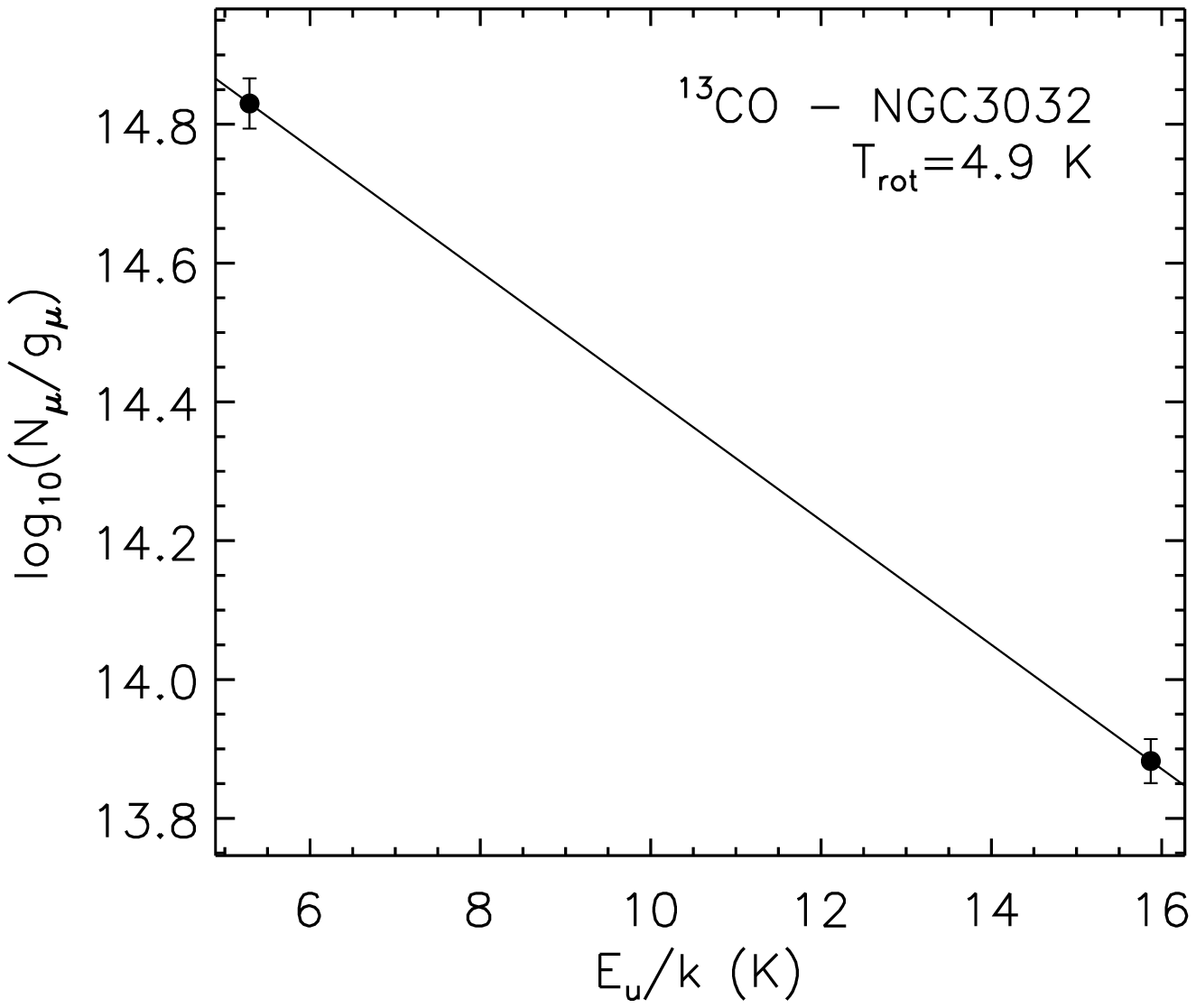}\\
\includegraphics[width=0.23\textwidth,angle=0,clip,trim=1.0cm 0cm 0.8cm 1.0cm]{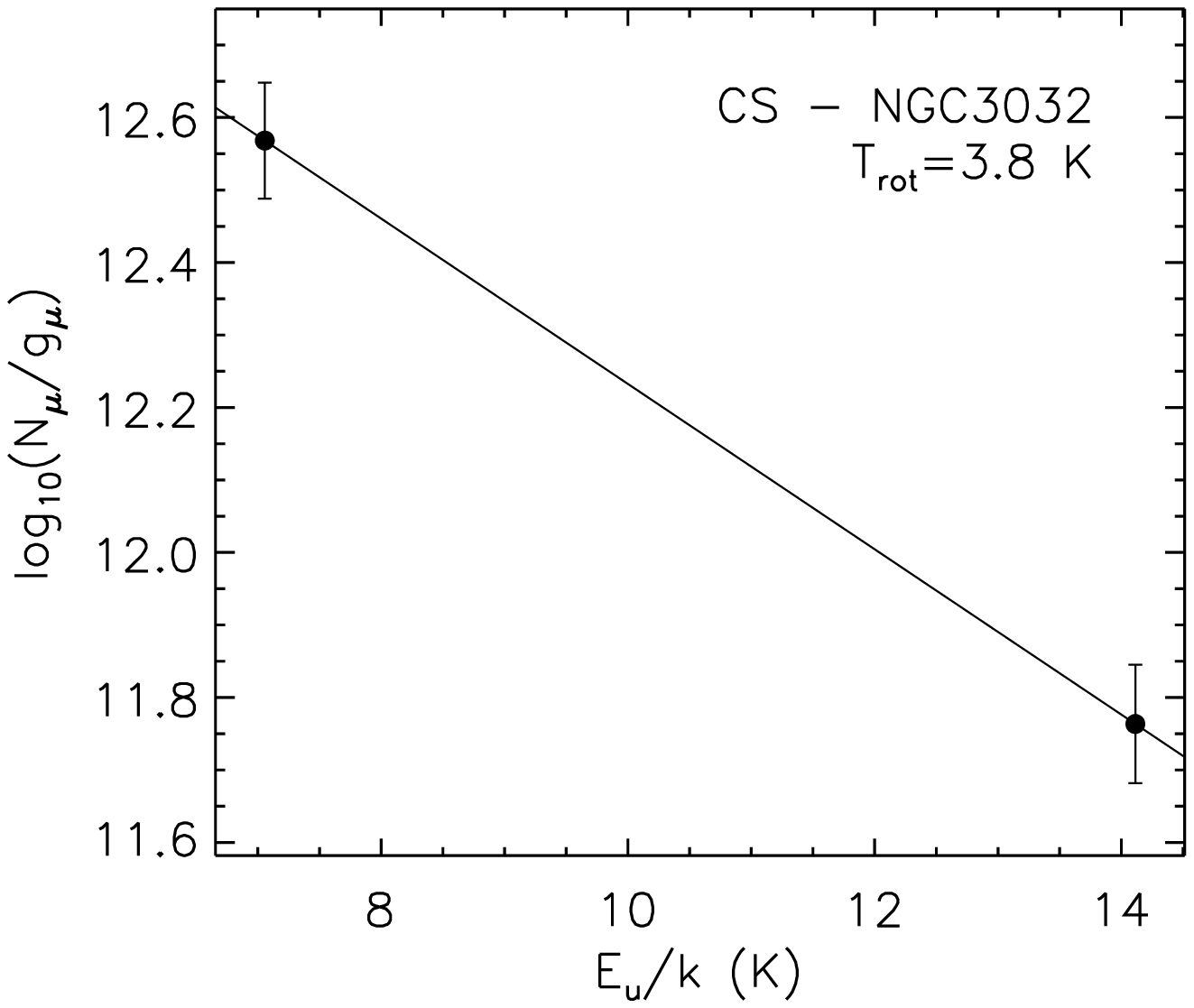}
 \end{center}
 %\caption{Rotation diagrams for NGC\,3032, details as in Figure \ref{NGC2764rotdiag}.}
 \contcaption{:- Rotation diagrams for NGC\,3032}{}
 \label{NGC3032rotdiag}
 \end{figure}

\begin{figure}
\begin{center}
\includegraphics[width=0.23\textwidth,angle=0,clip,trim=1.0cm 0cm 0.8cm 1.0cm]{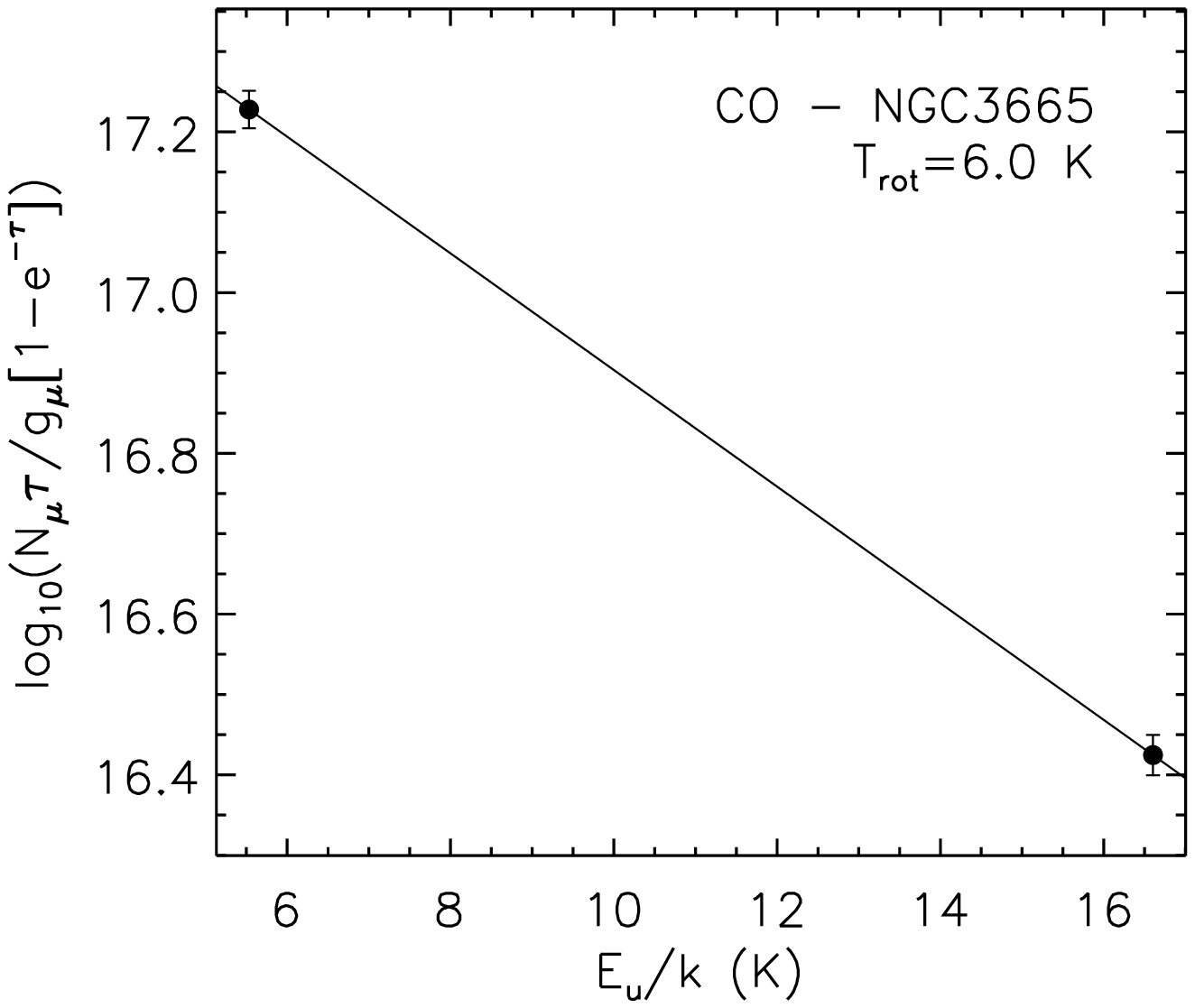}
\includegraphics[width=0.23\textwidth,angle=0,clip,trim=1.0cm 0cm 0.8cm 1.0cm]{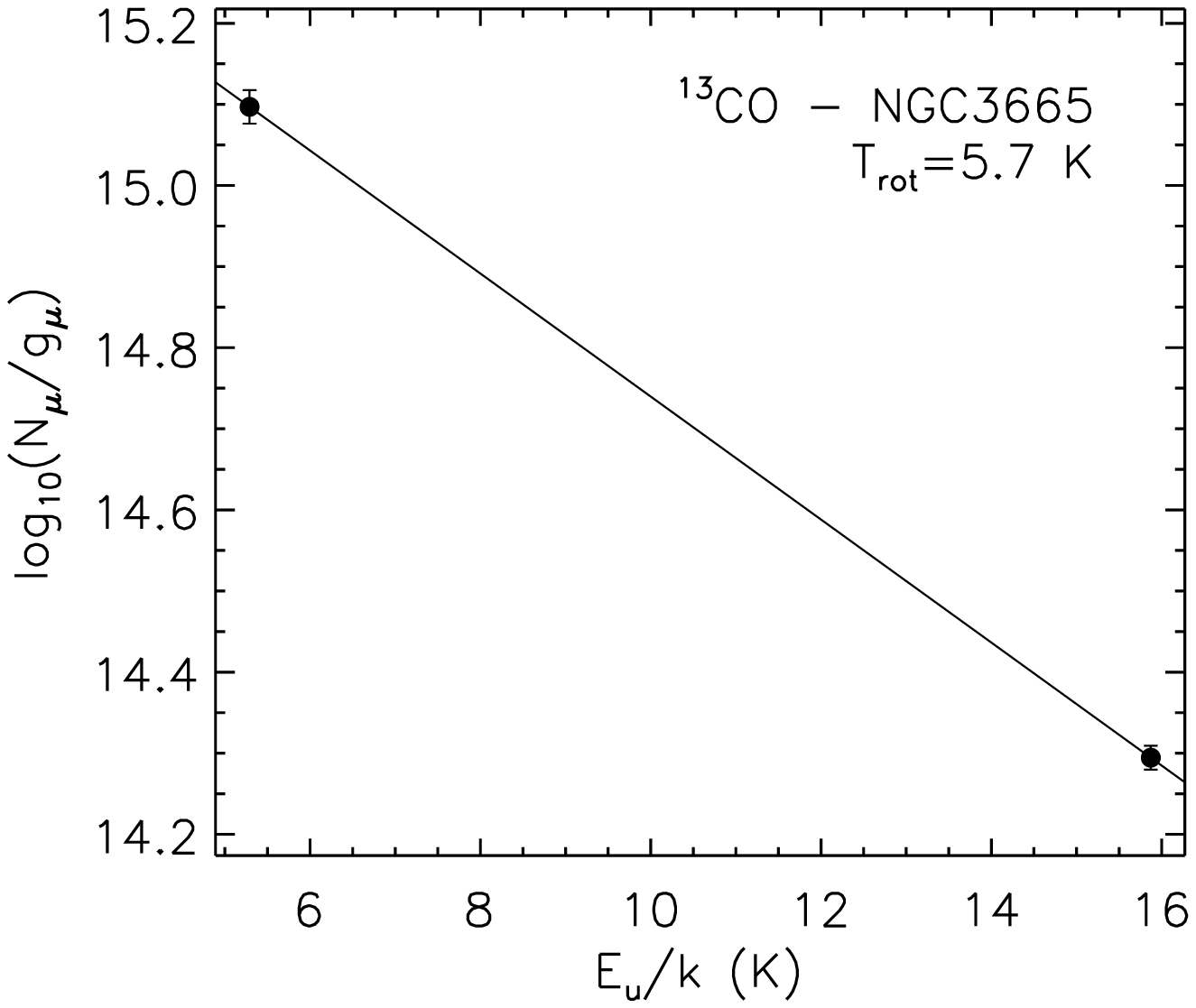}
 \end{center}
 \contcaption{:- Rotation diagrams for NGC\,3665}{}
 \label{NGC3665rotdiag}
 \end{figure}

\begin{figure}
\begin{center}
\includegraphics[width=0.23\textwidth,angle=0,clip,trim=1.0cm 0cm 0.8cm 1.0cm]{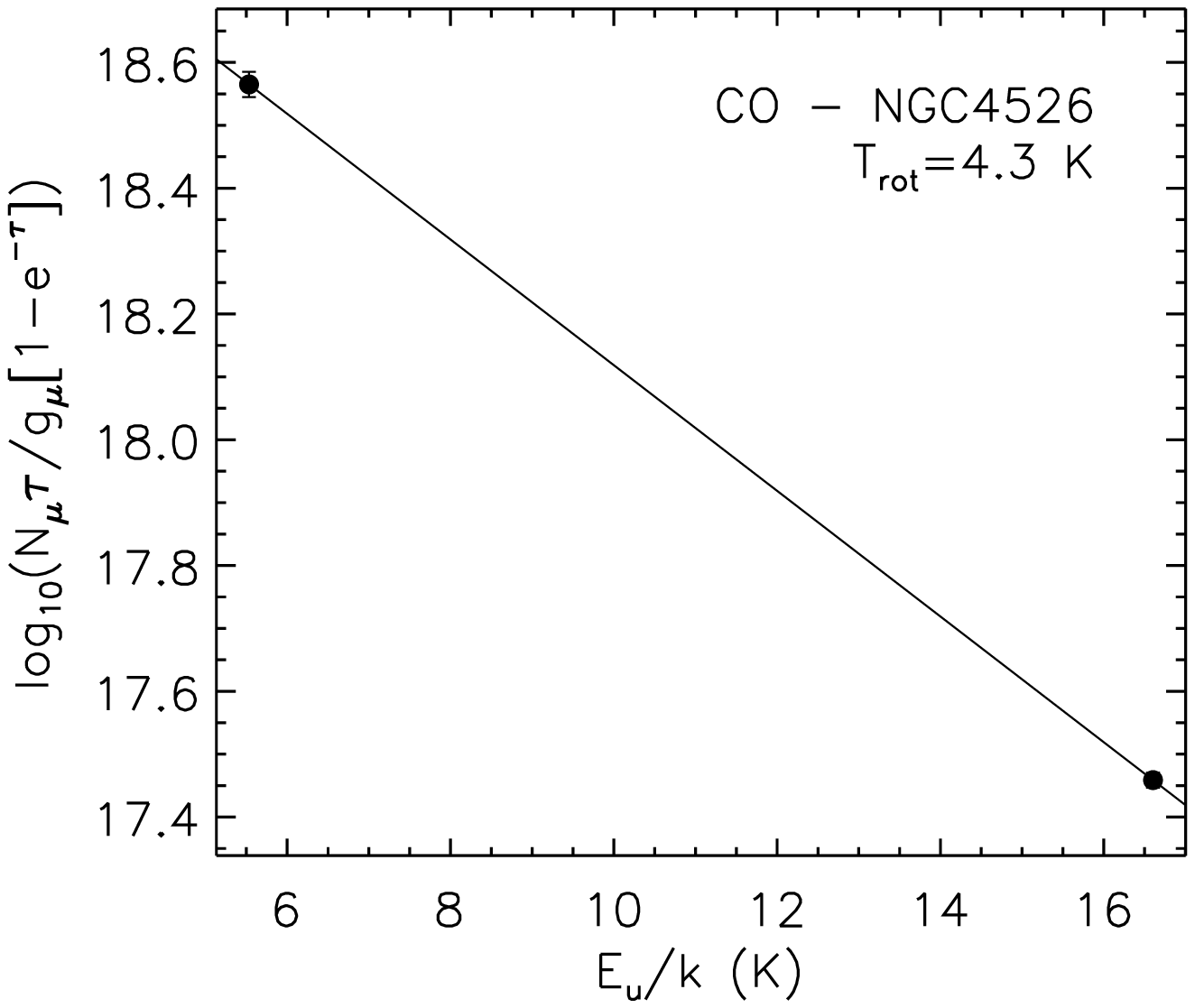}
\includegraphics[width=0.23\textwidth,angle=0,clip,trim=1.0cm 0cm 0.8cm 1.0cm]{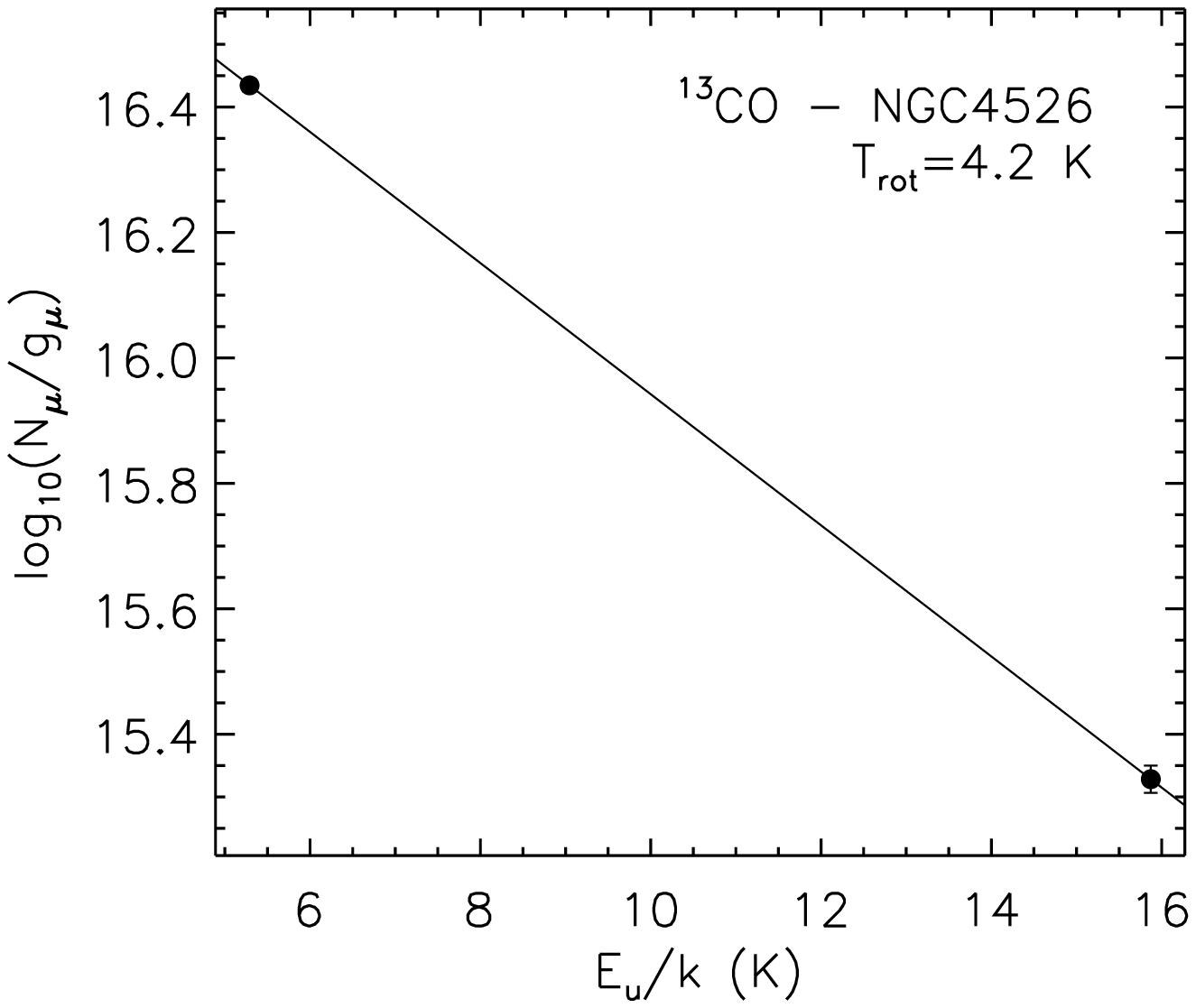}
 \end{center}
% \caption{Rotation diagrams for NGC\,4526, details as in Figure \ref{NGC2764rotdiag}.}
  \contcaption{:- Rotation diagrams for NGC\,4526}{}
 \label{NGC4526rotdiag}
 \end{figure}
 
 \begin{figure}
\begin{center}
\includegraphics[width=0.23\textwidth,angle=0,clip,trim=1.0cm 0cm 0.8cm 1.0cm]{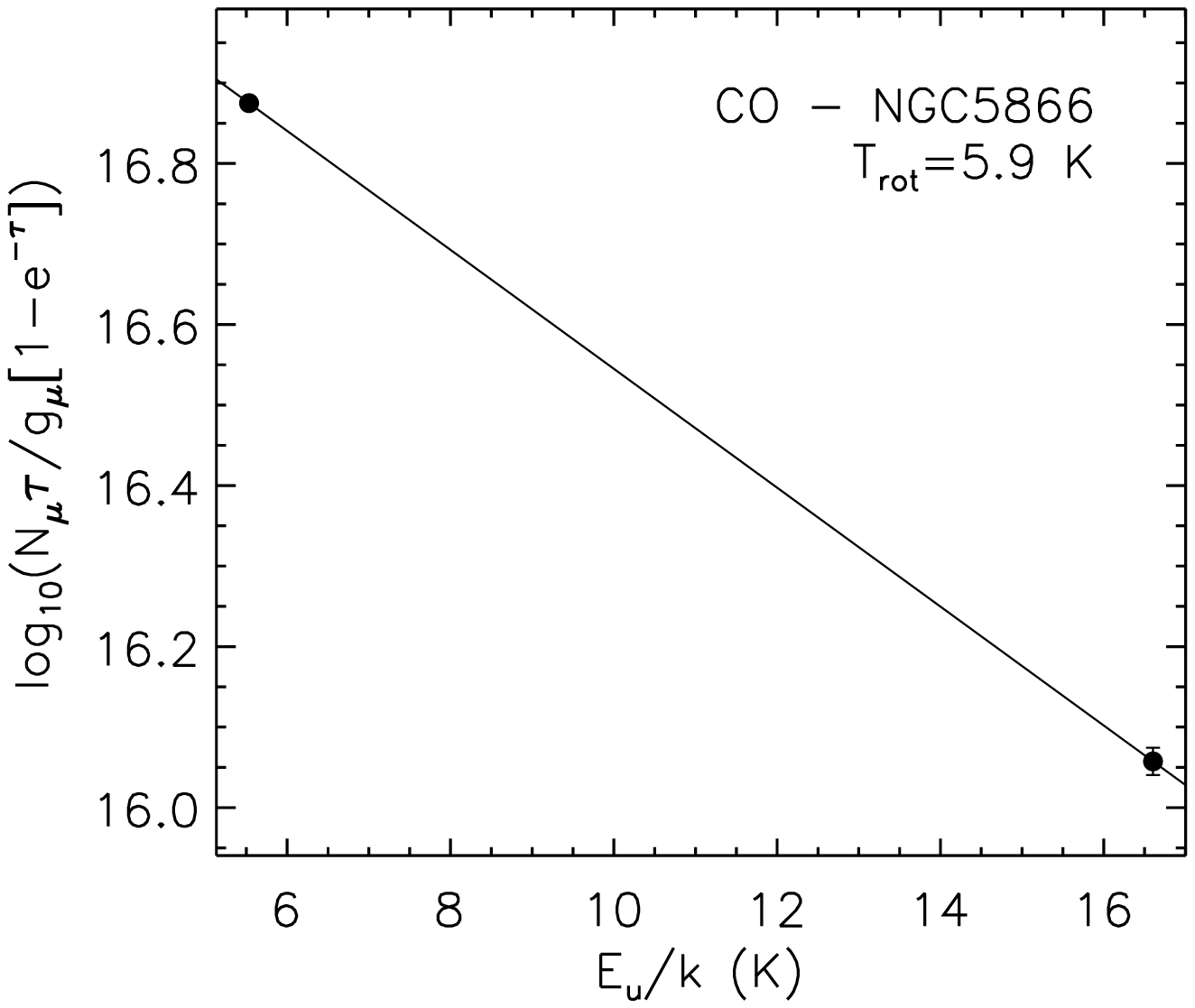}
\includegraphics[width=0.23\textwidth,angle=0,clip,trim=1.0cm 0cm 0.8cm 1.0cm]{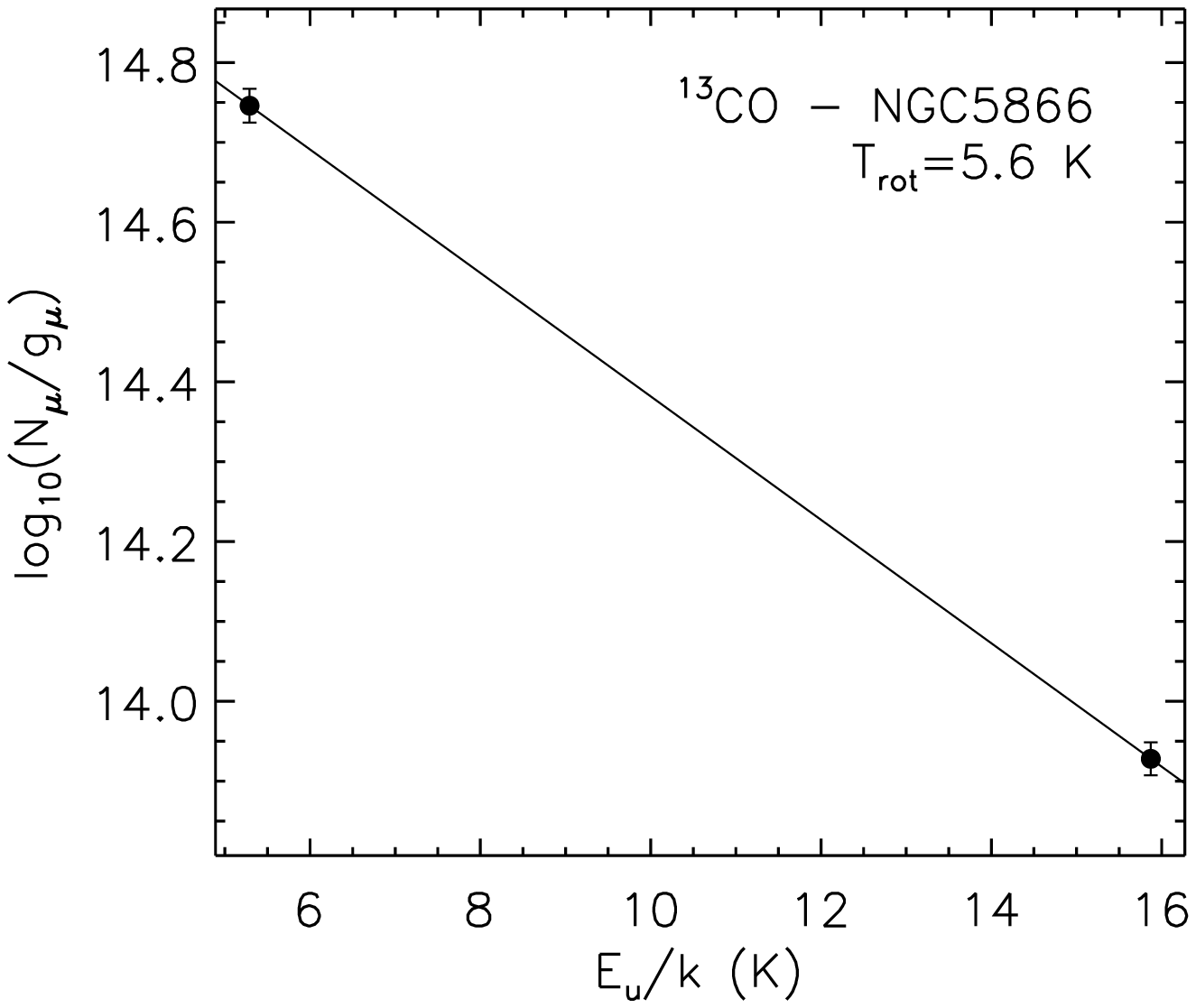}
\includegraphics[width=0.23\textwidth,angle=0,clip,trim=1.0cm 0cm 0.8cm 1.0cm]{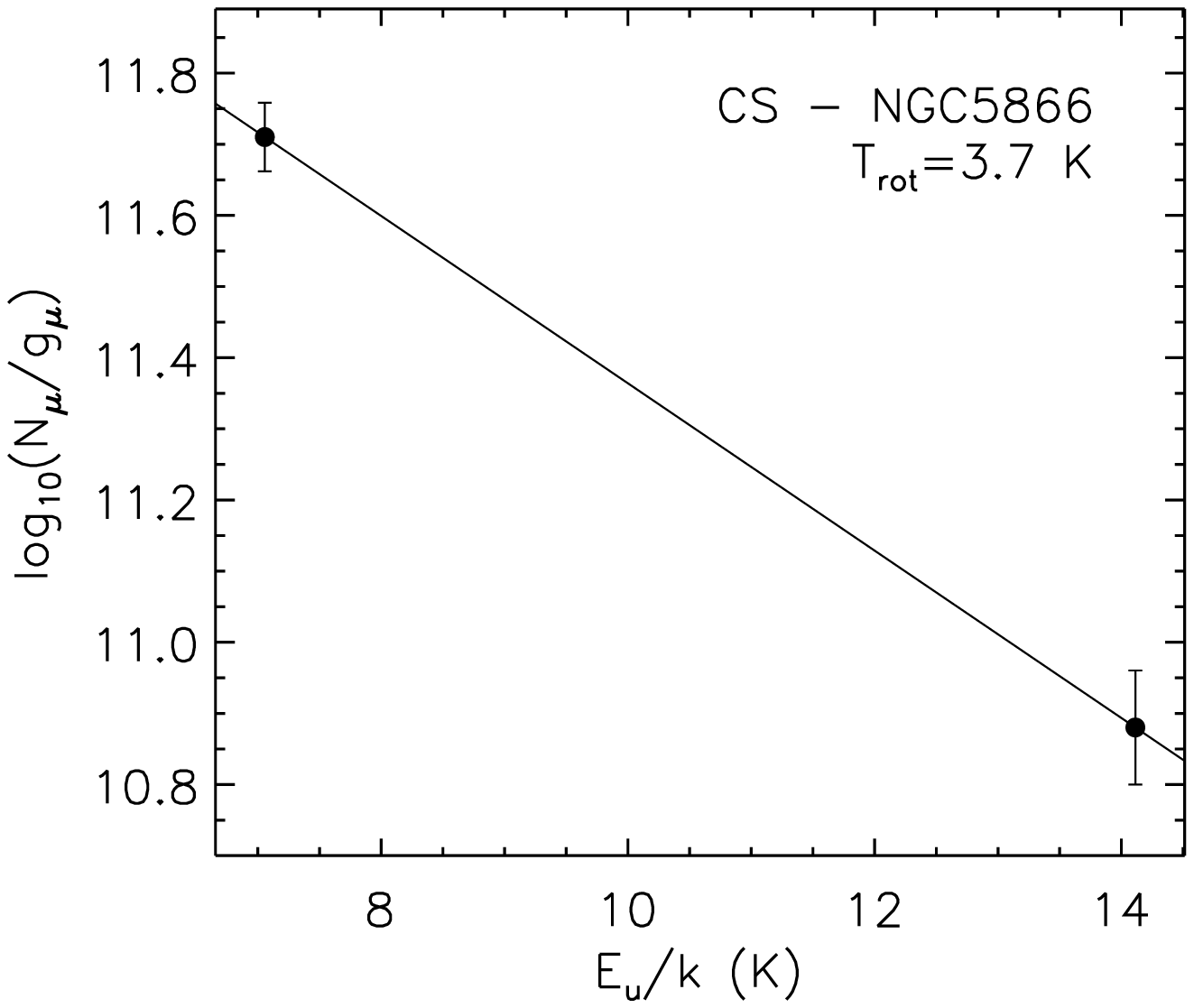}
\includegraphics[width=0.23\textwidth,angle=0,clip,trim=1.0cm 0cm 0.8cm 1.0cm]{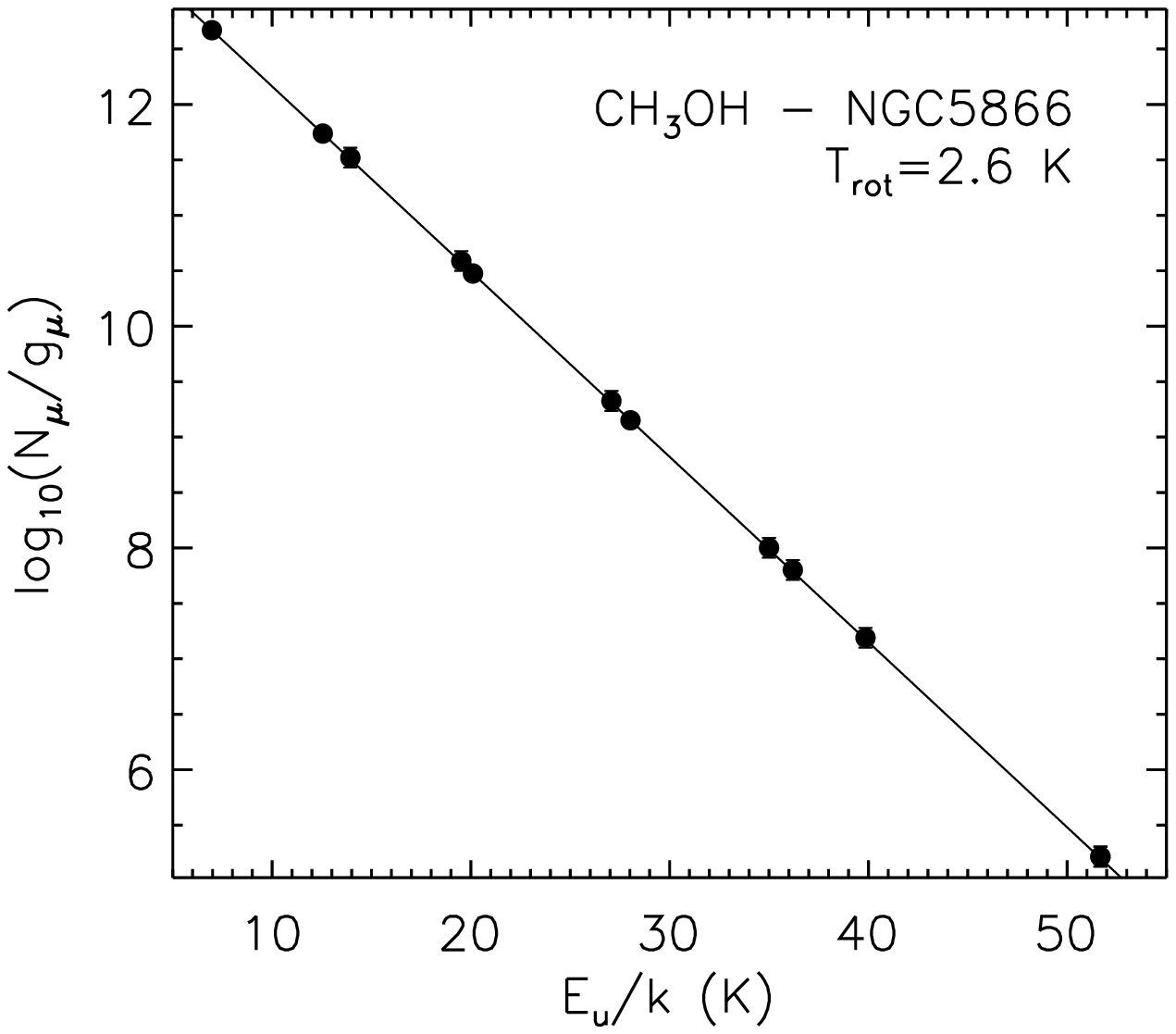}
 \end{center}
% \caption{Rotation diagrams for NGC\,5866, details as in Figure \ref{NGC2764rotdiag}. We split methanol blends statistically, as described in Section \ref{dataredux}.}
  \contcaption{:- Rotation diagrams for NGC\,5866. We split methanol blends statistically, as described in Section \ref{dataredux}.}{}
 \label{NGC5866rotdiag}
 \end{figure}

\begin{figure}
\begin{center}
\includegraphics[width=0.23\textwidth,angle=0,clip,trim=1.0cm 0cm 0.8cm 1.0cm]{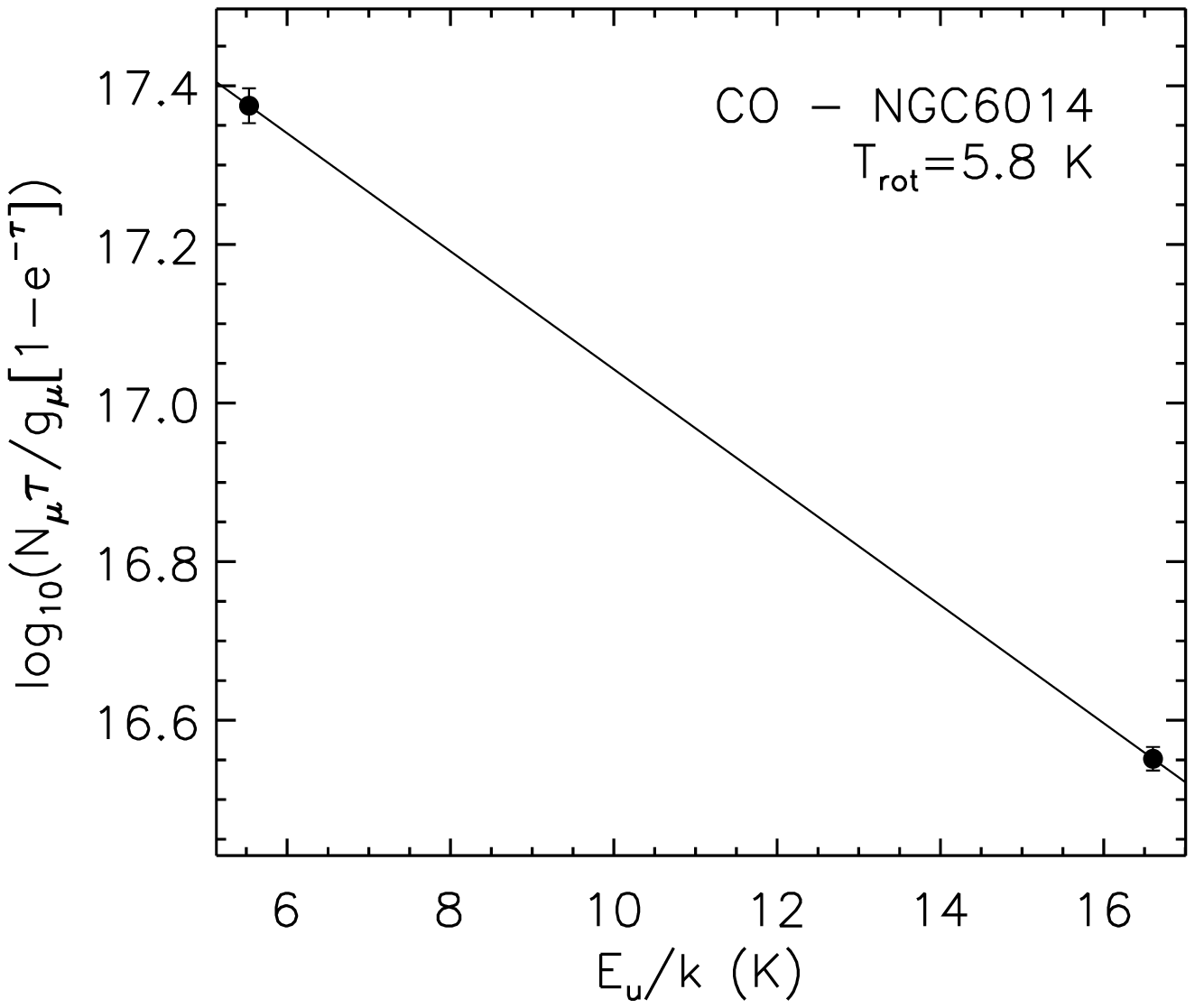}
\includegraphics[width=0.23\textwidth,angle=0,clip,trim=1.0cm 0cm 0.8cm 1.0cm]{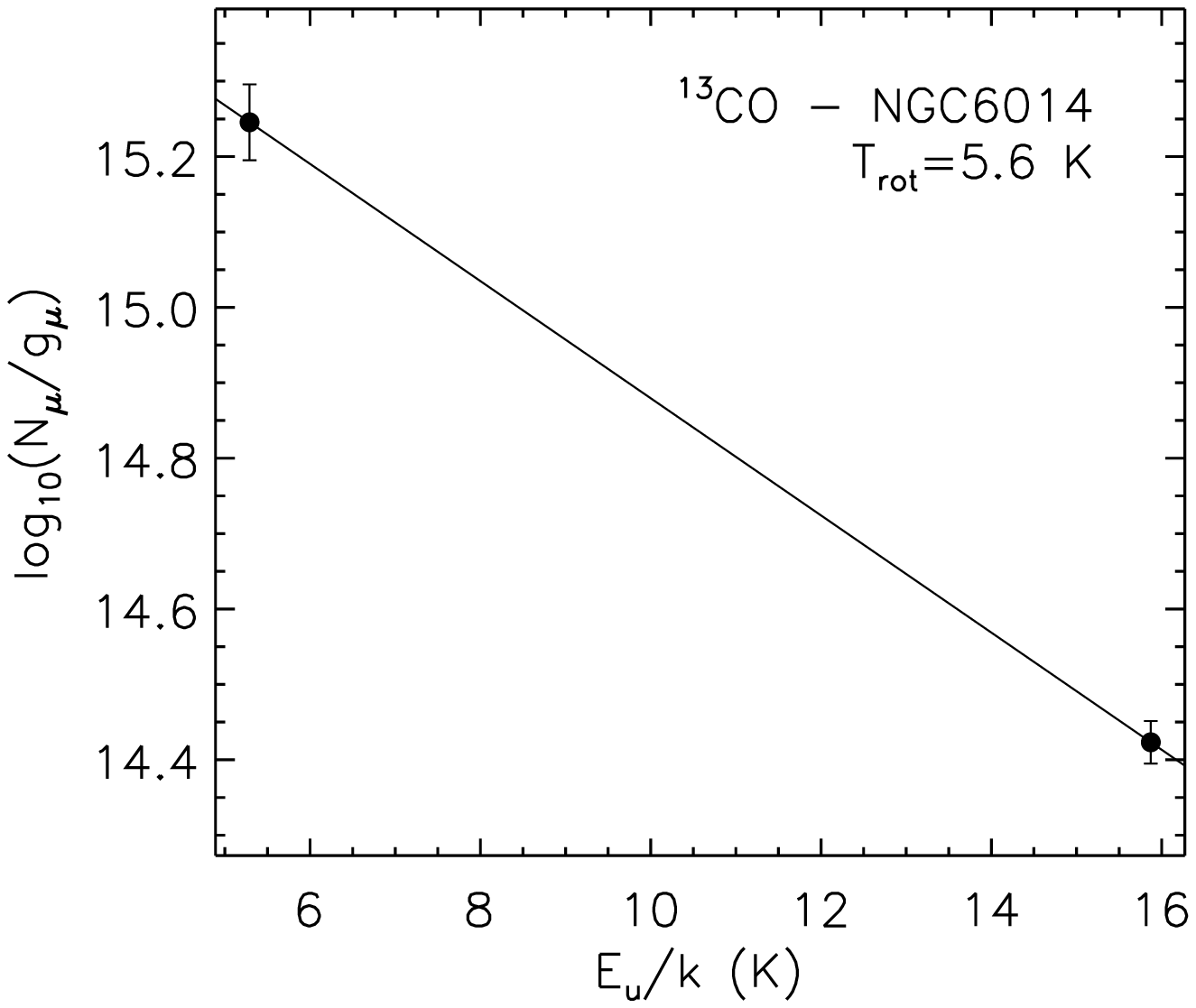}
 \end{center}
% \caption{Rotation diagrams for NGC\,6014, details as in Figure \ref{NGC2764rotdiag}.}
  \contcaption{:- Rotation diagrams for NGC\,6014}{}
 \label{NGC6014rotdiag}
 \end{figure}

\begin{figure}
\begin{center}
\includegraphics[width=0.23\textwidth,angle=0,clip,trim=1.0cm 0cm 0.8cm 1.0cm]{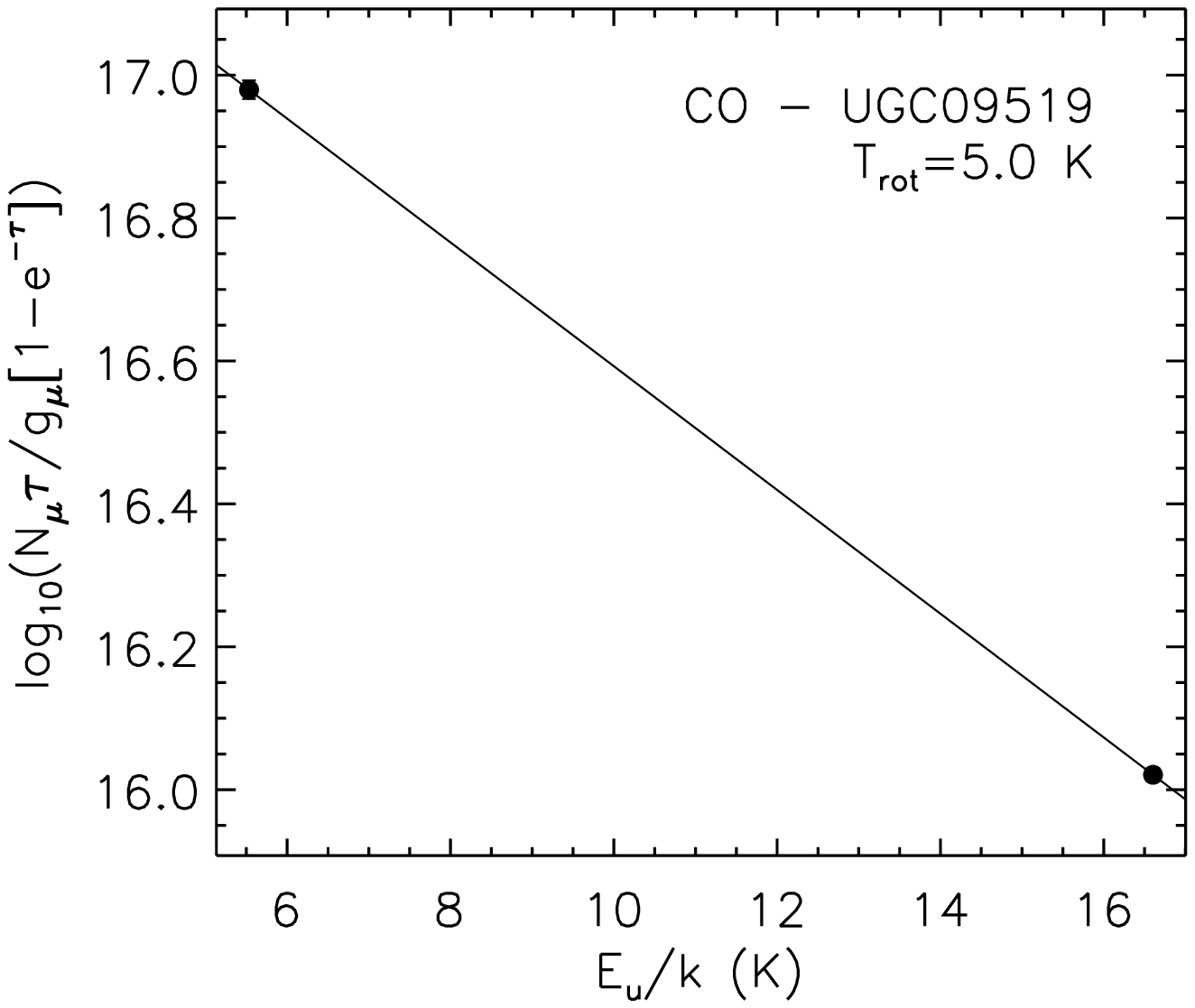}
\includegraphics[width=0.23\textwidth,angle=0,clip,trim=1.0cm 0cm 0.8cm 1.0cm]{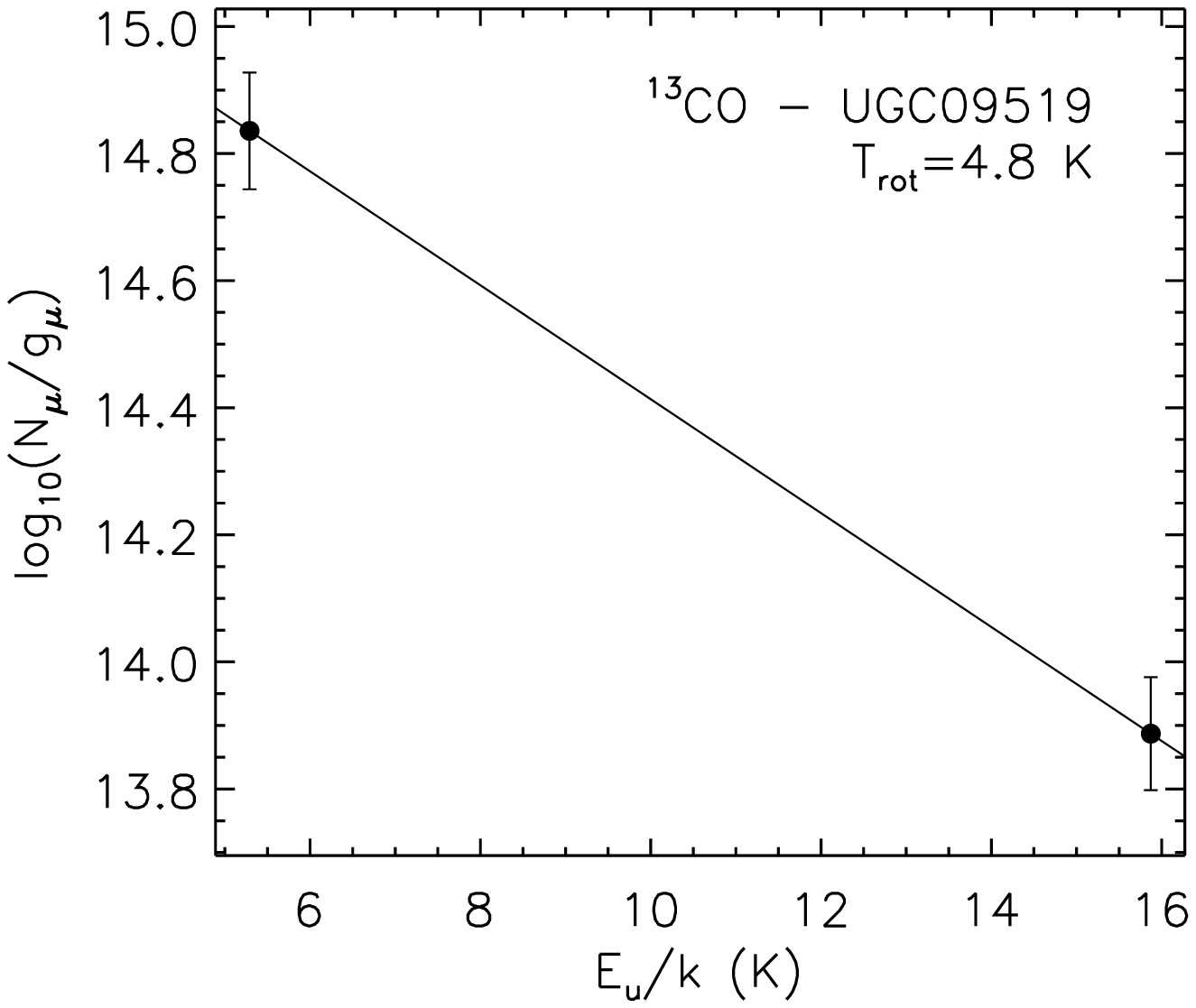}
 \end{center}
% \caption{Rotation diagrams for UGC\,09519, details as in Figure \ref{NGC2764rotdiag}.}
 \contcaption{:- Rotation diagrams for UGC\,09519}{}
 \label{UGCrotdiag}
 \end{figure}

\end{document}